%% file: 2-point_correlators_arXiV.tex
\documentclass[paper,11pt]{JHEP}
\usepackage[centertags]{amsmath}
\usepackage{cite}
\usepackage{amsfonts} \usepackage{amssymb} \usepackage{amsthm}
\allowdisplaybreaks[1]
\usepackage{graphicx}
 \usepackage{slashed}
 \usepackage{latexsym}
\usepackage{feynmp-auto}
\usepackage{bm}
\newcommand{\eq}[1]{Eq.~(\ref{#1})}

\newcommand{\ii}{\mathrm{i}}
\newcommand{\rme}{\mathrm{e}}

\newcommand{\vev}[1]{\langle #1 \rangle}

\newcommand{\tr}{\mathrm{tr}\,}

\newcommand{\cN}{{\mathcal{N}}}

\newcommand{\one}{{\rm 1\kern -.9mm l}}

\newcommand{\vphi}{\bm{\varphi}}

\newdimen\tableauside\tableauside=1.0ex
\newdimen\tableaurule\tableaurule=0.4pt
\newdimen\tableaustep
\def\phantomhrule#1{\hbox{\vbox to0pt{\hrule height\tableaurule
width#1\vss}}}
\def\phantomvrule#1{\vbox{\hbox to0pt{\vrule width\tableaurule
height#1\hss}}}
\def\sqr{\vbox{%
  \phantomhrule\tableaustep
\hbox{\phantomvrule\tableaustep\kern\tableaustep\phantomvrule\tableaustep}%
  \hbox{\vbox{\phantomhrule\tableauside}\kern-\tableaurule}}}
\def\squares#1{\hbox{\count0=#1\noindent\loop\sqr
  \advance\count0 by-1 \ifnum\count0>0\repeat}}
\def\tableau#1{\vcenter{\offinterlineskip
  \tableaustep=\tableauside\advance\tableaustep by-\tableaurule
  \kern\normallineskip\hbox
    {\kern\normallineskip\vbox
      {\gettableau#1 0 }%
     \kern\normallineskip\kern\tableaurule}%
  \kern\normallineskip\kern\tableaurule}}
\def\gettableau#1 {\ifnum#1=0\let\next=\null\else
  \squares{#1}\let\next=\gettableau\fi\next}
\newcommand{\nordg}[1]{:\! #1 \!:_g} 
\newcommand{\nord}[1]{:\! #1 \!:}

\def\XXint#1#2#3{{\setbox0=\hbox{$#1{#2#3}{\int}$}
     \vcenter{\hbox{$#2#3$}}\kern-.5\wd0}}

\def\beq{\begin{equation}}
\def\eeq{\end{equation}}
\def\be{\begin{equation}}
\def\be{\begin{equation}}
\def\ee{\end{equation}}
\def\ba{\begin{eqnarray}}
\def\ea{\end{eqnarray}}
\def\bea{\begin{eqnarray}}
\def\eea{\end{eqnarray}}
\def\eq{\begin{equation}}
\def\eqe{\end{equation}}
\def\eqa{\begin{eqnarray}}
\def\eqae{\end{eqnarray}}

\def\beqa{\begin{eqnarray}}
\def\eeqa{\end{eqnarray}}

\def\esp{\mathrm{e}}

\newcommand{\beqas}{\begin{eqnarray*}}
\newcommand{\eeqas}{\end{eqnarray*}}
\def\eeq{\end{equation}}
\def\be{\begin{equation}}
\def\ee{\end{equation}}
\def\ba{\begin{eqnarray}}
\def\ea{\end{eqnarray}}
\def\bea{\begin{eqnarray}}
\def\eea{\end{eqnarray}}
\def\eq{\begin{equation}}
\def\eqe{\end{equation}}
\def\eqa{\begin{eqnarray}}
\def\eqae{\end{eqnarray}}
\def\nn{\nonumber}
\def\beqa{\begin{eqnarray}}
\def\eeqa{\end{eqnarray}}

\title{\boldmath Two-point Correlators in $\mathcal{N}=2$ Gauge Theories}
\author{M. Bill\'o$^{1}$, F. Fucito$^{2}$, A. Lerda$^{3}$, J.F. Morales$^{2}$, 
Ya.S. Stanev$^{2}$, Congkao Wen$^{4,5}$
\\
\vskip 0.2cm
$^1$ Universit\`a di Torino, Dipartimento di Fisica \\
and I.N.F.N. - sezione di Torino, Via P. Giuria 1, I-10125 Torino, Italy
\vskip 0.2cm
$^2$ I.N.F.N - sezione di Roma Tor Vergata\\
Via della Ricerca Scientifica, I-00133 Roma, Italy
\vskip 0.2cm
$^3$Universit\`a del Piemonte Orientale, Dipartimento di Scienze e Innovazione Tecnologica \\
and I.N.F.N. - sezione di Torino, Via P. Giuria 1, I-10125 Torino, Italy\\
\vskip 0.2cm
$^4$ Walter Burke Institute for Theoretical Physics\\
California Institute of Technology, Pasadena, CA  91125, U.S.A.
 \vskip 0.2cm
$^5$ Mani L. Bhaumik Institute for Theoretical Physics\\
Department of Physics and Astronomy, UCLA, Los Angeles, CA 90095, U.S.A.\\
\vspace{0.35cm}
\email{billo,lerda@to.infn.it; fucito,morales,stanev@roma2.infn.it; cwen@caltech.edu}
}
\abstract{We consider two-point correlators in SU($N$) gauge theories on $\mathbb{R}^4$ with 
$\mathcal{N}=2$ supersymmetry and $N_f$ massless hypermultiplets in the fundamental representation. 
Using localization on $\mathcal{S}_4$, we compute the leading 
perturbative corrections to the two-point functions of chiral/anti-chiral operators made of scalar fields. 
The results are compared at two and three loops against direct field theory computations for some special operators whose correlators remain finite in perturbation theory at the specific loop order. In the conformal case, the match is shown up to two loops for a generic choice of operators and for arbitrary $N$.
}


\preprint{ROM2F/2017/02\\ CALT-TH-2017-011}

\begin{document}

\section{Introduction}
\label{secn:Intro}
Non-perturbative effects in gauge theories with $\mathcal{N}=2$ supersymmetry in four dimensions  
can be computed in many different ways. 
A method which has lately received a lot of attention is localization that provides exact results for the field theory path-integral by 
localizing it around a finite set of critical 
points\,\footnote{See for example
the collections \cite{Teschner:2014oja,Pestun:2016zxk} and references therein for a review 
on localization, and \cite{berline} for a mathematical discussion.}. 
In $\cN=2$ gauge theories this localization is typically achieved after giving a 
vacuum expectation value (vev) to the scalar field in the vector multiplet, 
and deforming the space-time geometry to break the gauge and Lorentz symmetries of the system. 
In the physics literature these deformations are usually called the $\Omega$-background 
\cite{Nekrasov:2002qd,Nekrasov:2003rj}. Proceeding in this way, one can explicitly evaluate the partition function ${Z}$ of the gauge theory on $\mathbb{R}^4$ 
and derive the celebrated Seiberg-Witten prepotential $\mathcal{F}$ that is identified as the free energy of
the system after the $\Omega$-background is turned off. 
Both ${Z}$ and $\mathcal{F}$, which receive non-perturbative contributions from instantons, 
are holomorphic functions of the complexified gauge coupling $\tau$ of the theory and of the scalar vevs.

The localization ideas can be extended also to more general space-times.
For example, the case of $\cN=2$ gauge theories defined on a four-sphere 
$\mathcal{S}_4$ was worked out in great detail in \cite{Pestun:2007rz}. 
In this case the path-integral receives contributions from both instantons and anti-instantons,
and localizes on configurations with purely imaginary scalar vevs. The corresponding
partition function, which we denote as $Z_{\mathcal{S}_4}$, is given 
by the modulus square of the one on 
$\mathbb{R}^4$ integrated over all possible imaginary vevs. In this picture, holomorphic and 
anti-holomorphic contributions are associated, respectively, to instantons and anti-instantons that 
localize around the north and south poles of the sphere.
This partition function is at the core of this paper, in which we want to test the predictions of 
localization against the results obtained by computing the Feynman diagrams in standard perturbation theory. To this aim, we focus on the weak coupling regime where instanton and anti-instanton contributions can be discarded.

Perturbative tests of the localization formulae have already been performed for Wilson loops in $\cN=4$ theories for which both one-loop and instanton corrections to the gauge partition function are absent. In this case, the localization formula reduces to a Gaussian matrix model integral counting the number of ladder 
diagrams, in full agreement with the field theory predictions \cite{Erickson:2000af}. Further investigations 
in this context have concerned the computation of the OPE coefficients between circular Wilson loops and 
chiral operators \cite{Berenstein:1998ij,Semenoff:2001xp,Fucito:2015ofa}, and the computation of 
half-BPS Wilson loops \cite{Andree:2010na} in superconformal $\cN=2$ super Yang-Mills theories. In all cases
a perfect agreement between the results of localization and those of perturbation theory has been found. 

More recently, a series of papers appeared in which two-point correlators of chiral/anti-chiral operators 
in $\cN=2$ superconformal theories are computed for low \cite{Baggio:2014sna,Baggio:2014ioa,Baggio:2015vxa,Gerchkovitz:2016gxx} and large \cite{Rodriguez-Gomez:2016ijh,Rodriguez-Gomez:2016cem,Baggio:2016skg} values of the rank of the gauge group. 
In a conformal theory, the two-point function between a chiral operator ${O}^i$ and an
anti-chiral operator $\bar{{O}}^j$ can be written as
 \begin{equation}
 \left \langle
{O}^{i} (x_1)  \, \bar{{O}}^j (x_2)  \right\rangle_{\mathrm{CFT}}  =
\frac{\mathcal{A}^{ij}(g)}{(4\pi^2 x_{12}^2)^{\Delta_0+\gamma(g)}\phantom{\Big|}} ~,
\label{2point00}
 \end{equation}
where $x_{12}=x_1-x_2$, $g$ is the gauge coupling, and 
$\Delta_0$ and $\gamma(g)$ are the engineering and anomalous dimensions of the operators. 
In \cite{Baggio:2014sna,Baggio:2014ioa,Baggio:2015vxa,Gerchkovitz:2016gxx}, 
it was shown that the functions $\mathcal{A}^{ij}(g)$ can be obtained by taking suitable derivatives of a deformed partition function on the sphere, or equivalently by computing two-point correlators in the associated matrix model. This correspondence was tested against the explicit result of standard perturbation theory up to two loops for half-BPS operators of low-dimensions in superconformal SU($N$) gauge theories with $N=2,3,4$.

In this paper we elaborate on this approach and extend it also to non-conformal models. For definiteness, we
consider $\cN=2$ theories with gauge group SU($N$) and $N_f$ massless hypermultiplets 
in the fundamental representation, and study their two-point correlators. Although
localization produces a finite result for any choice of operators, when the theory 
is non-conformal, {\it{i.e.}} when $N_f\neq 2N$, the two-point functions computed on the gauge theory side are in general divergent and have to be renormalized. 
To understand in detail the relation between the results in perturbative field theory and 
those obtained using localization is our ultimate goal. As a first step in this direction, here we focus on correlators that are already finite in field theory at a specific loop order. In this case we are able to provide some highly non-trivial tests of the correspondence. 
More precisely, we consider two-point functions that vanish up to a given loop order $L-1$. 
Under this assumption, we show that the leading contribution at $L$ loops is finite and the correlator takes the form
\begin{equation}
 \left \langle
{O}^{i} (x_1)  \, \bar{{O}}^j (x_2)  \right\rangle  =
\frac{\mathcal{A}_L^{ij} \,g^{2L}}{(4\pi^2 x_{12}^2)^{\Delta_0}\phantom{\Big|}}
+ \mathcal{O}\big(g^{2L+2}\big)
\label{2point000}
 \end{equation}
with specific coefficients $\mathcal{A}_L^{ij}$. We find a certain number of operators in this class 
for $L=2,3$ and for all of them we compare the results for $\mathcal{A}_L^{ij}$ coming from 
localization against those obtained by standard field theory methods, finding a perfect match.
In the superconformal case we prove the correspondence up to two loops for any operator 
and for any $N$, thus generalizing the results of \cite{Baggio:2014sna,Baggio:2014ioa,Baggio:2015vxa,Gerchkovitz:2016gxx}.
 
This paper is organized as follows. In Section~\ref{sec:correlmm} we derive the two-point correlators using localization methods. In Section~\ref{secn:fieldtheory} we compute the corrections to the two-point functions in perturbation theory at two and three loops. 
In Section~\ref{secn:corresp} we summarize our results and
discuss the correspondence between localization and field theory. Finally, in Section~\ref{secn:concl} we present our conclusions.
Some technical issues are discussed in the appendices.

\section{Two-point correlators in the matrix model}
\label{sec:correlmm}
We consider a four-dimensional $\mathcal{N}=2$ super Yang-Mills theory with gauge group 
SU($N$) and $N_f$ massless hypermultiplets in the fundamental representation. 
As discussed in the Introduction, 
we are interested in the two-point functions between chiral operators made out 
of the complex scalar field $\varphi$ sitting in the vector multiplet, and 
anti-chiral operators analogously defined in terms of the conjugate field $\bar{\varphi}$.
When the gauge theory is superconformal, {\it{i.e.}} when $N_f=2N$, building on the results of
\cite{Papadodimas:2009eu,Gerchkovitz:2014gta} it has been
shown in \cite{Baggio:2014sna,Baggio:2014ioa,Baggio:2015vxa,Gerchkovitz:2016gxx}
that these two-point functions can be obtained by placing the theory on $\mathcal{S}_4$ and taking suitable derivatives of a deformed partition function. The latter can be computed using localization in terms of a matrix model \cite{Pestun:2007rz}. Thus, matrix model correlators for the theory on $\mathcal{S}_4$ encode information on correlators of the gauge theory in $\mathbb{R}^4$. 
Here we  extract the field theory correlators  from matrix model correlators involving the insertion of
``normal-ordered" operators and argue that the correspondence extends, in a suitable sense, to non-conformal  models.  

We begin by briefly reviewing the interacting matrix model of \cite{Pestun:2007rz};
then we show how to efficiently compute correlators of matrix operators for arbitrary $N$.
In this way we do not only retrieve the results of 
\cite{Baggio:2014sna,Baggio:2014ioa,Baggio:2015vxa,Gerchkovitz:2016gxx}
for the SU($N$) superconformal theories with $N=2,3,4$, but we also extend them to generic $N$. 
Later we consider a special class of correlators in non-conformal models.

\subsection{The partition function on $\mathcal{S}_4$}
\label{subsec:partfunc}

As shown in \cite{Pestun:2007rz}, the partition function of an $\mathcal{N}=2$ super 
Yang-Mills theory with SU$(N)$ 
gauge group defined on a four-sphere $\mathcal{S}_4$ can be written in terms of 
a traceless $N\times N$ Hermitian matrix $a$ in the following way:
\begin{equation}
{Z}_{\mathcal{S}_4}=\int \prod_{u=1}^{N}da_u~\Delta(a)\,  \big| Z(\ii a,\tau)\big|^2 
\,\delta\Big(\sum_{v=1}^Na_v\Big)~. 
 \label{ws444}
\end{equation}
Here $a_u$ are the eigenvalues of $a$, the integration is over the real line, $\Delta$ is the Vandermonde determinant 
\begin{equation}
 \Delta(a) =   \prod _{u<v=1}^{N}a_{uv}^2 ~, 
 \label{vandermon}
\end{equation}
where $a_{uv}=a_u-a_v$, and $Z( \ii a,\tau)$ is the gauge theory partition function on 
$\mathbb{R}^4$. The latter is computed using the methods of \cite{Nekrasov:2003rj}, 
with the assumption that the adjoint scalar $\varphi(x)$ in the gauge multiplet has a purely imaginary 
vev given by $\vev\varphi=\ii a$, and that the $\Omega$-deformation is parameterized 
by $\epsilon_1=\epsilon_2=1/R$, where $R$ is the radius of $\mathcal{S}^4$ 
which, for simplicity, we take to be 1. 
Finally, the $\delta$-function in (\ref{ws444}) enforces the special unitarity condition, and 
$\tau$ is the complexified gauge coupling:
\begin{equation}
 \tau=\frac{\theta}{2\pi }+\ii \frac{4\pi}{g^2} ~.
\label{tau}
\end{equation}

The gauge theory partition function $Z$ can be written  
as a product of the classical, one-loop and instanton contributions, namely
\begin{equation}
  Z(\ii a,\tau)  =  Z_{\mathrm{class}}(\ii a,\tau)\,  Z_{\mathrm{one-loop}} (\ii a)\,  Z_{\mathrm{inst}} (\ii a,\tau) ~.
\end{equation}
The classical part produces a simple Gaussian term in the matrix model:
 \begin{equation}
 \left|Z_{\mathrm{class}}(\ii a,\tau)\right|^2  
 =  \esp^{-\frac{8\pi^2}{g^2}  \sum_u a_u^2  } 
\, = \, \esp^{-\frac{8\pi^2}{g^2} \,\tr a^2  }  ~,
 \end{equation}
while the one-loop contributions from the gauge multiplet and $N_f$ matter multiplets can be written as
\begin{equation}
\left|Z_{\mathrm{1-loop}}(\ii a)\right|^2 \,=\, 
\prod_{u < v=1}^N \!\! H( \ii a_{uv} )^2\,\,\prod_{u=1 }^N  H( \ii a_{u} )^{-N_f}~,
%\frac{\displaystyle{\prod_{u < v=1}^N \!\! H( \ii a_{uv} )^2}}
%{\displaystyle{ \prod_{u=1 }^N  H( \ii a_{u} )^{N_f} }  }~,
\label{zloops4}
\end{equation}
where
\begin{equation}
H(x)= G(1+x)\, G(1-x)
\label{H}
\end{equation}
with $G(x)$ being the Barnes $G$-function. 
In the weak-coupling limit $g\ll 1$ that we will consider, instantons can be discarded since
they are exponentially suppressed so that we can put
\begin{equation}
  \left|Z_{\mathrm{inst}}(\ii a,\tau)\right|^2  =1~. 
\end{equation}
At weak-coupling the integral (\ref{ws444}) is dominated by the region 
of small $a_u$, and thus we can expand the functions $H$ appearing in (\ref{zloops4}) 
for small values of their arguments using 
\begin{equation}
\log  H(x) =-(1+\gamma)\,x^2 -\sum_{n=2}^\infty \zeta(2n-1) \,\frac{x^{2n}}{n}~. 
\label{hsmall}
\end{equation}
Here $\zeta(n)$ is the Riemann zeta-function and $\gamma$ is the Euler-Mascheroni constant. 
In this way the one-loop contribution can be viewed as an interaction term in a free matrix model:
\begin{equation}
\left|Z_{\mathrm{1-loop}}(\ii a)\right|^2 = \esp^{-S(a) }~,
\end{equation}
where
\begin{equation}
\label{sint1}
  S(a) =-2 \sum_{u<v=1}^N \log H\left(\ii a_{uv}\right) 
  + N_f \sum_{u=1}^N \log H\left(\ii a_{u} \right)\,=\,S_2(a)  + S_4(a) +\cdots
\end{equation}
with $S_n(a)$ being homogeneous polynomials in $a$ of order $n$.  
The first few are:
 \begin{equation}
 \begin{aligned}
 S_2(a) &= -(1+\gamma) \bigg(\sum_{u,v=1}^n  a_{uv}^2-N_f  \sum_{u=1}^N  a_{u}^2 \bigg) 
=\, -(1+\gamma) \,(2N-N_f)\, \mathrm{tr}\,a^2 ~,\\
S_4(a) &=  \frac{\zeta(3)}{2} \, \bigg(\sum_{u,v=1}^N  
a_{uv}^4-N_f  \sum_{u=1}^N  a_{u}^4 \bigg) =\,\frac{\zeta(3)}{2} \, \Big[ (2 N-N_f) \,
{\mathrm{tr}}\,a^4 + 6 \left(\mathrm{tr}\,a^2\right)^2  \Big] ~,  \\
S_6(a) &=  -\frac{\zeta(5)}{3}\, \bigg(\sum_{u,v=1}^N  a_{uv}^6-N_f  \sum_{u=1}^N  a_{u}^6 \bigg)  
=\,- \frac{\zeta(5)}{3} \,  \Big[ (2 N-N_f) \, \mathrm{tr} \,a^6\\
&\qquad\qquad\qquad\qquad\qquad\qquad\qquad\qquad\qquad~
+30\, \mathrm{tr} \,a^4 \, \mathrm{tr} \,a^2 -20 \,\left(\mathrm{tr} \,a^3\right)^2
 \Big]~ .
 \end{aligned} 
 \label{Sintexp}
 \end{equation}
Performing the rescaling
\begin{equation}
\label{resca}
a \to \Big(\frac{g^2}{8\pi^2}\Big)^{\frac 12}\, a~,
\end{equation} 
the matrix model gets a canonically normalized Gaussian factor and the sphere partition function becomes
\begin{equation}
\label{rescaledmm}
Z_{\mathcal{S}_4} = \Big(\frac{g^2}{8\pi^2}\Big)^{\frac{N^2-1}{2}} \, \int \prod_{u=1}^{N} 
da_u~\Delta(a)\,\esp^{-\tr a^2 - S_{\mathrm{int}}(a)}\,\delta\Big(\sum_{v=1}^Na_v\Big)
\end{equation} 
with
\begin{equation}
\label{Sintresc}
S_{\mathrm{int}}(a) = \frac{g^2}{8\pi^2}~S_2(a) +  \Big(\frac{g^2}{8\pi^2}\Big)^2 \, S_4(a) + 
\Big(\frac{g^2}{8\pi^2}\Big)^3\, S_6(a) + \cdots ~.
\end{equation}
The overall $g$-dependent prefactor in (\ref{rescaledmm}) will be irrelevant when computing correlators, 
while the terms of order $g^{2L}$ in $S_{\mathrm{int}}$ will account for effects that take place at $L$ loops 
in the corresponding field theory computation. Therefore we will refer to the $g^2$-expansion 
of $S_{\mathrm{int}}$ as a loop expansion.

Note that the Vandermonde determinant $\Delta(a)$ in (\ref{rescaledmm})  arises from the Jacobian of the change of variables from a general $N\times N$ matrix to its  diagonal form in terms of its eigenvalues $a_u$, and thus 
the integral (\ref{rescaledmm}) can be alternatively expressed using a flat
integration measure $da$ over all matrix entries, namely 
\begin{equation}
\label{Zmmgauss}
Z_{\mathcal{S}_4}\, = c_N \, \Big(\frac{g^2}{8\pi^2}\Big)^{\frac{N^2-1}{2}} \,  \int da~
\esp^{-\tr a^2 - S_{\mathrm{int}}(a)}  
\end{equation}
where  $c_N$ is a $g$-independent constant and $da$ is defined by formula (\ref{decadja}) below. 

\subsection{Matrix model correlators}
Given any function $f(a)$ of the matrix $a$, its vev in the matrix model described above is defined as 
\begin{equation}
\label{vevmat}
\begin{aligned}
\vev{\,f(a)\,} & = \,
\frac{1}{Z_{\mathcal{S}_4}} 
\int \prod_{u=1}^{N}da_u~\Delta(a)\,  \big| Z(\ii a,\tau)\big|^2  
\,\delta\Big(\sum_{v=1}^Na_v\Big) \, f(a)\\
& = \,\frac{\displaystyle{ \int da ~\esp^{-\tr a^2-S_{\mathrm{int}}(a)}\,
f(a)}}
{ \displaystyle{\int da~\esp^{-\tr a^2-S_{\mathrm{int}}(a)}} }~,
\end{aligned}
\end{equation} 
where in the second step we used (\ref{Zmmgauss}). 
Since $S_{\mathrm{int}}(a)$ can be expressed as a series in $g^2$ (see  (\ref{Sintresc})), 
it is natural to view $\exp(-S_{\mathrm{int}}(a))$ as an interaction term in a Gaussian matrix model and write
\begin{equation}
\label{vevmattere}
\vev{\,f(a)\,}  = 
\frac{\big\langle
\esp^{- S_{\mathrm{int}}(a)}\,f(a)\big\rangle_0}
{\big\langle\,\esp^{- S_{\mathrm{int}}(a)}\big\rangle_0
\phantom{\Big|}}~.
\end{equation}
Here, we have denoted with a subscript 0 the expectation value in the Gaussian matrix model which, for any function $f(a)$, is defined as
\begin{equation}
\big\langle\,  f(a)\, \big\rangle_0 =
\frac{1}{Z_0}\,\int da\,\,\esp^{-\tr a^2} \, f(a)
\label{vevO}
\end{equation}
with
\begin{equation}
Z_0=\int da\,\,\esp^{-\tr a^2}~.
\label{Ztree}
\end{equation}

Having perturbatively reduced the computation of vevs in the interacting matrix model to vevs in a Gaussian model, we now give some details on the latter. 
If we write $a=a_b\,T^b$ where $T^b$ are the generators of SU($N$) in the fundamental representation,
normalized such that
\begin{equation}
\label{defgen}
\mathrm{tr}\, T^b \, T^c =\frac{1}{2}\, \delta^{bc}~,  \qquad   \mathrm{tr}\, T^b=0 ~,
\end{equation} 
and if we normalize the flat SU($N$) measure as
\begin{equation}
\label{decadja}
 da = \prod_{b=1}^{N^2-1} \frac{da_b}{\sqrt{2\pi}}~,
\end{equation}
then
\begin{equation}
Z_0=1
\end{equation}
and the basic Wick contraction becomes
\begin{equation}
 \langle \,a_b \, a_c \,\rangle_0    =   \delta_{bc} ~.
 \label{basic}
\end{equation}
For our later purposes, it is convenient to introduce the notation
\begin{equation}
t_{n_1,n_2,\ldots} \equiv \big\langle \mathrm{tr} \,a^{n_1}\,\mathrm{tr} \,a^{n_2}   \ldots  
\big\rangle_0 ~.
\label{amptree}
\end{equation}
For SU($N$) we evidently have 
\begin{equation}
\label{rrrecursion-2}
t_0 =\big\langle\,\mathrm{tr}\,1\,\big\rangle_0 =N~,~~~ t_1 = \big\langle\,\mathrm{tr}\, a\,\big\rangle_0=0~,
\end{equation}
while, after using (\ref{basic}), we get 
\begin{equation}
t_2=  \big\langle\,\mathrm{tr}\,a^2\,\big\rangle_0
=\mathrm{tr}\,T^bT^b\,=\,\frac{N^2-1}{2}~.
\end{equation}
Higher order correlators can be computed using the {fusion}/{fission}  identities
\begin{equation}
\begin{aligned}
\mathrm{tr}\,T^b B_1 T^b B_2 &=\frac{1}{2}\, \mathrm{tr}\,  B_1\,\mathrm{tr}\, B_2
-\frac{1}{2N}\,\mathrm{tr}\,B_1 B_2~,\\
\mathrm{tr}\,T^b B_1\,\mathrm{tr}\, T^b B_2&=\frac{1}{2}\,\mathrm{tr}\,B_1B_2
-\frac{1}{2N}\,\mathrm{tr}\,B_1\,\mathrm{tr}\,B_2~,
\end{aligned}
 \label{identities}
\end{equation}
which are valid for two arbitrary matrices $B_1$ and $B_2$. For example, we
have
\begin{equation}
\begin{aligned}
t_4&=\big\langle\,\mathrm{tr}\,a^4\,\big\rangle_0 =
2\,\big\langle\,\mathrm{tr}\,T^bT^ba^2\,\big\rangle_0+
\big\langle\,\mathrm{tr}\,T^b a T^b a\,\big\rangle_0=\Big(N-\frac{3}{2N}\Big)\,t_2~.
\end{aligned}
\end{equation}
In fact, one can recursively relate any correlator involving a total of $n$ matrices to the 
combination of correlators with $n-2$ matrices obtained after a single Wick contraction.
We find (for $n>2$)
\begin{align}
 t_{n}    &=   \frac12 \sum_{m=0}^{n-2}  \Big( t_{m,n-m-2}
  -\frac{1}{N}\, t_{n-2}  \Big)  ~,\notag\\
 t_{n,n_1} &=  \frac12 \sum_{m=0}^{n-2}  \Big( t_{m,n-m-2,n_1}
  -\frac{1}{N}\,   t_{n-2,n_1}  \Big)  
  + \frac{n_1}{2} \,\Big(  t_{n+n_1-2 } -\frac{1}{N} \,t_{n-1,n_1-1} \Big)~,\label{rrecursion}\\
  t_{n,n_1,n_2}  &=\frac12 \sum_{m=0}^{n-2}  \Big( t_{m,n-m-2,n_1,n_2}
  -\frac{1}{N} \,  t_{n-2,n_1,n_2}  \Big)  + \frac{n_1}{2} \Big(t_{n+n_1-2,n_2}
  -\frac{1}{N} \,  t_{n-1,n_1-1,n_2}\Big)  \notag\\
   &~~~~~~~~~~~+ \frac{n_2}{2} \Big( t_{n+n_2-2,n_1 }
   -\frac{1}{N}  t_{n-1,n_1,n_2-1} \Big) ~,\notag
\end{align}
and so on. 
For $n=0,1,2$, these expressions acquire a particularly simple form: 
\begin{equation}
\begin{aligned}
t_{0,n_1,n_2,\ldots} &= N \, t_{n_1,n_2,\ldots}~,\\
t_{1,n_1,n_2,\ldots} &= 0  ~,\\
t_{2,n_1,n_2,\ldots} &=\frac{N^2-1+n_1+n_2+\ldots}{2}\,   t_{n_1,n_2,\ldots}~ .  
\end{aligned}
\label{t2s}
\end{equation}
These recursive relations, together with the initial conditions (\ref{rrrecursion-2}), 
can be used to derive finite $N$ formulae in a very efficient way.
A few examples are given below:
\begin{equation}
\begin{aligned}
t_2  &=\frac{ N^2-1}{2}~, ~~~~
t_{2,2}  =    \frac{ N^4-1}{4}~, ~~~~   t_4  =  \frac{(N^2-1)(2N^2-3)}{4N}~, \\
t_6 &=  \frac{5(N^2-1)(N^4-3N^2+3)}{8N^2}~,~~~~
t_{3,3}  =    \frac{3(N^2-1)(N^2-4)}{8N}~,\\
t_{4,2}  &=    \frac{(N^2-1)(N^2+3)(2N^2-3)}{8N}~,~~~~
t_{2,2,2}=   \frac{(N^4-1)(N^2+3)}{8}~.
\end{aligned}
\label{chiralring}
\end{equation}

\subsection{Normal-ordered operators}
\label{subsecn:normord}
Another important ingredient is normal ordering. Indeed, since we are ultimately interested in establishing
a connection between correlators in the matrix model and correlators in the gauge theory, and since the latter
involve chiral and anti-chiral operators that do not have self-contractions, 
we have to introduce matrix model operators that do not have self-contractions either, 
{\it{i.e.}}, that are normal-ordered.  
Given an operator $O(a)$, subtracting all its self-contractions is equivalent to make it orthogonal to all the lower dimensional operators\,\footnote{Similar arguments were also carried out in \cite{Fucito:2015ofa} to compute the two point functions involving a Wilson loop and a chiral operator  (see also \cite{Semenoff:2001xp}).}. 
Let $\Delta$ be the dimension of $O(a)$, and $\{O_p\}$ be a basis of operators with dimensions lower 
than or equal to $(\Delta-2)$.  
We denote by $C^{pq}$ the inverse of the matrix of their correlators
\begin{equation}
C_{pq}= \langle O_p(a) \, O_q(a) \rangle ~,
\end{equation}
where the right hand side is defined via  (\ref{vevmat}) and (\ref{vevmattere}).
Then, the normal-ordered operator $\nordg{O(a)}$ is defined as
\begin{equation}
 \nordg{O(a)} \, \,
=\, O(a) - \sum_{p,q} \langle O(a)\, O_{p}(a) \rangle  \, C^{pq}\, O_q(a)  ~,
\label{normalo}
\end{equation}
This operator is orthogonal by construction to all operators $O_{p}(a)$ of lower dimension. 
In particular, its one-point function vanishes:
\begin{equation}
\label{opno}
\vev{\,\nordg{O(a)}\,}=0~,
\end{equation} 
 since the sum
 in the right hand side of (\ref{normalo}) precisely subtracts all self-contractions of $O(a)$.
We would like to stress that the definition (\ref{normalo}) of normal ordering
introduces an explicit $g$-dependence and the symbol $\nordg{~}$ we have used underlines this feature.

As an example, let us consider $O(a)=\mathrm{tr}\, a^2$. The only operator of lower dimension is the identity, so using (\ref{normalo}) one finds
\begin{equation}
\label{not2}
\nordg{\mathrm{tr}\, a^2} \, =\, \mathrm{tr}\,a^2\!  - \vev{\,\mathrm{tr}\,a^2\,}~.
\end{equation}
The second term precisely subtracts the self-contraction contributions inside a correlator. Indeed, 
\begin{equation}
\big\langle\!    \nordg{\mathrm{tr}\, a^2}\,\,\,\nordg{\mathrm{tr}\, a^2}    \!\big\rangle=   \big\langle \mathrm{tr}\, a^2\ \mathrm{tr}\, a^2  \big\rangle- \vev{\,\mathrm{tr}\,a^2\,}^2
\end{equation}
If we explicitly compute the right hand side of (\ref{not2}), we find 
\begin{equation}
\label{not2ex}
\nordg{\mathrm{tr}\, a^2} ~=~ \mathrm{tr}\,a^2\!  - \frac{N^2-1}{2} - 
\frac{(2N-N_f)(N^2-1)(1 + \gamma)}{2}\,\frac{g^2}{8\pi^2}+ \mathcal{O}(g^4)~.
\end{equation}
Notice that the $g^2$-term is proportional to $(2N-N_f)$ and thus vanishes in the conformal case.   
Actually, this is true for the $g^2$-terms of all normal-ordered operators, since these corrections originate from insertions of $S_2$ which is proportional to $(2N-N_f)$.

As will be more clear in the following, for the two-point functions studied in this paper the $g$-dependent terms in the normal-ordered operators will not contribute to the loop orders we will consider, {\it{i.e.}}
they will be always subleading in the gauge coupling and thus can be neglected. 
For this reason, we find convenient to introduce a specific notation 
for the $g\to 0$ limit of the normal order, namely
\begin{equation}
\nord{O(a)} \,\,=\,\lim_{g\to 0} \nordg{O(a)}
\label{doubledots}
\end{equation}
in such a way that most of the fomulae will look simpler. Note that 
%the normal ordering $\nord{~}$ 
the normal ordering of eq. (\ref{doubledots})
is the natural one in the Gaussian matrix model. For example, we have
\begin{equation}
\label{not2ex0}
\nord{\mathrm{tr}\, a^2} ~=~ \mathrm{tr}\,a^2\!  - \frac{N^2-1}{2}~.
\end{equation}
Applying the definition (\ref{normalo}) to operators of the form 
$\big(\mathrm{tr} \,a^{n_1}\,\mathrm{tr} \,a^{n_2}   \cdots\big)$,
and using (\ref{chiralring}), it is quite straightforward to obtain the explicit expressions
of the leading terms of other normal-ordered operators. For operators of dimensions up 
to six, beside (\ref{not2ex0}), one finds the following results:
\begin{subequations}
\begin{align}
\nord{\left(\mathrm{tr}\, a^2\right)^2}&~= \left(\mathrm{tr}\, a^2\right)^2-(N^2+1)\,
\mathrm{tr}\, a^2+\frac{N^4-1}{4}~,\phantom{\bigg|}\label{ordera41}\\
\nord{\left(\mathrm{tr}\, a^2\right)^3}& ~=~\left(\mathrm{tr}\, a^2\right)^3
 - \frac{3(N^2+3)}{2}\left(\mathrm{tr}\, a^2\right)^2+
 \frac{3(N^2+3)(N^2+1)}{4}\,\mathrm{tr}\, a^2~,\phantom{\bigg|}\notag\\
 &\qquad-\frac{ (N^2+3)(N^4-1)}{8}\phantom{\bigg|}\\
 \nord{\mathrm{tr}\, a^4}& ~=~\mathrm{tr}\, a^4
 - \frac{2N^2-3}{ N} \,\mathrm{tr}\, a^2+\frac{(N^2-1)(2N^2-3)}{4N}
 ~, \phantom{\bigg|}\\
 \nord{\mathrm{tr}\, a^6}& ~=~\mathrm{tr}\, a^6
 -\frac{3}{2}\,\left(\mathrm{tr}\, a^2\right)^2-\frac{3( 2N^2-5)}{2N}\,
 \mathrm{tr}\, a^4+ \frac{ 15(N^4-3 N^2+3)}{ 4N^2}\,\mathrm{tr}\, a^2\phantom{\bigg|}\notag\\
 &\qquad-\frac{5(N^2-1)(N^4-3\, N^2+3)}{ 8N^2}~,\phantom{\bigg|}
 \phantom{\bigg|}\\
\nord{\left(\mathrm{tr}\, a^3\right)^2}&~=~\left(\mathrm{tr}\, a^3\right)^2+\frac{9}{2N}
\left(\mathrm{tr}\, a^2\right)^2-\frac{9}{2}\,\mathrm{tr}\, a^4+ \frac{ 9( N^2-4)}{4N} 
\,\mathrm{tr}\, a^2\notag\\
&\qquad-\frac{3(N^2-4)(N^2-1)}{ 8N}~,\phantom{\bigg|}\\
\nord{\mathrm{tr}\, a^4\,\mathrm{tr}\, a^2}&~=~\mathrm{tr}\, a^4\,\mathrm{tr}\, a^2
-\frac{2N^2-3}{N} \left(\mathrm{tr}\, a^2\right)^2- \frac{N^2+7}{2}\,\mathrm{tr}\, a^4
\notag\\
&\qquad+ \frac{ 3(2N^2-3)(N^2+3)}{ 4N} \,\mathrm{tr}\, a^2
-\frac{(N^2-1)(N^2+3)(2 N^2-3)}{ 8N} ~. \phantom{\bigg|}
\end{align}
\label{ordera4}
\hspace{-0.25cm}
\end{subequations} 
For each of the expressions considered above, it is easy to check that all subsequent terms in the right hand 
side are obtained as self-contractions of the first one. 
For instance, for $\left(\mathrm{tr}\, a^3\right)^2$ 
the terms proportional to $ \left(\mathrm{tr}\, a^2\right)^2 $ and $\mathrm{tr}\, a^4$ 
correspond to a single Wick contraction of $\left(\mathrm{tr}\, a^3\right)^2$, 
the term proportional to $\mathrm{tr}\, a^2$ is the result of two Wick contractions and the last one arises from three contractions.

As we mentioned above, for the two-point functions considered in this paper the 
$g$-dependent terms in the normal-ordered operators are not really needed 
at the loop orders we study. To see this, we observe that in the conformal case
the one-loop corrections vanish and therefore the first $g$-dependent terms are of order $g^4$. These 
corrections lead
to modifications of the two-point correlators at order $g^8$, {\it{i.e.}} at four loops, which is beyond the approximation considered here. In the non-conformal theories, instead, we will focus on correlators that vanish at weak coupling up to a given loop order (specifically two- and three-loops) and restrict ourselves to the leading contributions (of order $g^4$ and $g^6$, respectively). 
In these cases one can show that the $g$-dependent terms in the normal order definition yield subleading contributions to the correlator and thus can be neglected. 
To check this, it will be useful to observe that 
\begin{equation}
\nordg{\left(\mathrm{tr}\, a^2\right)^n}\,\,=\,\, 
\nord{\left(\mathrm{tr}\, a^2\right)^n}\,-\,\frac{g^2}{8\pi^2}\,{\frac{n(2N\!-\!N_f)(N^2\!+\!2n\!-\!3)
(1\!+\!\gamma)}{2}} \nord{\left(\mathrm{tr}\, a^2\right)^{n-1}}+\,\mathcal{O}(g^{4})~.
\end{equation}
Another useful property is that in the correlator between two normal-ordered operators, the normal ordering 
can be removed from one of them, namely
\begin{equation}
 \big\langle\! \nordg{O(a)} \,\, \nordg{\widetilde{O}(a)} \!\big\rangle=
 \big\langle O(a)\,\, \nordg{\widetilde{O}(a)}\! \big\rangle ~.
 \label{2nord}
\end{equation}

\subsection{Two-point  correlators}
\label{subsec:2point}

A generic operator of dimension $n$ can be written as 
\begin{equation}
{O}^i_{(n)}(a)= \sum_{n_1\geq n_2\geq \cdots } c^i_{n_1,n_2,\ldots}\,\, 
{\,\tr a^{n_1}~\tr a^{n_2} \cdots\,}~,
\label{Oni}
\end{equation}
where the sum is over the partitions of $n=\sum_s n_s$ and the index $i$ labels the various combinations that 
one can make using coefficients $c^i_{n_1,n_2,\ldots}$.  
Of course, for SU($N$) we can restrict to $n_i>1$ since $\tr a=0$; thus
the independent operators of dimension $n$ are as many as the partitions of $n$ 
in which the number 1 does not appear.

Alternatively, using the adjoint decomposition $a=a_b \,T^b$, we can write (\ref{Oni}) as
\begin{equation}
{O}^i_{(n)}(a)= R^i_{b_1b_2\cdots b_n}\, {a^{b_1}a^{b_2}\cdots a^{b_n} }~,
\label{Onj}
\end{equation}
where $R^i$ is a gauge tensor. 
Thus, to any operator ${O}^i_{(n)}(a)$ we can associate a completely symmetric $n$-index tensor $R^i$.

We now define the two-point correlator $\mathcal{A}^{ij}_{(n)}(g) $
as the expectation value of two normal-ordered operators in the interacting matrix model, namely
\begin{equation}
\mathcal{A}^{ij}_{(n)}(g) \,\equiv\,\big\langle\nordg{{O}^i_{(n)}(a)} \,\,\nordg{
{O}^j_{(n)}(a)}\big\rangle
\,\,=\, 
\frac{\big\langle\,
\esp^{- S_{\mathrm{int}}(a)}\,
\nordg{{O}^i_{(n)}(a)} \,\,\nordg{
{O}^j_{(n)}(a)}\big\rangle_0}
{\big\langle\,\esp^{- S_{\mathrm{int}}(a)}\big\rangle_0\phantom{\Big|}}~.
\label{Aismmc}
\end{equation}
The correlator $\mathcal{A}^{ij}_{(n)}(g)$ 
can thus be computed perturbatively in $g$ using (\ref{Sintresc}) and performing Wick contractions in the free matrix model. In this way we 
%obtain a perturbative expansion like
write
\begin{equation}
\mathcal{A}^{ij}_{(n)}(g)= \sum_{L=0} \mathcal{A}^{ij}_{(n)}(g)\Big|_{L-\mathrm{loop}}~.
\label{Aijexp}
\end{equation}
The first term of this expansion is simply given by
\begin{equation}
\begin{aligned}
 \mathcal{A}^{ij}_{(n)}(g)\Big|_{0-\mathrm{loop}}
 &= \,\big\langle
\nord{{O}^i_{(n)}(a)}\,\,\, \nord{{O}^j_{(n)}(a)}
\big\rangle_0\,=\,n!\, R^i_{b_1b_2\cdots b_n}\,R^{j\,b_1b_2\cdots b_n}~.
\end{aligned}
 \label{Aij0}
\end{equation}
Indeed, at tree level we can put $S_{\mathrm{int}}=0$ and neglect all $g$-dependent terms in the 
normal-ordered operators; moreover, due to normal ordering, the only possibility we have is to contract each
matrix of ${O}^i_{(n)}(a)$ with a matrix of ${O}^j_{(n)}(a)$ in all possible ways
using (\ref{basic}); in this way we immediately find the above expression.

At higher loops the calculations are less straightforward, and the results depend on the
features of the model considered. In the next subsection we give some explicit examples, starting with the superconformal case.

\subsection{The conformal case $N_f=2N$}
When $N_f=2N$, the expansion (\ref{Sintresc}) simplifies since (\ref{Sintexp}) becomes
\begin{equation}
\begin{aligned}
S_2(a) &= 0 ~,  \\
S_4 (a)&= \frac{\zeta(3)}{2} \, \Big[6 \left(\mathrm{tr}\,a^2\right)^2  \Big] ~, \\
S_6(a) &=- \frac{\zeta(5)}{3} \,  \Big[
30\, \mathrm{tr} \,a^4 \, \mathrm{tr} \,a^2 -20 \,\left(\mathrm{tr} \,a^3\right)^2\Big]~ .
\end{aligned}
\end{equation}
Using $S_2=0$ in (\ref{Aismmc}), we obtain
\begin{equation}
\begin{aligned}
\mathcal{A}^{ij}_{(n)}(g)\Big|_{1-\mathrm{loop}}&=0~,\\
\mathcal{A}^{ij}_{(n)}(g)\Big|_{2-\mathrm{loop}}&=
-\Big(\frac{g^2}{8\pi^2}\Big)^2
  \Big\langle  \nord{{O}^i_{(n)} (a)}\,\,\,
  \nord{{O}^j_{(n)}(a)}\,\left[S_4(a) -\big\langle S_4(a) \big\rangle_0\,\right]\Big\rangle_0~,\\
\mathcal{A}^{ij}_{(n)}(g)\Big|_{3-\mathrm{loop}}&=-\Big(\frac{g^2}{8\pi^2}\Big)^3
  \Big\langle \nord{{O}^i_{(n)} (a)}\,\,\,
  \nord{{O}^j_{(n)}(a)}\,\left[S_6(a) -\big\langle S_6(a) \big\rangle_0\,\right]\Big\rangle_0~.
\end{aligned}
\label{confexpansion}
\end{equation}
The correlators that appear in the right hand sides can be evaluated in the free Gaussian model
for any $N$ using the recursion relations (\ref{rrecursion}) in a straightforward manner.

Let us consider some explicit examples involving the operators 
of low dimensions, which we list here:
\begin{equation}
 {O}^1_{(2)} =\,{{\mathrm{tr}}\,a^2}~, \quad {O}^1_{(3)}=\,\,{{\mathrm{tr}}\,a^3}~,  
 \quad  {O}^1_{(4)}=\,\,
 {\left(\mathrm{tr}\, a^2\right)^2}~, \quad 
 {O}^2_{(4)}=\,\,
 {\mathrm{tr}\, a^4}~.
\label{operators}
\end{equation}
Up to three loops, we find the following correlators:
\begin{subequations} 
 \begin{align}
 \mathcal{A}^{11}_{(2)}(g)&=\frac{N^2\!-\!1}{2}
    \!-\!\frac{9(N^4\!-\!1)}{2}\,\zeta(3) \Big(\frac{g^2}{8 \pi^2}\Big)^2\notag\\
&~~~+\frac{15 (N^4\!-\!1) (2 N^2\!-\!1)}{N} \, \zeta(5)  \Big(\frac{g^2}{8 \pi^2}\Big)^3
+ \mathcal{O}(g^8)~,\phantom{\Bigg|}\\
\notag\\
\mathcal{A}^{11}_{(3)}(g)&=\frac{ 3(N^2\!-\!1)(N^2\!-\!4)}{8 N}
-\frac{ 27 (N^2\!-\!1)(N^2\!-\!4)(N^2\!+\!3)}{8N} \, \zeta(3)\Big(\frac{g^2}{8 \pi^2}\Big)^2\notag\\
&~~~+\!\frac{15 (N^2\!-\!1)^2 ( N^2\!-\!4)(22N^2\!+\!53)}{16N^2} \, \zeta(5)\Big(\frac{g^2}{8 \pi^2}\Big)^3
+ \mathcal{O}(g^8)~,\phantom{\Bigg|}\\
\notag\\
\mathcal{A}^{11}_{(4)}(g)&=\frac{ (N^4\!-\!1)}{2}-9 (N^4\!-\!1)(N^2\!+\!3) \, \zeta(3)\Big(\frac{g^2}{8 \pi^2}\Big)^2\notag\\
&~~~+\!\frac{30 (N^4\!-\!1) ( N^2\!+\!6)(2N^2\!-\!1)}{N} \, \zeta(5) \Big(\frac{g^2}{8 \pi^2}\Big)^3
+ \mathcal{O}(g^8)~,\phantom{\Bigg|}\\
\notag\\
\mathcal{A}^{22}_{(4)}(g)&=\frac{ (N^2\!-\!1)(N^4\!-\!6N^2\!+\!18)}{4N^2}\!
    -\!\frac{ 3 (N^2\!-\!1)(N^6\!+\!2N^4\!-\!18 N^2\!+\!81)}{N^2}\, \zeta(3)\Big(\frac{g^2}{8 \pi^2}\Big)^2\notag\\
&~~~+\!\frac{5 (N^2\!-\!1) ( 8N^8\!+\!19 N^6 \!-\!9 N^4 \!+\!432 N^2\!-\!648) }{2N^3} \, \zeta(5) \Big(\frac{g^2}{8 \pi^2}\Big)^3
+ \mathcal{O}(g^8)~,\phantom{\Bigg|}\\
\notag\\
\mathcal{A}^{12}_{(4)}(g)&=\frac{( N^2\!-\!1)(2N^2\!-\! 3)}{2N}-\frac{9( N^2\!-\!1) (2N^2\!-\! 3)(N^2\!+\!3)}{N}\,\zeta(3) \Big(\frac{g^2}{8 \pi^2}\Big)^2\notag\\
&~~~+\!\frac{5( N^2\!-\!1)(29N^6\!+\!31 N^4\!-\!90N^2\!+\!108)}{N^2}\, \zeta(5) \Big(\frac{g^2}{8 \pi^2}\Big)^3+ \mathcal{O}(g^8)~\phantom{\Bigg|}~.
 \end{align}
 \label{resconf}
 \hspace{-0.35cm}
\end{subequations}
One can check that these formulae reproduce the results in \cite{Baggio:2014sna,Baggio:2014ioa,Baggio:2015vxa,Gerchkovitz:2016gxx} for $N=2, 3, 4$ and generalize them to any $N$. 
In Section~\ref{secn:fieldtheory} we will compare these expressions with 
the perturbative computation of the corresponding correlators in field theory at two loops and find perfect agreement.
 
\subsection{Special correlators}
\label{sec:special}
When $N_f\not= 2N$ things are more complicated. For example, the one-loop interaction term $S_2$ is no
longer vanishing and the perturbative expansion has many more terms. We restrict our attention to a particular class of  correlators that vanish up to a given loop order
$L-1$, {\it{i.e.}} that are of the form
\begin{equation}
\mathcal{A}^{ij}_{(n)}(g)=\mathcal{A}^{ij}_{(n)}(g)\Big|_{L-\mathrm{loop}}+O(g^{2L+2})
\label{Asp}
\end{equation}
with the first non-vanishing contribution of order $g^{2L}$.

To make the analysis as simple as possible, we fix one of the operators to be
\begin{equation}
\label{choicetr2m}
{\left(\tr a^2\right)^m}\,\,=\frac{1}{2^m}\,\delta_{b_1b_2}\cdots\delta_{b_{2m-1}b_{2m}}\,
{a^{b_1}a^{b_2} \cdots a^{b_{2m-1}}a^{b_{2m}}}~.  
\end{equation}
This choice greatly simplifies the calculations in the matrix model but, as we will see 
in Section~\ref{secn:fieldtheory}, it also helps in the perturbative computation of the 
corresponding correlator in field theory which becomes tractable without being trivial%
\footnote{For a more general
choice of the operators, corrections to the normal order definition
contribute to (\ref{Asp}) and the match against field theory results requires
more care. We have performed some two-loop checks for correlators in
this more general class and found perfect agreement with field theory. }.

The two-point correlator we study is then
\begin{equation}
\mathcal{A}^{i}_{(2m)}(g)  =\big\langle
\nordg{{O}^i_{(2m)}(a)}\,\,\, \nordg{\left(\tr a^2\right)^m} \big\rangle 
=
\frac{\big\langle
\nordg{{O}^i_{(2m)}(a)}\,\,\, \nordg{\left(\tr a^2\right)^m}\,\, \esp^{- S_{\mathrm{int}}(a)}
\big\rangle_0}
{\big\langle\,\esp^{- S_{\mathrm{int}}(a)}\big\rangle_0\phantom{\Big|}}~,
\label{Aig}
\end{equation}
where the operator ${O}^i_{(2m)}(a)$ is chosen in such a way 
that  
\begin{equation}
\label{tr2mmmnc}
\mathcal{A}^{i}_{(2m)}(g)\Big|_{\ell-\mathrm{loop}}=0 \qquad \mathrm{for} \qquad \ell=0,1,\ldots,L-1~. 
\end{equation}
To evaluate this correlator one makes use of the expansion of $S_{\mathrm{int}}$ given in (\ref{Sintresc}),
which implies
\begin{equation}
\begin{aligned}
\esp^{-S_{\mathrm{int}}(a)}  &=
1 - \Big(\frac{g^2}{8\pi^2}\Big) \,S_2(a) - \Big(\frac{g^2}{8\pi^2}\Big)^2 \,
\Big(S_4(a) -\frac{1}{2} S_2(a)^2 \Big)  \\
&\qquad- \Big(\frac{g^2}{8\pi^2}\Big)^3 \,
\Big(S_6(a) - S_4(a) S_2(a) + \frac{1}{6} S_2(a)^3\Big) +\ldots~.
\end{aligned}
\label{esint}
\end{equation}
Notice that if one already knows that all terms of order less than $L$ vanish, then the denominator in the
right hand side of (\ref{Aig}) does not contribute at order $g^{2L}$ and thus can be neglected.

The first requirement is the vanishing of the zero-loop term, {\it{i.e.}}
\begin{equation}
\label{g0cond}
\mathcal{A}^i_{(2m)}(g)\Big|_{0-\mathrm{loop}}=
\big\langle \nord{{O}^i_{(2m)}(a)}\,\,\, \nord{\left(\tr a^2\right)^m}\big\rangle_0 = 0~. 
\end{equation}
Using (\ref{Aij0}) and (\ref{choicetr2m}), we immediately see that this condition is equivalent to
\begin{equation}
R^i_{b_1b_2\cdots b_{2m-1}b_{2m}}\,\delta^{b_1b_2}\cdots\delta^{b_{2m-1}b_{2m}}=0~,
\label{notrace}
\end{equation}
{\it{i.e.}} the $R^i$ tensor associated to ${O}^i_{(2m)}(a)$ must be totally traceless. 

It is easy to show that if the zero-loop condition (\ref{g0cond}) holds, then also the one-loop contribution vanishes. Indeed, we have
\begin{equation}
\begin{aligned}
\mathcal{A}^i_{(2m)}(g)\Big|_{1-\mathrm{loop}}~
&=- \frac{g^2}{8\pi^2}
\,\big\langle
 \nord{{O}^i_{(2m)}(a)}\,\,\, \nord{\left(\tr a^2\right)^m}\, S_2(a) \,\big\rangle_0
 \\ &
 \propto\, \big\langle
 \nord{{O}^i_{(2m)}(a)}\,\,\, \nord{\left(\tr a^2\right)^m}\, \tr a^2 \,\big\rangle_0
 \phantom{\Big|}\\
 & \propto\,
 \big\langle
 \nord{{O}^i_{(2m)}(a)}\,\,\, \nord{\left(\tr a^2\right)^m} \big\rangle_0\,=\,\,
 \mathcal{A}^i_{(2m)}\Big|_{0-\mathrm{loop}}=0\phantom{\Big|}~.
\end{aligned}
\label{1loopcond}
\end{equation}
To obtain the last line, one first writes the product $\nord{\left(\tr a^2\right)^m}\, \tr a^2$ as a linear combination of normal-ordered operators $\nord{\left(\tr a^2\right)^k} $ with $k\leq (m+1)$, and then uses the fact that only the term with $k=m$ in this combination can contribute to the correlator, yielding in the end a vanishing result because of the condition (\ref{g0cond}). The first non-trivial contributions occur then at two or more loops.

\subsubsection{Two loops}
\label{subsec:twolr}
Assuming the vanishing of the zero- and one-loop terms, 
we consider the two-loop contribution to the correlator $\mathcal{A}^{i}_{(2m)}(g)$,
which reads 
\begin{equation}
\begin{aligned}
\mathcal{A}^i_{(2m)}(g)\Big|_{2-\mathrm{loop}} ~
&= -\Big(\frac{g^2}{8\pi^2}\Big)^2 \,
\big\langle
\nord{{O}^i_{(2m)}(a)}\,\,\, \nord{\left(\tr a^2\right)^m}\, \big[S_4(a) - \frac{1}{2}\, S_2(a)^2 \big]\, \big\rangle_0\phantom{\bigg|}\\
&= -\Big(\frac{g^2}{8\pi^2}\Big)^2 \,
\big\langle
\nord{{O}^i_{(2m)}(a)}\,\,\, \nord{\left(\tr a^2\right)^m}\, S_4(a) \,\big\rangle_0
\phantom{\bigg|}\\
&= - \frac{6\,\zeta(3)}{2}  \, \Big(\frac{g^2}{8\pi^2}\Big)^2  \,  
\mathcal{C}^{i}_{(2m)}~,\phantom{\bigg|}
\end{aligned}
\label{2loopterm}
\end{equation}
where
\begin{equation}
\mathcal{C}^{i}_{(2m)}  =\frac{1}{6}\,
\big\langle
\nord{{O}^i_{(2m)}(a)}\,\,\, \nord{\left(\tr a^2\right)^m}\, \big[(2N-N_f)\,\tr a^4+6\left(\tr a^2\right)^2
\big]\,\big\rangle_0 ~.
 \label{loc02}
\end{equation}
Above we used the fact that multiple insertions of $S_2$ automatically vanish given the vanishing of the correlator at the previous orders, as we showed in considering the one-loop contribution, and took into account the explicit form of
$S _4$ given in (\ref{Sintexp}). Let us now consider some explicit examples. 

At dimension two ($m=1$), there is a single operator, namely
$\tr a^2$ , which does not satisfy (\ref{notrace}). So we cannot construct special correlators
when $m=1$.

At dimension four ($m=2$), we have two independent operators, $\tr a^4$ and 
$\left(\tr a^2\right)^2$ , none of which satisfies the tracelessness condition (\ref{notrace}).
However their linear combination
\begin{equation}
\label{O14}
{O}^1_{(4)}(a) = \,{\tr a^4} + \,c^1_{2,2}\,{\left(\tr a^2\right)^2}
\end{equation} 
with
\begin{equation}
c^1_{2,2}= -\frac{t_4}{t_{2,2}}= -\frac{2N^2-3}{N(N^2+1)}   \label{c122def}
\end{equation}
does. To see this, we first observe that
\begin{equation}
\begin{aligned}
\big\langle\nord{{O}^1_{(4)}(a)}\,\,\, \nord{\left(\tr a^2\right)^2}\big\rangle_0\,&=
\big\langle \,\big[\tr a^4 + \,c^1_{2,2}\,\left(\tr a^2\right)^2\big]\, \nord{\left(\tr a^2\right)^2}\big\rangle_0\\
&=\sum_{k=0}^2c_k\,\big\langle\, \big[\tr a^4 + \,c^1_{2,2}\,\left(\tr a^2\right)^2\big]\,\left(\tr a^2\right)^k\big\rangle_0~,
\end{aligned}
\label{c122}
\end{equation}
where in the first step we have removed the normal ordering from ${O}^1_{(4)}(a) $ according to (\ref{2nord}), and in the second step we have written $\nord{\left(\tr a^2\right)^2}\,\,=\sum_{k=0}^2 c_k \left(\tr a^2\right)^k$ according to (\ref{ordera41}). 
Then, using the definitions (\ref{amptree}), the zero-loop correlator becomes
\begin{equation}
\big\langle\nord{{O}^1_{(4)}(a)}\,\, \, \nord{\left(\tr a^2\right)^2}\big\rangle_0\,=\,
\sum_{k=0}^2c_k\,\big[t_{4,\,\underbrace{\scriptstyle{2,\cdots,2}}_{k}}
+\,c^1_{2,2}\,t_{2,2,\,\underbrace{\scriptstyle{2,\cdots,2}}_{k}}\,\big]~.
\label{c122a}
\end{equation}
Exploiting the recursion relations (\ref{rrecursion}) we see that each term in the above
square bracket is proportional to $\big(t_4+\,c^1_{2,2}\,t_{2,2}\big)$, which
vanishes because of (\ref{c122def}). Thus, the condition (\ref{g0cond}) is satisfied.
Alternatively, one can check that the $R$ tensor associated to ${O}^1_{(4)}$, namely
\begin{equation}
\label{QforO4}
\tr T^{(a_1} T^{a_2} T^{a_3} T^{a_4)} + c^1_{2,2}\, \delta^{(a_1 a_2} \delta^{a_3 a_4)}~,
\end{equation} 
is traceless in any pair of indices for the choice of $c^1_{2,2}$ given in (\ref{c122def}).

Using the above findings, we then obtain
\begin{equation}
\mathcal{A}^1_{(4)}(g)= \big\langle \nordg{{O}^1_{(4)}(a)}\,\,\,\nordg{\left(\tr a^2\right)^2}\big\rangle=
- \frac{6\,\zeta(3)}{2}  \, \Big(\frac{g^2}{8\pi^2}\Big)^2  \,  
\mathcal{C}^{1}_{(4)} + \mathcal{O}(g^6)
\label{A14g}
\end{equation}
where
\begin{equation}
\label{C14two}
\mathcal{C}^1_{(4)} =
 \frac{ (2N-N_f)(N^2-1)(N^2-4)(N^2-9)}{2(N^2+1)}~.
\end{equation}
This expression follows directly from (\ref{loc02}), and vanishes in the conformal case when $N_f=2N$.

This analysis can be done also for operators of higher dimensions. For example, when $m=3$ we
find three operators of dimension six that satisfy the condition (\ref{g0cond}). They are
\begin{equation}
\begin{aligned}
{O}^1_{(6)} (a) &=\,{\tr a^6}+\,\,c_{2,2,2}^1 \,\,{\left(\tr  a^2 \right)^3} ~, \\
{O}^2_{(6)} (a) &= \,{\tr a^6}+\,\,c_{3,3}^1  \,\,{\left(\tr  a^3 \right)^2} ~, \\
{O}^3_{(6)} (a) &=\,{\tr a^6}+\,\,c_{4,2}^1   \,\,{\tr a^4\,\tr  a^2}~,
\end{aligned}
\label{Opm3}
\end{equation}
where the coefficients, determined with the method described above, are
\begin{equation}
\label{c6}
\begin{aligned}
c_{2,2,2}^1&=-\frac{t_6}{t_{2,2,2}}=-\frac{5(N^4-3N^2+3)}{N^2(N^2+1)(N^2+3)}~,\\
c_{3,3}^1 &=-\frac{t_6}{t_{3,3}}=-\frac{5( N^4-3N^2+3)}{3N(N^2-4)} ~,\\
c_{4,2}^1 &=-\frac{t_6}{t_{4,2}}=-\frac{5(N^4-3N^2+3)}{N(N^2+3)(2N^2-3)}~.
\end{aligned}
\end{equation}
Using these results in (\ref{2loopterm}), we find
\begin{equation}
\mathcal{A}^i_{(6)}(g)= \big\langle \nordg{{O}^i_{(6)}(a)}\,\,\,\nordg{\left(\tr a^2\right)^2}\big\rangle=
-\frac{6\,\zeta(3)}{2}  \, \Big(\frac{g^2}{8\pi^2}\Big)^2  \,  
\mathcal{C}^{i}_{(6)} + \mathcal{O}(g^6)
\label{Ai6g}
\end{equation}
with
\begin{equation}
\label{Ci6two}
\begin{aligned}
{\mathcal C}_{(6)}^1 &=
 \frac{ 9(2N-N_f)(N^2-1)(N^2-4)(N^2-9)(2N^2-5)}{ 4N (N^2+1)} ~, \\
{\mathcal C}_{(6)}^2 &=
-\frac{9(2N-N_f) (N^2-1)(N^2-9)(3N^2-5)}{4N}  ~,\\
{\mathcal C}_{(6)}^3 &=
 \frac{3(2N-N_f) (N^2-1)(N^2-4)(N^2-9)(7N^4-39N^2+30)}{4N(N^2+3)(2N^2-3)} ~.
\end{aligned}
\end{equation} 
We notice again that these correlators vanish in the superconformal theories.

In the next section we will compare the correlators $\mathcal{A}^i_{(4)}(g)$ and 
$\mathcal{A}^i_{(6)}(g)$ with the corresponding ones computed in field theory using standard Feynman diagrams, finding perfect agreement.

\subsubsection{Three loops}
\label{subsec:threelr}
Our previous analysis shows that it is possible to find two independent linear combinations 
$\widehat{{O}}^i_{(6)}$, with $i=1,2$, of the operators (\ref{Opm3}) 
such that also the two-loop term of the corresponding correlator vanishes. 
Using (\ref{2loopterm}), we see that this amounts to require
\begin{equation}
\label{Ch6v}
\big\langle\nord{\widehat{{O}}^i_{(6)}(a)}\,\,\, \nord{\left(\tr a^2\right)^3}\, S_4(a)  
\big\rangle_0 = 0~.
\end{equation}
The two independent operators can be chosen as 
\begin{equation}
\begin{aligned}
\widehat{{O}}^1_{(6)}(a) 
&=\,{\tr a^6} \,+\,\, \widehat{c}^{\,1}_{4,2} \,\,{\tr a^4 \, \tr a^2}
\,+\,\,\widehat{c}^{\,1}_{2,2,2} \,\,{\left( \tr a^2 \right)^3} ~, \phantom{\Big|} \\
\widehat{{O}}^2_{(6)} (a) &=\,{\tr a^6} \,+\, \,\widehat{c}^{\,2}_{3,3}  \,\,{\left( \tr a^3 \right)^2}
\, +\,\,\widehat{c}^{\,2}_{2,2,2}   \, \,{\left( \tr a^2 \right)^3}  ~, \phantom{\Big|}
\end{aligned}
\label{ohat3}
\end{equation}
with
\begin{equation}
\begin{aligned}
\widehat{c}^{\,1}_{4,2} &= -\frac{3(2N^2-5)}{N(N^2+7)}~,  \qquad  \quad
 \widehat{c}^{\,1}_{2,2,2}= \frac{(7\, N^4-39 \, N^2 +30)}{N^2(N^2+3)(N^2+7)}~,\\
  \widehat{c}^{\,2}_{3,3} &= -\frac{2N^2-5}{3N} ~,  \qquad\quad\quad~
  \widehat{c}^{\,2}_{2,2,2}=- \frac{3N^2-5}{N^2(N^2+3)} ~.
\end{aligned}
\end{equation}
When a correlator vanishes up to order $g^4$, as is the case of the operators 
$\widehat{{O}}^i_{(6)}$ we have just introduced, the first non-trivial contribution 
occurs at three loops and takes the form
\begin{eqnarray}
\widehat{\mathcal{A}}^i_{(6)}(g)\Big|_{3-\mathrm{loop}}
&=& -\Big(\frac{g^2}{8\pi^2}\Big)^3 
\big\langle
\nord{\widehat{{O}}^i_{(6)}(a)}\,\,\, \nord{\left(\tr a^2\right)^3}\, \big[
S_6(a) - S_4(a) S_2(a) + \frac{1}{6} S_2(a)^3 \big]\,\big\rangle_0\phantom{\bigg|}
\nonumber\\
& =& -\Big(\frac{g^2}{8\pi^2}\Big)^3
\big\langle
\nord{\widehat{{O}}^i_{(6)}(a)}\,\,\, \nord{\left(\tr a^2\right)^3}\, S_6(a)\,\big\rangle_0
\phantom{\bigg|}\label{3loopterm}\\&
=&\frac{20\,\zeta(5)}{3}  \,   \Big(\frac{g^2}{8\pi^2} \Big)^3 \,  
\widehat{\mathcal{C}}^{\,i}_{(6)}\phantom{\bigg|}\nonumber
\end{eqnarray}
with 
\begin{equation}
\widehat{\mathcal{C}}^{\,i}_{(6)}   =\frac{1}{20}\,
 \big\langle
 \nord{\widehat{{O}}_{(6)}^i(a)}\,\, \,  \nord{\left(\tr a^2\right)^3} \,   
 \big[ (2N -N_f) \tr a^6  +30\,  \tr a^4\,\tr a^2 - 20 \left(\tr a^3\right)^2 \big]\,
 \big\rangle_0~.
 \label{loc03}
\end{equation}
In (\ref{3loopterm}) we used again the fact that terms containing $S_2$ vanish in this situation
and, in the last step, inserted the explicit form of $S_6$ from (\ref{Sintexp}) to obtain
$\widehat{\mathcal{C}}^{\,i}_{(6)}$.
The explicit evaluation of this expression for the operators (\ref{ohat3}) gives the following result:
\begin{equation}
\begin{aligned}
\widehat{\mathcal{C}}^{\,1}_{(6)}  &=\frac{ 9 (N^2-1)(N^2-4)}{16 N^2(N^2+3)(N^2+7)}
\,\Big[2N^3  (N^6-118N^4+897N^2-4620)\\
&\qquad\qquad\qquad -N_f (N^8 -28 N^6+477 N^4-1890 N^2-2400) \Big]~,\\
\widehat{\mathcal{C}}^{\,2}_{(6)}  &=\frac{45 (N^2-1)(N^2-4)}{16 N^2(N^2+3)}\,
\Big[2N^3 (3N^4+10N^2-109) \\
&\qquad\qquad\qquad+N_f (N^6 -10 N^4+25 N^2+ 80)\Big]~.
\end{aligned}
\label{ex33}
\end{equation}
In the next section these findings will be shown to match with the corresponding ones obtained with a perturbative field theory computation at three loops.

\section{Two-point correlators in field theory}
\label{secn:fieldtheory}
In this section we compute two-point correlation functions
in $\mathcal{N}=2$ gauge theories using standard Feynman diagram techniques.
In order to be self-contained we begin by briefly describing the set-up and the Feynman rules
in the superspace formalism.

\subsection{Superfield actions and Feynman rules}

We use the $\mathcal{N}=1$ superfield formulation of $\mathcal{N}=2$ theories
in Euclidean space and collect our conventions and notations in Appendix~\ref{secn:conventions}.  In this formulation, the $\mathcal{N}=2$ vector multiplet consists of a $\mathcal{N}=1$ vector
multiplet $V$ and a chiral multiplet $\Phi$ in the adjoint representation of the gauge group.
The corresponding gauge theory action is
\begin{equation}
\begin{aligned}
S_{\mathrm{gauge}}&=\frac{1}{8g^2}\bigg(\!\int\!d^4x\,d^2\theta\,\mathrm{tr}\big(W^\alpha W_\alpha\big)+ \mbox{h.c.}\bigg)\!+
2 \!\int\!d^4x\,d^2\theta\,d^2\bar{\theta}~\mathrm{tr}
\big(\rme^{-2gV}\,\Phi^\dagger\,\rme^{2gV}\,\Phi\big)\\
&\quad-\frac{\xi}{4}\!\int\!d^4x\,d^2\theta\,d^2\bar{\theta}~\mathrm{tr}\big(\bar{D}^2V\,D^2V\big)~,
\end{aligned}
\label{Sgauge}
\end{equation}
where $g$ is the gauge coupling (for simplicity we have set to zero the Yang-Mills $\theta$ angle) 
and $W_\alpha$ is the chiral superfield-strength of $V$, namely
\begin{equation}
W_\alpha=-\frac{1}{4}\bar{D}^2\Big(\rme^{-2gV} D_\alpha\,\rme^{2gV}\Big)
\label{Wsusy}
\end{equation}
with
\begin{equation}
D_\alpha=+\frac{\partial}{\partial \theta^\alpha}+\ii\,\sigma^\mu_{\alpha\dot{\alpha}}
\bar{\theta}^{\dot{\alpha}}\partial_\mu~,~~
\bar{D}_{\dot{\alpha}}=-\frac{\partial}{\partial \bar{\theta}^{\dot{\alpha}}\phantom{\big|}}
-\ii\,\theta^\alpha\sigma^\mu_{\alpha\dot{\alpha}} \partial_\mu~.
\label{DbarD}
\end{equation}
The last term in (\ref{Sgauge}) implements the gauge fixing, with $\xi=1$ corresponding to 
the Fermi-Feynman gauge\,\footnote{While in general the gauge choice ({\it{i.e.}} different values of $\xi$ 
in (\ref{Sgauge})) is a matter of taste, in a supersymmetric field theory the choice of the Fermi-Feynman gauge ($\xi=1$) is obligatory, since otherwise the auxiliary field in the vector multiplet with zero canonical dimension
has a logarithmic propagator. Note that $S_{\mathrm{gauge}}$ should contain also a ghost contribution, 
which however is not relevant for our calculations, and thus we omit it.}. 
With our conventions, the action $S_{\mathrm{gauge}}$, as well as the actions we are going 
to write hereinafter, is negative defined and thus it appears in the path integral as 
$\rme^{S_{\mathrm{gauge}}}$.

Writing $V= V^aT^a$, $\Phi=\Phi^aT^a$ where $T^a$ are the generators of SU($N$) in the fundamental representation normalized as in (\ref{defgen}), and expanding up to second order in $V$ the action (\ref{Sgauge}) with $\xi=1$ becomes
\begin{equation}
\begin{aligned}
S_{\mathrm{gauge}}&=\!\int\!d^4x\,d^2\theta\,d^2\bar{\theta}\,\Big(
\!\!-V^a\square V^a+\Phi^{\dagger a}\Phi^a+2\ii g f^{abc}\,\Phi^{\dagger a}V^b\Phi^c
\\&\qquad\qquad\qquad\qquad-2g^2 f^{abe}f^{ecd}\,\Phi^{\dagger a}V^b V^c \Phi^d
+\cdots\Big)~,
\end{aligned}
\label{Sgaugeexp}
\end{equation}
where $f^{abc}$ are the structure constants of SU($N$).

In $\mathcal{N}=1$ language, a fundamental $\mathcal{N}=2$ massless hypermultiplet 
consists of two chiral multiplets $Q$ and $\tilde Q$ transforming in 
the fundamental and anti-fundamental representation, respectively, together with their 
anti-chiral partners $Q^{\dagger}$ and $\tilde Q^\dagger$. The action for $N_f$ 
such hypermultiplets is
\begin{equation}
\begin{aligned}
S_{Q}&=\sum_{A=1}^{N_f}\bigg[\!\int\!d^4x\,d^2\theta\,d^2\bar{\theta}\,\Big(Q_A^\dagger\,
\rme^{2gV} Q_A+ 
\widetilde{Q}_A \,\rme^{-2gV}\widetilde{Q}^\dagger_A\Big)
\\&\qquad\qquad
+\Big(\ii\sqrt{2}g\!\int\!d^4x\,d^2\theta\,\widetilde{Q}_A\Phi\,Q_A+\mathrm{h.c.}\Big)\bigg]
\end{aligned}
\label{SQ}
\end{equation}
which, up to second order in $V$, explicitly reads
\begin{align}
S_{Q}&=\sum_{A=1}^{N_f}\!\int\!d^4x\,d^2\theta\,d^2\bar{\theta}\,\Big(Q_{A u}^\dagger Q_{A u}\!+\!
2g\, Q_{A u}^\dagger V^aT^a_{uv}\,Q_{Av}\!+\! 2g^2\,Q_{A u}^\dagger
V^aT^a_{uw} V^bT^b_{wv} Q_{Av}
\notag\\
&~\qquad\quad+\widetilde{Q}_{Au}\widetilde{Q}^\dagger_{Au}-2g\,\widetilde{Q}_{Au} V^aT^a_{uv}
\widetilde{Q}^\dagger_{Av}+2g^2\,\widetilde{Q}_{Au}V^aT^a_{uw} V^bT^b_{wv}
\widetilde{Q}^\dagger_{Av}+\cdots\notag\phantom{\sum_A}\\
&~\qquad\quad+\ii\sqrt{2}g\,\widetilde{Q}_{Au}\Phi^aT^a_{uv} Q_{A v}\,\bar{\theta}^2-
\ii\sqrt{2}g\,Q^\dagger_{Au} \Phi^{\dagger a}T^a_{uv} \widetilde{Q}^\dagger_{A v}\,\theta^2
\Big)~.
\label{SQexp}
\end{align}
Here we have understood the summation over $u,v=1,\cdots,N$.

The action of the $\mathcal{N}=2$ super Yang-Mills theory 
with $N_f$ massless hypermultiplets can therefore be written as
\begin{equation}
S_{\mathcal{N}=2}^{(N_f)}=  S_{\mathrm{gauge}} +S_{Q} ~.
\label{SN2nf}
\end{equation}
For our purposes it is convenient to view the pure $\mathcal{N}=2$ theory as a truncation of the maximally supersymmetric $\mathcal{N}=4$ theory. Indeed, the latter can be obtained by adding to the 
pure $\mathcal{N}=2$ theory the contributions of two adjoint chiral multiplets $\Phi_2$ and $\Phi_3$, 
which together with their conjugate $\Phi^{\dagger}_2$ and 
$\Phi^{\dagger}_3$ build an adjoint $\mathcal{N}=2$ hypermultiplet
$H$. Thus the action of the $\mathcal{N}=4$ theory is
\begin{equation}
S_{\mathcal{N}=4}=S_{\mathrm{gauge}}+S_H
\label{SN4}
\end{equation}
with
\begin{equation}
\begin{aligned}
S_{H}&=2\sum_{I=2,3} \int\!d^4x\,d^2\theta\,d^2\bar{\theta}\,\,
\mathrm{tr}\big(\rme^{-2gV}\Phi_I^\dagger
\,\rme^{2gV}\,\Phi_I\big)
\\&\qquad
+\frac{g\sqrt{2}}{3}\sum_{I,J,K=1}^3\bigg[\int\!d^4x\,d^2\theta\,\,\epsilon_{IJK}\,
\mathrm{tr}\big(\Phi_I[\Phi_J,\Phi_k]\big)+\mbox{h.c.}\bigg]~,
\end{aligned}
\label{SH}
\end{equation}
where in the second line we have used the notation $\Phi_1=\Phi$. Up to second order in $V$, the action 
for the adjoint hypermultiplet is
\begin{align}
S_{H}&=\int\!d^4x\,d^2\theta\,d^2\bar{\theta}\,\,\bigg[\!\sum_{I=2,3} 
\big(
\Phi^{\dagger a}_I\Phi^a_I+2\ii g\,f^{abc}\,\Phi^{\dagger a}_IV^b\Phi^c_I
-2g^2\,f^{abe} f^{ecd}\,\Phi^{\dagger a}_IV^b V^c \Phi^d_I+\cdots\big)\notag\\
&\qquad\qquad\qquad\qquad+\ii g\sqrt{2}\,f^{abc}\,\Phi^a\,\Phi_2^b\,\Phi_3^c\,\bar{\theta}^2
-\ii g\sqrt{2}
\,f^{abc}\,\Phi^{\dagger a}\,\Phi_2^{\dagger b}\,\Phi_3^{\dagger c}\,\theta^2\bigg]~.
\label{SHexp}
\end{align}
For later convenience, we now display the Feynman rules for the various superfields 
that can be obtained from the actions (\ref{Sgaugeexp}), (\ref{SQexp}) and (\ref{SHexp}).
In configuration space, the superpropagators can be compactly written, using the 
following notations \cite{Wess:1992cp}:
\begin{equation}
x_{ij}=x_i-x_j~,~~ \theta_{ij}=\theta_i-\theta_j~,~~
\bar{\theta}_{ij}=\bar{\theta}_i-\bar{\theta}_j~,~~\xi_{ij}=\ii\,\theta_i\sigma\bar{\theta}_j~,
\end{equation}
 as indicated in Fig.~1.
 \begin{fmffile}{propagators}
\begin{eqnarray}
 \mbox{$\Phi\,\Phi^\dagger$ - propagator:}\qquad\quad ~~ &&   \parbox{30mm}{
 \begin{fmfgraph*}(80,40)
 \fmfleft{i1}
\fmfright{o1}
\fmflabel{$I,a,x_1,\theta_1$}{i1}
\fmflabel{$J,b,x_2,\theta_2$}{o1}
\fmf{fermion}{i1,o1}
     \end{fmfgraph*}
 }  \qquad ~~~= \,~ \delta_{IJ} \,  \delta^{ab} \, \rme^{(\xi_{11}+\xi_{22}-2\xi_{12})\cdot\partial_{x_1}} \frac{1}{4\pi^2x_{12}^2}\nn  \\\nn
  \mbox{$Q\,Q^\dagger$ - propagator:}\qquad\quad ~~  &&   \parbox{30mm}{
 \hspace{10pt}\begin{fmfgraph*}(80,40)
  \fmfleft{i1}
\fmfright{o1}
\fmflabel{$ A,i,x_1,\theta_1$}{i1}
\fmflabel{$B,j,x_2,\theta_2$}{o1}
\fmf{dashes_arrow}{i1,o1}
     \end{fmfgraph*}
 }  \qquad ~~~= ~
 \delta_{AB}\, \delta_{uv} \, \rme^{(\xi_{11}+\xi_{22}-2\xi_{12})\cdot\partial_{x_1}} 
 \frac{1}{4\pi^2 x_{12}^2} \nn \\ \nn
  \mbox{$\widetilde{Q}\,\widetilde{Q}^\dagger$ - propagator:}\qquad\quad ~~  &&   \parbox{30mm}{
 \begin{fmfgraph*}(80,40)
  \fmfleft{i1}
\fmfright{o1}
\fmflabel{$ A,i,x_1,\theta_1$}{i1}
\fmflabel{$B,j,x_2,\theta_2$}{o1}
\fmf{dots_arrow}{i1,o1}
     \end{fmfgraph*}
 }  \qquad ~~~=  ~\delta_{AB}\, \delta_{uv}  \, \rme^{(\xi_{11}+\xi_{22}-2\xi_{12})\cdot\partial_{x_1}} \frac{1}{4\pi^2 x_{12}^2} \nn \\\nn
\mbox{$V\,V$ - propagator:}\qquad\quad ~~ &&  \parbox{30mm}{
\begin{fmfgraph*}(80,40)
\fmfleft{i1}
\fmfright{o1}
\fmflabel{$a,x_1,\theta_1$}{i1}
\fmflabel{$b,x_2,\theta_2$}{o1}
\fmf{photon}{i1,o1}
   \end{fmfgraph*}
} \qquad ~~~=  ~- \delta^{ab}\,
\frac{\theta_{12}^2 \, \bar \theta_{12}^2}{8\pi^2 x_{12}^2}\nn
 \end{eqnarray}
  \end{fmffile}
\centerline{\small{\bf{Figure~1}}: {\small{The superpropagators in the $\mathcal{N}=2$ theory in configuration superspace.}}}
\vskip 0.5cm
\noindent
The cubic interaction vertices that can be read from the actions (\ref{Sgaugeexp}), (\ref{SQexp}) and (\ref{SHexp}) are listed in Fig.~2.
\begin{fmffile}{vertices}
\vspace{0.5cm}
\begin{eqnarray}
\mbox{$(\Phi^{\dagger a}_I\,\Phi^{\dagger b}_J\,\Phi^{\dagger c}_K)$ - vertex:}\qquad~~
&&       \parbox{30mm}{
 \begin{fmfgraph*}(70,45)
 \fmfleft{i1,i2}
\fmfright{o1}
\fmflabel{\small{$K,c$}}{i2}
\fmflabel{\small{$I,a$}}{i1}
\fmflabel{\small{$J,b$}}{o1}
\fmf{fermion}{i1,v1}
\fmf{fermion}{i2,v1}
\fmf{fermion}{o1,v1}
     \end{fmfgraph*}
 }
\quad ~ = ~- \frac{ {\ii} g \sqrt{2}}{3!}  \, \epsilon_{IJK} \,  f^{abc}\, \theta^2
\nn\\
\nn\\
\nn\\
\nn\\
\mbox{$(\Phi^a_I\,\Phi^b_J\,\Phi^c_K)$ - vertex:}\qquad~~
&&     \parbox{30mm}{
 \begin{fmfgraph*}(70,45)
\fmfleft{i1,i2}
\fmfright{o1}
\fmflabel{\small{$K,c$}}{i2}
\fmflabel{\small{$I,a$}}{i1}
\fmflabel{\small{$J,b$}}{o1}
\fmf{fermion}{v1,i1}
\fmf{fermion}{v1,i2}
\fmf{fermion}{v1,o1}
     \end{fmfgraph*}
 } 
\quad ~  = ~~ \frac{\ii g \sqrt{2}}{ 3!} \, \epsilon_{IJK} \,  f^{abc}\, \bar\theta^2  
\nn\\
\nn\\
\nn\\
\nn\\
\mbox{$(Q^\dagger_u\,\Phi^{\dagger a}_I\,\widetilde{Q}^\dagger_v)$ - vertex:}\qquad~~
 &&       \parbox{30mm}{
 \begin{fmfgraph*}(70,45)
 \fmfleft{i1,i2}
\fmfright{o1}
\fmflabel{\small{$v$}}{i2}
\fmflabel{\small{$u$}}{i1}
\fmflabel{\small{$I,a$}}{o1}
\fmf{dashes_arrow}{i1,v1}
\fmf{dots_arrow}{i2,v1}
\fmf{fermion}{o1,v1}
     \end{fmfgraph*}
 }  
 \quad ~  = ~-     {\ii} g \sqrt{2}   \, \delta_{I1}\,T^{a}_{uv}\, \theta^2  
\nn\\
\nn\\
\nn\\
\nn\\
\mbox{$(\widetilde{Q}_u\,\Phi^a_I\,Q_v)$ - vertex:}\qquad~~
&&    \parbox{30mm}{
 \begin{fmfgraph*}(70,45)
\fmfleft{i1,i2}
\fmfright{o1}
\fmflabel{\small{$v$}}{i2}
\fmflabel{\small{$u$}}{i1}
\fmflabel{\small{$I,a$}}{o1}
\fmf{dots_arrow}{v1,i1}
\fmf{dashes_arrow}{v1,i2}
\fmf{fermion}{v1,o1}
     \end{fmfgraph*}
 } 
\quad ~  = ~~  {\ii} g \sqrt{2} \, \delta_{I1} \,  T^{a}_{uv} \, \bar\theta^2
\nn\\
\nn\\
\nn\\
\nn\\
\mbox{$(\Phi^{\dagger a}_I\,V^b\,\Phi^c_J)$ - vertex:}\qquad~~
 &&   \parbox{30mm}{
 \begin{fmfgraph*}(70,45)
\fmfleft{i1,i2}
\fmfright{o1}
\fmflabel{\small{$b$}}{o1}
\fmflabel{\small{$J,c$}}{i2}
\fmflabel{\small{$I,a$}}{i1}
\fmf{fermion}{i1,v1,i2}
\fmf{photon}{o1,v1}
     \end{fmfgraph*}
 } 
 \quad ~  = ~~  2{\ii}g  \, \delta_{IJ}\, f^{abc} 
 \nn\\
\nn\\
\nn\\
\nn\\
\mbox{$(Q^{\dagger}_u\,V^a\,Q_v)$ - vertex:}\qquad~~
&&   \parbox{30mm}{
 \begin{fmfgraph*}(70,45)
\fmfleft{i1,i2}
\fmfright{o1}
\fmflabel{\small{$a$}}{o1}
\fmflabel{\small{$v$}}{i2}
\fmflabel{\small{$u$}}{i1}
\fmf{dashes_arrow}{i1,v1,i2}
\fmf{photon}{o1,v1}
     \end{fmfgraph*}
 } 
 \quad ~  = ~~ 2g\,  T^{a}_{uv} 
 \nn\\
\nn\\
\nn\\
\nn\\
\mbox{$(\widetilde{Q}_u\,V^a\,\widetilde{Q}^\dagger_v)$ - vertex:}\qquad~~
  &&  \parbox{30mm}{
 \begin{fmfgraph*}(70,45)
\fmfleft{i1,i2}
\fmfright{o1}
\fmflabel{\small{$a$}}{o1}
\fmflabel{\small{$v$}}{i2}
\fmflabel{\small{$u$}}{i1}
\fmf{dots_arrow}{i2,v1,i1}
\fmf{photon}{o1,v1}
     \end{fmfgraph*}
 }
  \quad ~  = ~- 2g\,  T^{a}_{uv}\nn
  \end{eqnarray}
   \end{fmffile}
\\\\
\centerline{\small{\bf{Figure~2}}: {\small{The cubic interaction vertices of the actions (\ref{Sgaugeexp}), (\ref{SQexp}) and (\ref{SHexp}).}}}
\vskip 0.5cm
\noindent
There are also higher order interaction vertices, but they are not needed for our calculations 
and thus we do not write them explicitly here.

In this set-up, a correlator in the $\mathcal{N}=2$ theory can be written in the form
\begin{equation}
\mathcal{A}^{(N_f)}_{\mathcal{N}=2}\,=\,\mathcal{A}_{\mathcal{N}=4}
+\mathcal{A}_{Q}-\mathcal{A}_{H}~,
\label{ampldec}
\end{equation}
where $\mathcal{A}_{H}$ corresponds to the contribution of the diagrams in the $\mathcal{N}=4$  theory in which the adjoint
chiral multiplets $\Phi_2$ and $\Phi_3$ that build the hypermultiplet $H$ run in the internal lines, and $\mathcal{A}_{Q}$ corresponds to the diagrams in which matter
chiral multiplets $Q$ and $\widetilde{Q}$ run in the internal lines. We call them $H$- and $Q$-diagrams, respectively. Furthermore, if one considers correlators among BPS protected states, one can simplify (\ref{ampldec}) because such correlators do not receive corrections in the $\mathcal{N}=4$ theory, namely
$\mathcal{A}_{\mathcal{N}=4}=\mathcal{A}_0$, so that we have
\begin{equation}
\mathcal{A}^{(N_f)}_{\mathcal{N}=2}\,=\,\mathcal{A}_0
+\mathcal{A}_{Q}-\mathcal{A}_{H}~.
\label{ampldec1}
\end{equation}
We will extensively make use of this formula to study the two-point functions of 
half-BPS operators in $\mathcal{N}=2$ theories and compare them with the matrix model results of the
previous sections. In this respect it is important to realize that the $H$-diagrams that give rise 
to $\mathcal{A}_{H}$ do not really exist in the $\mathcal{N}=2$ theories under consideration,
but they provide a simple book-keeping device which greatly simplifies our analysis.

In the following we are going to consider two-point functions between chiral and anti-chiral operators.
The former are constructed only with the lowest component of the chiral superfield 
$\Phi_1=\Phi$ of the $\mathcal{N}=2$ vector multiplet, that is the adjoint scalar
\begin{equation}
\varphi(x) = \Phi(x,\theta,\bar\theta)\big|_{\theta=\bar\theta=0} \,=\, \varphi^a(x)\,T^a~,
\label{phix}
\end{equation}
while the latter are made only with the conjugate scalar 
\begin{equation}
\bar\varphi(x) = \Phi^\dagger(x,\theta,\bar\theta)\big|_{\theta=\bar\theta=0} \,=\, \bar\varphi^a(x)\,T^a~.
\label{barphix}
\end{equation}
Gauge invariant chiral operators of dimension $n$ can be generically written as
\begin{equation}
\begin{aligned}
{O}_{(n)}^i \big(\varphi(x)\big) &=  R_{a_1 a_2 \cdots a_{n}}^i\, \varphi^{a_{1}}(x) \,
\varphi^{a_{2}}(x) \cdots \varphi^{a_{n}}(x)~,
\end{aligned}
\label{Ooperators}
\end{equation}
where $R^i$ is a totally symmetric $n$-index tensor of SU($N$). 
The superscript $i$ labels the different  such tensors and hence operators 
that can be considered. The anti-chiral operators can be defined in a similar way by 
replacing $\varphi(x)$ with $\bar\varphi(x)$. To simplify
the formulae in the following, we adopt the notation
\begin{equation}
{O}_{(n)}^i(x) \,\equiv\, {O}_{(n)}^i \big(\varphi(x)\big)\quad
\mbox{and}\quad
\bar{{O}}_{(n)}^i(x) \,\equiv\, {O}_{(n)}^i \big(\bar\varphi(x)\big)~.
\label{OObaroperators}
\end{equation}
Examples of these are the multi-trace operators
\begin{equation}
\mathrm{tr}\,\varphi^{n_1}(x)\cdots\mathrm{tr}\,\varphi^{n_\ell}(x)\qquad
\mbox{and}\qquad\mathrm{tr}\,\bar\varphi^{n_1}(x)\cdots\mathrm{tr}\,\bar\varphi^{n_{k}}(x)~,
\end{equation}
or combinations thereof.

The two-point functions we will consider are
\begin{equation}
\big\langle \,{O}_{(n)}^i(x_1)\,
\bar{{O}}_{(n)}^j(x_2)
\,\big\rangle
\label{2point}
\end{equation}
for various choices of $i$ and $j$. At tree-level we can easily compute this correlator using the
propagator
\begin{equation}
\big\langle\varphi^{a}(x_1)\,\bar\varphi^b(x_2)\big\rangle=\frac{\delta^{ab}}{4\pi^2 x_{12}^2\phantom{\Big|}}\,
\label{phiprop}
\end{equation}
which is obtained from the $\Phi\Phi^\dagger$ superpropagator in Fig.~1 by setting to zero all fermionic superspace coordinates, and fixing $I=J=1$. Performing all contractions, we find
\begin{equation}
\big\langle \,{O}_{(n)}^i(x_1)\,
\bar{{O}}_{(n)}^j(x_2)
\,\big\rangle_{\mathrm{0-loop}} \,=\frac{\,n!\,R^i_{a_1\cdots a_n}\,{R}^{j\,a_1\cdots a_n} }{(4\pi^2 x_{12}^2)^{n}\phantom{\Big|}}~.
\label{2point0}
\end{equation}
We can pictorially describe this result as shown in Fig.~3 where each oriented line
represents the tree-level propagator (\ref{phiprop}).
\begin{fmffile}{tree}
\vspace{0.2cm}
 \begin{eqnarray}
  &&\!\!\! \!\!\! \!\!\! \!\!\! \!\!\! \!\!\! \!\!\!  \!
  \parbox{15mm}{
 \begin{fmfgraph*}(120,20)
  \fmfleft{i1}
\fmfright{o1}
\fmflabel{\small{$a_1$}}{i1}
\fmflabel{\small{$b_1$}}{o1}
\fmf{fermion}{i1,o1}
 \end{fmfgraph*}
 }\nn \\
   &&\!\!\! \!\!\! \!\!\! \!\!\! \!\!\! \!\!\! \!\!\!  \!  \parbox{15mm}{
 \begin{fmfgraph*}(120,20)
  \fmfleft{i1}
\fmfright{o1}
\fmflabel{\small{$a_2$}}{i1}
\fmflabel{\small{$b_2$}}{o1}
\fmf{fermion}{i1,o1}
     \end{fmfgraph*}
 } \nn \\
  && \qquad\!\!\vdots  \nn \\
    && \!\!\! \!\!\! \!\!\! \!\!\! \!\!\! \!\!\! \!\!\!  \! \parbox{15mm}{
 \begin{fmfgraph*}(120,20)
  \fmfleft{i1}
\fmfright{o1}
\fmflabel{\small{$a_n$}}{i1}
\fmflabel{\small{$b_n$}}{o1}
\fmf{fermion}{i1,o1}
     \end{fmfgraph*}
 } \nn 
  \end{eqnarray}
   \end{fmffile}
\centerline{\small{\bf{Figure~3}}: {\small{The diagram representing the tree-level correlator (\ref{2point0}).}}}
\vskip 0.7cm
%%%%%%%%%%%%%%%%%%%%%%%%%%%%%%%%%%%%%%%%%%%%%%%%%%%%%%%%%%%
\noindent
At higher orders the calculation of the two-point function (\ref{2point}) is less straightforward and
crucially depends on the type of $\mathcal{N}=2$ theory one deals with. 
In the next subsection we start by considering the superconformal theories corresponding to $N_f=2N$.
 
\subsection{The conformal case $N_f=2N$}
\label{subsecn:conf}
When $N_f=2N$, the $\mathcal{N}=2$ theory is superconformal invariant.
In this case many simplifications occur: for example, due to the vanishing of the $\beta$-function, 
the gauge coupling is not renormalized and the chiral/anti-chiral operators retain their engineering 
scaling dimensions without anomalous terms. Furthermore, the one-loop corrections to the propagators 
and to the three-point coupling exactly cancel between the $Q$- and $H$-diagrams, 
so that any diagram containing this difference as a subdiagram vanishes for arbitrary choices of the operators 
${O}^i_{(n)}$ and ${O}^j_{(n)}$. For example, the one-loop diagram represented in
Fig.~4 vanishes. 
 %%%%%%%%%%%%%%%%%%%%%%%%%%%%%%%%%%%%%%%%%%%%%%%%%%%%%%%%%%
\begin{fmffile}{1loop}
\vspace{0.5cm}
 \begin{eqnarray}
    && \!\!\!\! \!\!\!\!  \!\!\!\!  \!\!\!\! \!\!\!\!\parbox{30mm}{
 \begin{fmfgraph*}(120,20)
 \fmfleft{i1}
   \fmfright{o1}
   \fmflabel{\small{$b_1$}}{o1}
\fmflabel{\small{$a_1$}}{i1}
   \fmf{fermion}{i1,v1,o1}
    \fmfv{decor.shape=circle,decor.filled=30, decor.size=35 }{v1}
\fmf{phantom,label.dist=0,label=\small{$\mathbf{Q-H}$}}{i1,o1}
     \end{fmfgraph*}
 } \nn \\
   &&  \!\!\!\! \!\!\!\!  \!\!\!\!  \!\!\!\!  \!\!\!\!\parbox{30mm}{
 \begin{fmfgraph*}(120,20)
  \fmfleft{i1}
\fmfright{o1}
\fmflabel{\small{$a_2$}}{i1}
\fmflabel{\small{$b_2$}}{o1}
\fmf{fermion}{i1,o1}
     \end{fmfgraph*}
 } \nn \\ && \qquad \qquad \!\!\!\! \!\!\!\! \!\!\!\!\vdots \nn \\
    &&  \!\!\!\! \!\!\!\!   \!\!\!\! \!\!\!\! \!\!\!\!\parbox{15mm}{
 \begin{fmfgraph*}(120,20)
  \fmfleft{i1}
\fmfright{o1}
\fmflabel{\small{$a_n$}}{i1}
\fmflabel{\small{$b_n$}}{o1}
\fmf{fermion}{i1,o1}
     \end{fmfgraph*}
 } \nn 
  \end{eqnarray}
   \end{fmffile}
 \\
{\small{\bf{Figure~4}}}: {\small{The diagram representing the one-loop contribution to the correlator (\ref{2point}). The label $Q-H$ in the loop means that this is the difference between the $Q$ and $H$ contributions.}}
\vskip 0.7cm
\noindent
This implies that
\begin{equation}
\big\langle \,{O}_{(n)}^i(x_1)\,
\bar{{O}}_{(n)}^j(x_2)
\,\big\rangle_{\mathrm{1-loop}} \,=\,0~.
\label{2point1}
\end{equation}
Similarly, the two-loop subdiagram of Fig.~5 vanishes when the $Q-H$ difference
is computed for $N_f=2N$ and thus all diagrams containing it can be discarded.
\vskip .5cm
\begin{fmffile}{Hdiagram}
 \begin{eqnarray}
    &&~~~~~
  \parbox{50mm}{
 \begin{fmfgraph*}(120,60)
 \fmfleft{i1,i2}
   \fmfright{o1,o2}
   \fmflabel{$b_j$}{o1}
\fmflabel{$a_i$}{i2}
\fmflabel{$b_i$}{o2}
\fmflabel{$a_j$}{i1}
   \fmf{fermion}{i1,v1,o1}
    \fmfv{decor.shape=circle,decor.filled=30, decor.size=35 }{v1}
\fmf{phantom,label.dist=0,label=$~{\bf Q}-{\bf H}$ }{i1,o1}
   \fmf{fermion}{i2,v2,o2}
  \fmf{photon,tension=0}{v1,v2}
     \end{fmfgraph*}
 }
\nn
  \end{eqnarray}
   \end{fmffile}
   \\\\
   \noindent
{\small{\bf{Figure~5}}}: {\small{A two-loop subdiagram containing the one-loop correction to the gauge coupling
that vanishes in the superconformal theory with $N_f=2N$.}}
\vskip 0.5cm
\noindent
The only diagrams that can contribute at two loops are those containing
the irreducible corrections to the propagator represented in Fig.~6,
or the two-loop diagrams drawn in Fig.~7.
 \\
  \begin{fmffile}{2loop}
 \begin{eqnarray}
 \hspace{-3.3cm}&
 \parbox{20mm}{  \begin{fmfgraph*}(150,80)
 \fmfleft{w}\fmfright{e}
 \fmf{fermion}{w,vw}\fmf{fermion}{ve,e}
  \fmf{dashes_arrow,tension=.5}{ve,vn,vw}
    \fmf{dots_arrow,tension=.5}{ve,vs,vw}   \fmf{photon}{vn,vs}
 \fmffixed{(0,h)}{vn,vs}
% \fmfdot{vw,vn,ve,vs}
  \fmflabel{\small{$a$}}{w}
\fmflabel{\small{$b$}}{e}
 \end{fmfgraph*}
 }   
 \qquad \qquad \qquad  \qquad\qquad ~,\qquad\qquad
 \parbox{20mm}{  \begin{fmfgraph*}(150,80)
 \fmfleft{w}\fmfright{e}
 \fmf{fermion}{w,vw}\fmf{fermion}{ve,e}
  \fmf{fermion,tension=.5}{ve,vn,vw}
  \fmf{fermion,tension=.5}{ve,vs,vw}
   \fmf{photon}{vn,vs}
 \fmffixed{(0,h)}{vn,vs}
% \fmfdot{vw,vn,ve,vs}
  \fmflabel{\small{$a$}}{w}
\fmflabel{\small{$b$}}{e}
 \end{fmfgraph*}
 } 
\label{v4}
\nn
  \end{eqnarray}
   \end{fmffile}
   \\\\
{\small{\bf{Figure~6}}}: {\small{The irreducible two-loop correction to the scalar propagator. 
The left diagram describes the loop of the fundamental superfields $Q$ and $\widetilde{Q}$, while the right 
one accounts for the loop of the adjoint hypermultiplet $H$.}}
\vskip 0.5cm
\begin{figure}[h]
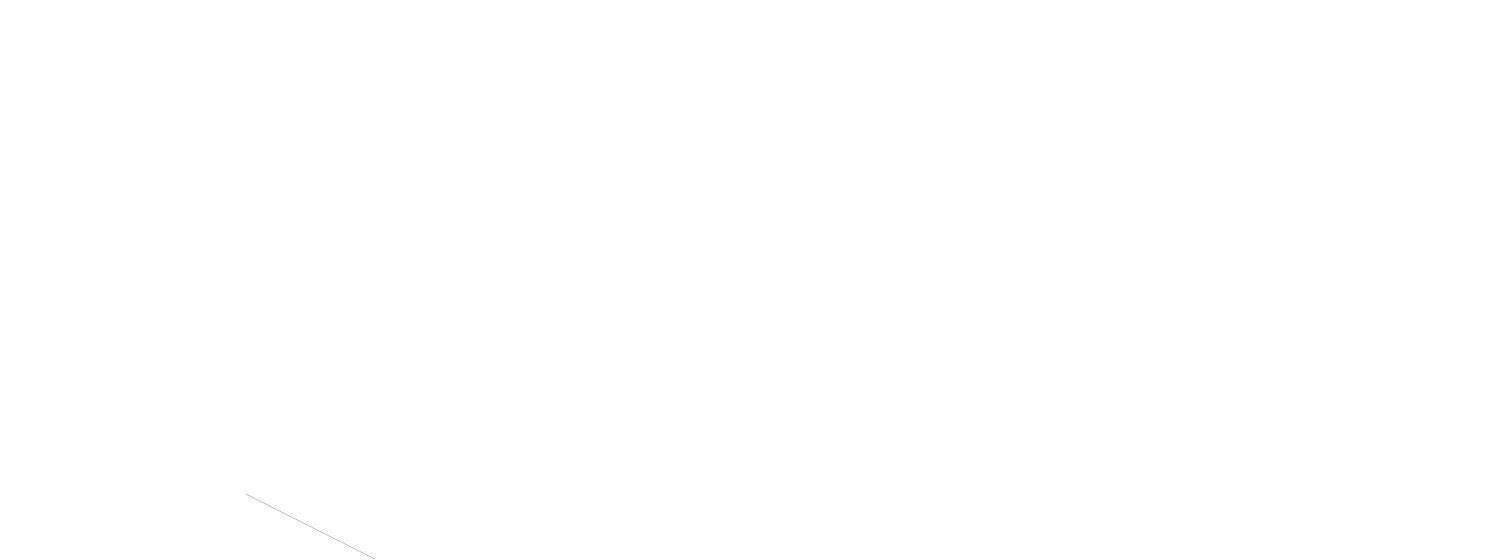
\vskip -1.5cm
\noindent
  {\small{\bf{Figure~7}}}: {\small{The two-loop diagrams that can contribute to the two-point function (\ref{2point}).
The left diagram refers to the contribution of the matter hypermultiplets while the right diagram refers
to the adjoint hypermultiplet $H$.}}
\end{figure}
%%%%%%%%%%%%%%%%%%%%%%%%%%%%%%%%%%%%%%%%%%%%%%%%%%%%%%%%%%%%
\vskip 0.2cm
\noindent
Their sum yields the two-loop correlator, which we write as
\begin{equation}
\big\langle \,{O}_{(n)}^i(x_1)\,
\bar{{O}}_{(n)}^j(x_2)
\,\big\rangle_{\mathrm{2-loop}} \,=
 \,\frac{\mathcal{A}^{ij}_{(n)}(g) \big|_{2-\mathrm{loop}}} {(4\pi^2 x_{12}^2)^{n}\phantom{\Big|}}~.
 \label{2point2}
\end{equation}
Here we adopted the same symbol $\mathcal{A}^{ij}_{(n)}(g)$ used for the matrix model correlators in the previous section since, as we will show, the two coincide.

We start by analyzing the contribution of the irreducible two-loop subdiagrams of Fig.~6. 
Since the propagators of the superfields that make the hypermultiplets $Q$ and $H$ differ only in their gauge index structure, the difference between the $Q$- and $H$-subdiagrams 
can be written in a factorized form as\,\footnote{The
overall prefactor is $-(2g)^2(\ii\sqrt{2}g)(-\ii\sqrt{2}g)=-8g^4$ and comes from the normalization of
the interaction vertices involved in these diagrams.}
\begin{equation}
-8g^4\,W_2(x_{12})\,C_2^{ab}~,
\label{A2}
\end{equation}
where $W_2(x_{12})$ is the result of the integral over the superspace internal variables
and the colour factor $C_2^{ab}$, following from the Feynman rules of Fig.~2, is given by
\begin{equation}
\begin{aligned}
C_2^{ab}&=N_f\,T^{a}_{u_4u_1}T^{c}_{u_1u_2}T^{b}_{u_2u_3}T^{c}_{u_3u_4}
-f^{ad_4d_1}f^{c d_1d_2}f^{bd_2d_3}f^{c d_3d_4}\phantom{\Big|}\\
&=N_f\,\mathrm{tr}\,T^{a}T^{c}T^{b}T^{c}-\mathrm{Tr}_{\mathrm{adj}}\,T^{a}T^{c}T^{b}T^{c}
\phantom{\Big|}
\end{aligned}
\label{C2}
\end{equation}
with $\mathrm{Tr}_{\mathrm{adj}}$ being the trace in the adjoint representation. Expressing this trace
in terms of traces in the fundamental representation and using 
the fusion/fission rules (\ref{identities}), we can
rewrite this factor as follows
\begin{equation}
C_2^{ab}=-\Big(N^2+\frac{N_f}{2N}\Big)\,\mathrm{tr}\,T^{a}T^{b}=-(N^2+1)
\,\mathrm{tr}\,T^{a}T^{b}~,
\label{C2ab}
\end{equation}
where in the last step we used the superconformal value $N_f=2N$.

The total contribution of these diagrams to the correlator (\ref{2point2}) is obtained by performing all contractions between ${O}^i$ and $\bar{{O}}^j$ as in the
tree-level computation but with one propagator replaced by (\ref{A2}). 
To perform these contractions in an efficient way, we introduce 
the notation $\vphi=\vphi^aT^a$ and $\bar\vphi=\bar\vphi^aT^a$ 
to denote auxiliary scalars with the Wick contraction
\begin{equation}
\big\langle   \vphi^a \,  \bar\vphi^b   \big\rangle=  \delta^{ab} ~.
\end{equation}
We also write the operators as ${O}^i_{(n)}(\vphi)$ and ${O}^j_{(n)}(\bar\vphi)$
by replacing $\varphi(x)$ and $\bar\varphi(x)$ with $\vphi$ and $\bar\vphi$ in 
${O}_{(n)}^i(x)$ and $\bar{{O}}^j_{(n)}(x)$, respectively.
Then, one can easily realize that replacing one propagator with (\ref{A2}) 
corresponds to inserting in the tree-level correlator the following effective 
vertex
 \begin{equation}
8g^4\, W_2(x_{12}) \, V_2( \vphi,\bar\vphi)~,
\label{effective2}
 \end{equation}
where
\begin{equation}
\begin{aligned}
V_2( \vphi,\bar\vphi)  &= -C_2^{ab} 
 :\vphi_{a}  \bar\vphi_{b}\!: \,\,\, = (N^2+1) :\mathrm{tr}\,\vphi \bar\vphi:~.
\end{aligned}
\label{vertexV2} 
\end{equation}
Here and below we will always represent effective vertices by normal-ordered products (denoted by $:\,:$) of  
$\vphi$ and $\bar\vphi$ in which all self-contractions are discarded. 
The scalars $\vphi$ and $\bar\vphi$ of $V_2$ will contract with, respectively, one $\bar\vphi$ of ${{O}}^j$ 
and one $\vphi$ of ${O}^i$; 
so in this way the replacement of one propagator with the expression in (\ref{A2}) is correctly implemented.
The contribution of the diagrams of Fig.~6 to the two-loop correlator (\ref{2point2}) is therefore
\begin{equation}
8g^4\,
\frac{W_2(x_{12})}{(4\pi^2x_{12}^2)^{n-1}\phantom{\Big|}}\,
\left\langle
{O}^i_{(n)} (\vphi) \, {{O}}^j_{(n)} (\bar\vphi) \,
V_2( \vphi,\bar\vphi) \right\rangle ~.
\label{2point2a}
\end{equation}
The last ingredient is the superspace integral $W_2(x_{12})$ which, as shown in Appendix~\ref{secn:W2}, is
\begin{equation}
 W_2(x_{12}) = -\frac{3\,\zeta(3)}{\big(16\pi^2\big)^2}\,\frac{1}{\big(4\pi^2x_{12}^2\big)\phantom{\Big|}}~.
\label{W2}
\end{equation}
Using this, we see that (\ref{2point2a}) becomes
\begin{equation}
-6\,\zeta(3)\,\Big(\frac{g^2}{8\pi^2}\Big)^2\,
\frac{1}{(4\pi^2x_{12}^2)^{n}\phantom{\Big|}}\,
\left\langle
{O}^i_{(n)} (\vphi) \,{{O}}^j_{(n)} (\bar\vphi) \,
V_2( \vphi,\bar\vphi) \right\rangle ~.
\label{2point2aa}
\end{equation}

Let us now turn to the diagrams in Fig.~7. Focusing on the irreducible subdiagrams, we see that just as before, the $Q-H$ difference can be written in a factorized
form as\,\footnote{The overall numerical factor $2g^4$ is explained as follows: we have a factor of $\frac{(\ii\sqrt{2}g)^2(-\ii\sqrt{2}g)^2}{(2!)^2\phantom{\big|}}$ coming from the insertions 
of the four three-point vertices, 
times a factor of 2 coming from the permutations of the fields $\Phi_2$ and $\Phi_3$ (or $Q$ and $\widetilde{Q}$) inside the loop.
Indeed one can make two independent loops with two $\Phi_2$ and two $\Phi_3$ (or two $Q$ and two 
$\widetilde{Q}$) in alternating positions as required by the Feynman rules. In general, the symmetry factor of a 
loop diagram with $2k$ three-point vertices is given by $\frac{(\sqrt{2}g)^{2k}}{k!^2\phantom{\big|}}$ 
times $k!(k-1)!$ which is the number of independent loops that can be made with $k$ $\Phi_2$ and $k$ $\Phi_3$ (or $k$ $Q$ and $k$ $\widetilde{Q}$) in alternating positions. The net result is therefore $\frac{2^kg^{2k}}{k}$. \label{footnotefactor}}
\begin{equation}
2g^4\,W_4(x_{12})\,C_4^{a_1a_2b_1b_2}
\label{A4}
\end{equation}
where $W_4(x_{12})$ is the result of the integral over the superspace internal variables
and the colour factor following from the Feynman rules is
\begin{equation}
\begin{aligned}
C_4^{a_1a_2b_1b_2}&=N_f\,T^{a_1}_{u_4u_1}T^{b_1}_{u_1u_2}T^{a_2}_{u_2u_3}T^{b_2}_{u_3u_4}-
f^{a_1d_4d_1}f^{b_1d_1d_2}f^{a_2d_2d_3}f^{b_2d_3d_4}\phantom{\Big|}\\
&=N_f\,\mathrm{tr}\,T^{a_1}T^{b_1}T^{a_2}T^{b_2}-
\mathrm{Tr}_{\mathrm{adj}}\,T^{a_1}T^{b_1}T^{a_2}T^{b_2}\phantom{\Big|}\\
&=-\Big[(2N-N_f)\,\mathrm{tr}\,T^{a_1}T^{b_1}T^{a_2}T^{b_2}
+2\,\mathrm{tr}\,T^{a_1}T^{b_1}\,\mathrm{tr}\,T^{a_2}T^{b_2} \\
&\quad\quad~+
2\,\mathrm{tr}\,T^{a_1}T^{a_2}\,\mathrm{tr}\,T^{b_1}T^{b_2} 
+2\,\mathrm{tr}\,T^{a_1}T^{b_2}\,\mathrm{tr}\,T^{a_2}T^{b_1}\Big]~.
\end{aligned}
\label{C4}
\end{equation}
The last equation follows from rewriting the adjoint trace in terms of the traces in the fundamental representation.

The total contribution of these subdiagrams to the correlator is obtained 
by performing all contractions as in the tree-level computation but with two propagators 
replaced by the sub-correlator (\ref{A4}). This amounts to
inserting in the tree-level correlator the following effective vertex
\begin{equation}
-2g^4\, W_4(x_{12}) \, V_4( \vphi,\bar\vphi)~,
\label{effective4}
 \end{equation}
where
\begin{equation}
\begin{aligned}
V_4( \vphi,\bar\vphi)  &= -C_4^{a_1 a_2 b_1 b_2} 
\, :\vphi_{a_1}  \bar\vphi_{b_1} \vphi_{a_2}   \bar \vphi_{b_2}: \phantom{\Big|} \\
&=(2N-N_f):\!\mathrm{tr}(\vphi \bar\vphi)^2\!:
+4 :\!\left( \mathrm{tr}\,\vphi \bar\vphi\right)^2 \!:+\,
2:\!\mathrm{tr}\,\vphi^2\,\mathrm{tr}\,\bar\vphi^2\!:~.\phantom{\Big|}
\end{aligned}
\label{V4} 
\end{equation}
In the superconformal case $N_f=2N$ the first term of $V_4$ is absent.
Notice again that the normal ordering in $V_4$ is necessary to correctly implement 
the replacement of two propagators with (\ref{A4}).
Using the effective vertex (\ref{effective4}), the two-loop correlator corresponding to the
sub-diagrams of Fig.~7 is 
\begin{equation}
-2g^4\,
\frac{W_4(x_{12})}{(4\pi^2x_{12}^2)^{n-2}\phantom{\Big|}}\,
\left\langle
{O}^i_{(n)} (\vphi) \,{{O}}^j_{(n)} (\bar\vphi) \,
V_4( \vphi,\bar\vphi) \right\rangle~.
\label{2point2b}
\end{equation}
The superspace integral $W_4(x_{12})$, which is evaluated in Appendix~\ref{app:W4}, is
\begin{equation}
W_4(x_{12})=\frac{6\,\zeta(3)}{\big(16\pi^2\big)^2}\,\frac{1}{\big(4\pi^2x_{12}^2\big)^2\phantom{\Big|}}~.
\label{W4fin}
\end{equation}
Thus, (\ref{2point2b}) becomes
 \begin{equation}
-6\,\frac{\zeta(3)}{2}\,\Big(\frac{g^2}{8\pi^2}\Big)^2\,
\frac{1}{(4\pi^2x_{12}^2)^{n}\phantom{\Big|}}\,
\left\langle
{O}^i_{(n)} (\vphi) \, {{O}}^j_{(n)} (\bar\vphi) \,
V_4( \vphi,\bar\vphi) \right\rangle~.
\label{2point2bb}
\end{equation}
Adding (\ref{2point2aa}) and (\ref{2point2bb}), we find
\begin{equation}
\mathcal{A}^{ij}_{(n)}(g)\Big|_{\mathrm{2-loop}}= -6\,\frac{\zeta(3)}{2}
\Big(\frac{g^2}{8\pi^2}\Big)^2\,\mathcal{C}^{ij}_{(n)}~,
\label{final20}
\end{equation}
where
\begin{align}
\mathcal{C}^{ij}_{(n)}&=\left\langle
{O}^i_{(n)} (\vphi) \, {{O}}^j_{(n)} (\bar\vphi) \,\Big[V_4(\vphi,\bar\vphi)
+2V_2(\vphi,\bar\vphi)\Big]\, \right\rangle\phantom{\bigg|}\label{Cijconf}\\
& =\left\langle
{O}^i_{(n)} (\vphi) \, {{O}}^j_{(n)} (\bar\vphi) \,\Big[
4 :\!\left( \mathrm{tr}\,\vphi \bar\vphi\right)^2\!:+\,
2:\!\mathrm{tr}\,\vphi^2\,\mathrm{tr}\,\bar\vphi^2\!:+2(N^2+1):\!\mathrm{tr}\,\vphi\bar\vphi \!:\!
\Big] \,\right\rangle.\phantom{\bigg|}\nn
\end{align}
We have evaluated this expression for the various operators listed in (\ref{operators}) and, at two loops, found
perfect agreement with the matrix model results (\ref{resconf}).

\subsection{Special correlators}
When $N_f\neq 2N$, the $\mathcal{N}=2$ theories are not superconformal invariant and many of the above simplifications no longer occur. More in general we need to implement a renormalization procedure.
For example, the gauge coupling has a non-vanishing
one-loop $\beta$-function and the chiral/anti-chiral operators acquire anomalous dimensions. 
It is nevertheless possible to find a set-up where the calculations of the two-point functions
remain simple and where a contact with the matrix model results of the previous sections can be 
established in a direct manner. 
To do so, we choose operators such that their two-point function 
vanishes up to a given loop order $(L-1)$. The $L$-loop contribution is thus finite and does not need to be renormalized.
Therefore, we restrict our attention to special correlators 
that take the following form
\begin{equation}
\big\langle \,{O}_{(n)}^i(x_1)\,\bar{{O}}_{(n)}^j(x_2)
\,\big\rangle =  \frac{\mathcal{A}^{ij}_{(2m)}(g) \big|_{L-\mathrm{loop}}} {(4\pi^2 x_{12}^2)^{n}\phantom{\Big|}}
+\mathcal{O}\big(g^{2L+2}\big)
\label{amplspecial0}
\end{equation} 
for $L\geq 1$. Then, in order to keep the discussion and the calculations as simple as possible,
we fix the anti-chiral operator to be
\begin{equation}
\big(\mathrm{tr}\,\bar\varphi^2(x)\big)^m =  \frac{1}{2^m}\,
 \delta_{b_1 b_2} \,\cdots \,\delta_{b_{2m-1}b_{2m}} \,
{\bar\varphi}^{b_{1}}(x)\,  {\bar\varphi}^{b_{2}}(x) \cdots {\bar\varphi}^{b_{2m-1}}(x)
\,{\bar\varphi}^{b_{2m}}(x)~.
\label{trphi2n}
\end{equation}
Thus,
we will study the following special correlators
\begin{equation}
\big\langle \,{O}_{(2m)}^i(x_1)
\big(\mathrm{tr}\,\bar\varphi^2(x_2)\big)^m\,
\big\rangle =  \frac{\mathcal{A}^{i}_{(2m)}(g) \big|_{L-\mathrm{loop}}} {(4\pi^2 x_{12}^2)^{2m}\phantom{\Big|}}
+\mathcal{O}\big(g^{2L+2}\big)
\label{amplspecial}
\end{equation} 
for $L\geq 1$ and various choices of $i$.

Of course, the first condition we have to require is the vanishing of the tree-level correlator 
\begin{equation}
\mathcal{A}^{i}_{(2m)}(g) \Big|_{0-\mathrm{loop}}=0~.
\label{2pointtree}
\end{equation}
Using (\ref{trphi2n}), it is immediate to realize that (\ref{2pointtree}) is equivalent to the ``tracelessness'' condition of the $R$-tensor, namely
\begin{equation}
R^{i}_{a_1a_2\cdots a_{2m}}\, \delta^{a_1a_2}\,\cdots \,\delta^{a_{2m-1} a_{2m} } =0  ~. 
\label{traceless}
\end{equation}
This condition ensures that also the one-loop correction to the correlator (\ref{amplspecial})
vanishes in the $\mathcal{N}=2$ theory, {\it{i.e.}}
\begin{equation}
\mathcal{A}^{i}_{(2m)}(g) \Big|_{\mathrm{1-loop}}=0~.
\label{2point1loop}
\end{equation}
Indeed, the only $Q$- and $H$-diagrams that can contribute 
correspond to the one-loop correction of the chiral propagator (see Fig.~4). When $N_f\neq 2N$, the $Q-H$ difference is not vanishing but it is  still diagonal in the color indices, and thus leads to a 
vanishing result after the tracelessness condition (\ref{traceless}) is used.

Therefore, the first non-trivial corrections occur at two or higher loops ($L\geq2$). 
We are going to consider them in detail in the next subsections.

\subsubsection{Two loops}
\label{subsecn:2loop}
At two loops there are several $Q$- and $H$-diagrams that \emph{a priori} can contribute to the
correlator
\begin{equation}
\mathcal{A}^i_{(2m)}(g)\Big|_{\mathrm{2-loop}}\,=\,(4\pi^2 x_{12}^2)^{2m}\,\big\langle 
\,{O}_{(2m)}^i(x_1)\,\big(\mathrm{tr}\,\bar\varphi^2(x_2)\big)^m\,\big\rangle_{\mathrm{2-loop}}~,
\label{2point2loop}
\end{equation}
but many of them have a vanishing colour factor. For example,
it is easy to realize that all diagrams accounting for the two-loop corrections of the chiral propagator 
can be disregarded since these corrections are  diagonal in the color indices
and give a vanishing contribution to the correlator (\ref{2point2loop}) upon using the tracelessness condition (\ref{traceless}). In particular, we can discard the two-loop diagrams of Fig.~6.
Similarly, we can disregard the $Q$- and $H$-diagrams involving corrections to the gauge coupling represented in Fig.~5. Indeed, these diagrams have a colour factor of the form
\begin{equation}
f^{a_1b_1c}\,f^{a_2b_2c}
\end{equation}
and their contribution to the full correlator will lead either to
\begin{equation}
\begin{aligned}
R^i_{a_1a_2\cdots a_{2m}}\,f^{a_1b_1c}\,f^{a_2b_2c}\,&\delta^{b_1b_2}\,\delta^{a_3a_4}\cdots
\delta^{a_{2m-1}a_{2m}}\\
&\sim~
R^i_{a_1a_2\cdots a_{2m}}\,\delta^{a_1a_2}\,\delta^{a_3a_4}\cdots
\delta^{a_{2m-1}a_{2m}}~,
\end{aligned}
\end{equation}
which vanish because of the tracelessness condition (\ref{traceless}), or to 
\begin{equation}
\begin{aligned}
R^i_{a_1a_2\cdots a_{2m}}\,f^{a_1b_1c}\,f^{a_2b_2c}\,&\delta^{b_1a_3}\,\delta^{b_2a_4}\cdots
\delta^{a_{2m-1}a_{2m}}\\
&\sim~
R^i_{a_1a_2\cdots a_{2m}}\,f^{a_1a_3c}\,f^{a_2a_4c}\,\cdots
\delta^{a_{2m-1}a_{2m}}~,
\end{aligned}
\label{zeroone}
\end{equation}
which vanish due to the symmetry/anti-symmetry properties of $R$ and $f$.

The only two-loop diagrams that can contribute to our special correlators are those represented in Fig.~7.
Proceeding as above, we can use the effective vertex (\ref{effective4})
and show that the two-loop correlator becomes
\begin{equation}
\begin{aligned}
\big\langle \,{O}_{(2m)}^i(x_1)\,\big(\mathrm{tr}\,\bar\varphi^2(x_2)\big)^m\,
\big\rangle_{\mathrm{2-loop}} = -2g^4\,
\frac{W_4(x_{12})}{(4\pi^2x_{12}^2)^{2m-2}\phantom{\Big|}}\,\mathcal{C}^i_{(2m)}~,
\end{aligned}
\end{equation}
where
\begin{equation}
 \mathcal{C}^i_{(2m)}  =   \left\langle
{O}^i_{(2m)} (\vphi) \, \left(\mathrm{tr}\,\bar\vphi^2 \right)^m\,
V_4( \vphi,\bar\vphi) \right\rangle
 \label{final222}
\end{equation}
with $V_4$ given in (\ref{V4}). Using the explicit expression (\ref{W4fin}) for 
the superspace integral $W_4(x_{12})$, after collecting all terms we find
\begin{equation}
\mathcal{A}^i_{(2m)}(g)\Big|_{\mathrm{2-loop}}= -\frac{6\,\zeta(3)}{2}\,\Big(\frac{g^2}{8\pi^2}\Big)^2\,
\mathcal{C}^i_{(2m)}~.
\label{final2nonconf}
\end{equation}
Specifying to the operators  (\ref{O14}) and  (\ref{Opm3}) that correspond respectively 
to $m=2$ and $m=3$, one finds perfect agreement between the field theory correlator (\ref{final2nonconf}) 
and the matrix model results (\ref{C14two}) and (\ref{Ci6two}).

\subsubsection{Three loops}
\label{subsecn:3loop}
The computation of the three-loop correlators goes along the same lines. In this case we consider
operators
\begin{equation}
\widehat{{O}}_{(2m)}^i(x) = \widehat{R}^i_{a_1a_2\cdots a_{2m}}\,
\varphi^{a_{1}}(x) \,
\varphi^{a_{2}}(x) \cdots \varphi^{a_{2m}}(x)
\label{OOOoperators}
\end{equation}
such that the two-point function
\begin{equation}
\big\langle \,\widehat{{O}}_{(2m)}^i(x_1)
\,\big(\mathrm{tr}\,\bar\varphi^2(x_2)\big)^m\,\big\rangle
\label{corr3loop}
\end{equation}
vanishes up to two loops. This condition is equivalent to require that
\begin{equation}
\begin{aligned}
&\widehat{R}^i_{a_1 a_2\cdots a_{2m}} \,\delta^{a_1a_2}\cdots \delta^{a_{2m-1}a_{2m}}  = 0~,\\
&\widehat{R}^i_{a_1 a_2\cdots a_{2m}} \, C_4^{a_1 a_2 a_3 a_4} \,\delta^{a_5 a_6}
\cdots \delta^{a_{2m-1} a_{2m}}  =0~,
\label{traceless2}
\end{aligned}
\end{equation}
where the tensor $C_4$ is defined in (\ref{C4}). Under these conditions, the first contribution to the two-point function is represented by the three-loop correlator
\begin{equation}
\widehat{\mathcal{A}}^i_{(2m)}(g)\Big|_{\mathrm{3-loop}}\,=\,(4\pi^2 x_{12}^2)^{2m}\,\big\langle \,
\widehat{{O}}_{(2m)}^i(x_1)\,
\big(\mathrm{tr}\,\bar\varphi^2(x_2)\big)^m\,\big\rangle_{\mathrm{3-loop}}~.
\label{2point3loop}
\end{equation}
For $m=3$ we have verified that this three-loop correlator receives contributions only from
the $Q$- and $H$-diagrams 
represented in Fig.~8.
%%%%%%%%%%%%%%%%%%%%%%%%%%%%%%%%%%%%%%%%%%%%%%%%%%%%%%%%%%%%
\begin{figure}[t]
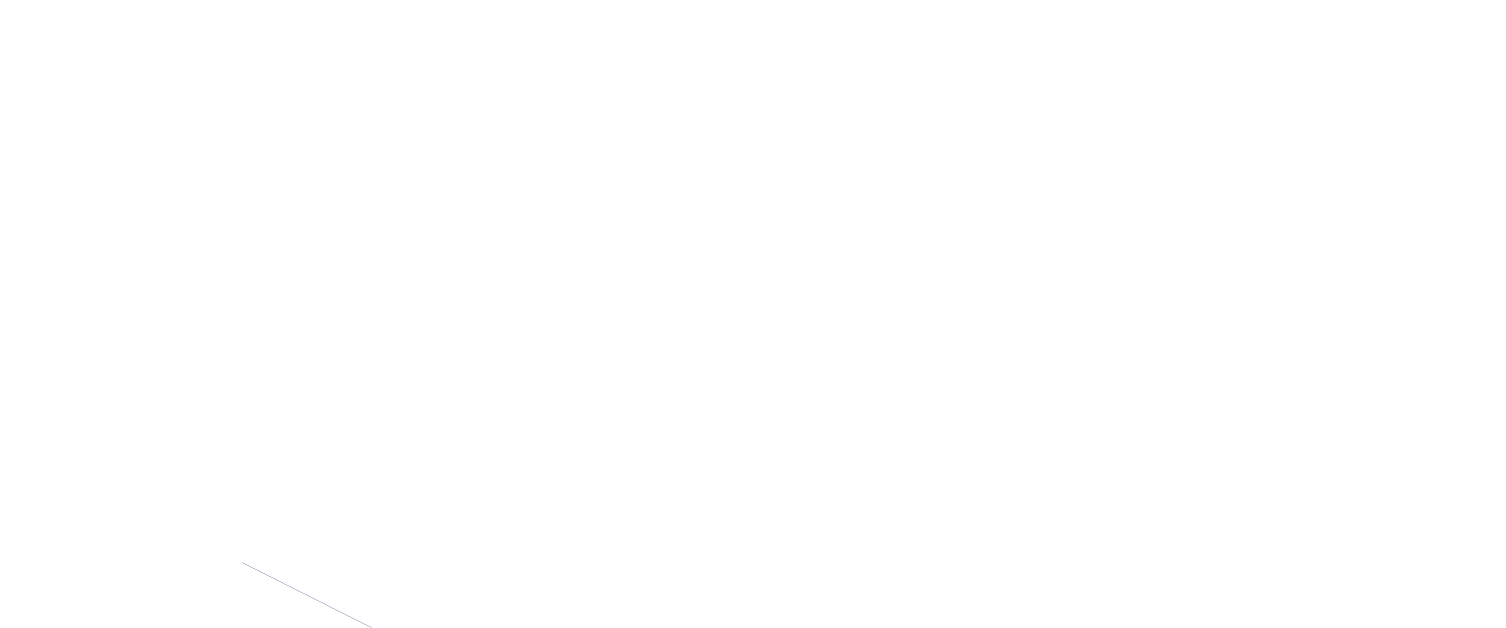
\vskip -1.5cm
{\small{\bf{Figure~8}}}: {\small{The three-loop diagrams that can contribute to the two-point function (\ref{corr3loop}) for $m=3$.
The left diagram refers to the contribution of the matter hypermultiplets while the right diagram refers
to the adjoint hypermultiplet $H$.}}
\end{figure}
%%%%%%%%%%%%%%%%%%%%%%%%%%%%%%%%%%%%%%%%%%%%%%%%%%%%%%%%%%%%

Following the same steps as before, one finds that the three-loop correlator is obtained 
by performing the same contractions as in the tree-level computation but with three propagators
replaced by the following effective vertex\,\footnote{The overall factor of $8g^6/3$ follows from
footnote~\ref{footnotefactor} choosing $k=3$.}
\begin{equation}
\frac{8}{3}\,g^6\, W_6(x_{12}) \, C_6^{a_1b_1a_2b_2a_3b_3}~,
\end{equation}
where $W_6(x_{12})$ comes from  the integration of the superspace internal variables and
the colour factor is
\begin{equation}
C_6^{a_1b_1a_2b_2a_3b_3}=N_f\,\mathrm{tr}\,T^{a_1}T^{b_1}T^{a_2}T^{b_2}T^{a_3}T^{b_3}-
\mathrm{Tr}_{\mathrm{adj}}\,T^{a_1}T^{b_1}T^{a_2}T^{b_2}T^{a_3}T^{b_3}~.
\end{equation}
This amounts to insert in the tree-level correlator the expression
\begin{equation}
-\frac{8}{3}\,g^6\, W_6(x_{12}) \, V_6( \vphi,\bar\vphi)~,
\end{equation}
where
\begin{equation}
\begin{aligned}
 V_6( \vphi,\bar\vphi)  &=-C_6^{a_1b_1a_2b_2a_3b_3}\,:\!\vphi_{a_1}\bar\vphi_{b_1}
 \vphi_{a_2}\bar\vphi_{b_2}\vphi_{a_3}\bar\vphi_{b_3}\!:\phantom{\Big|}\\
&=(2N-N_f):\!\mathrm{tr}(\vphi \bar\vphi)^3\!:
+6 \,\mathrm{tr}:\!\vphi^3 \bar\vphi \,\,\mathrm{tr}\, \bar\vphi^2\!:
 + 6 \,\mathrm{tr}:\!\bar\vphi^3 \vphi \,\,\mathrm{tr}\,\vphi^2\!:\phantom{\Big|}\\
 &\quad+ 6\,\mathrm{tr}:\!\vphi^2  \bar\vphi^2\,\,\mathrm{tr}\,\vphi  \bar\vphi\!:
 +12\,\mathrm{tr}:\!(\vphi  \bar\vphi)^2\,\,\mathrm{tr}\,\vphi  \bar\vphi \!:
 - 2 \,\mathrm{tr}:\!\vphi^3 \,\,\mathrm{tr}\bar\vphi^3\!:\phantom{\Big|}\\
&\quad-18\,\mathrm{tr}:\!\vphi^2  \bar \vphi \,\,\mathrm{tr}\bar\vphi^2 \vphi \!:\phantom{\Big|}~.
\label{vertexV6}
\end{aligned}
\end{equation}
The three-loop contribution to the correlator (\ref{corr3loop}) for $m=3$ can therefore be written as
\begin{equation}
\begin{aligned}
\big\langle \,\widehat{{O}}_{(6)}^i(x_1)
\,\big(\mathrm{tr}\,\bar\varphi^2(x_2)\big)^3\,
\big\rangle_{\mathrm{3-loop}} = -\frac{8}{3}\,g^6\,
\frac{W_6(x_{12})}{(4\pi^2x_{12}^2)^{3}\phantom{\Big|}}\,\widehat{\mathcal{C}}^i_{(6)}
\end{aligned}
\end{equation}
where
\begin{equation}
\widehat{ \mathcal{C}}^i_{(6)}  =   \left\langle
\widehat{{O}}^i_{(6)} (\vphi) \, \left(\mathrm{tr}\,\bar\vphi^2 \right)^3\,
V_6( \vphi,\bar\vphi) \right\rangle~.
 \label{final333}
\end{equation}
The integrals over the superspace variables of the internal points are computed in App.~\ref{app:W6} and yield 
the following result
\begin{equation}
W_6(x_{12})=-\frac{20\,\zeta(5)}{\big(16\pi^2\big)^3}\,\frac{1}{\big(4\pi^2x_{12}^2\big)^3\phantom{\Big|}}~.
\label{W6fin}
\end{equation}
Thus, collecting all factors 
%and comparing with eq. (\ref{amplspecial0}) 
we obtain 
\begin{equation}
\widehat{\mathcal{A}}^i_{(6)}(g)\Big|_{\mathrm{3-loop}}= \frac{20\,\zeta(5)}{3}\,\Big(\frac{g^2}{8\pi^2}\Big)^3\,
\widehat{\mathcal{C}}^i_{(6)}~.
\label{final30}
\end{equation}
Specifying to the operators (\ref{ohat3}),
we again find perfect agreement with the matrix model results (\ref{ex33}).

\section{Correspondence between matrix model and field theory}
\label{secn:corresp}
In the previous sections we have provided explicit evidence for the localization formula
  \begin{equation}
  \left \langle
 {O}_{(n)}^i (\varphi(x_1)) \, \bar{{O}}_{(n)}^j(\bar\varphi(x_2))
  \right\rangle_\mathrm{QFT}  =    \, \frac{1}{(4\pi^2 x_{12}^2)^n\phantom{\Big|}}~ \left \langle
 \nordg{{O}_{(n)}^i (a)}\,\,\,  \nordg{\bar{{O}}_{(n)}^j  (a)}    \right\rangle_\mathrm{matrix~model}~,
 \label{2point012}
 \end{equation}
where the correlator in the quantum field theory side is assumed to be finite at a specific loop order.
As it stands, this relation has been checked in the superconformal case for any half-BPS operators at two loops,
while in the non-conformal case it has been proven for certain class of operators at two and three loops. However, we think 
it should be possible to generalize and suitably extend this relation to any two-point function.  
We postpone the interesting question of how to deal with divergences and renormalization to a future work. To set the stage for these developments, let us further comment our results.

At tree level, the relation (\ref{2point012}) is quite straightforward to prove
since the matrix model effectively counts the same tree-level contractions of the field theory.
However at higher orders, as we have seen, the correspondence is less obvious.

\subsection{The conformal case $N_f=2N$}
In the superconformal case, for arbitrary choices of the operators ${O}^i_{(n)}$ and 
$\bar{{O}}^j_{(n)}$, there are no corrections at one loop, while at two loops we have
shown that the field theory correlator (\ref{final20}) matches the one coming from the matrix model using the localization formula (\ref{confexpansion}). This confirms the results of 
\cite{Baggio:2014sna,Baggio:2014ioa,Baggio:2015vxa,Gerchkovitz:2016gxx} and extends them to generic
values of $N$ and generic operators. Such a match can be made even more compelling if we
recall that the two-loop corrections originate in the matrix model from the insertion of 
$-\big[ S_4(a)- \langle S_4(a) \rangle \big]$ in the tree-level correlator (see (\ref{confexpansion})). 
By rewriting this in terms of normal-ordered operators as
 \begin{equation}
  -\big[S_4(a) - \langle S_4(a) \rangle_0\big]= -\frac{6\,\zeta(3)}{2}\,
 \Big[ \nord{ (\mathrm{tr} \,a^2)^2} + (N^2+1) \nord{\mathrm{tr} \,a^2}  \Big]
 \end{equation}
and making the replacement $a\to \vphi +\bar \vphi$, we obtain
\begin{equation}
\begin{aligned}
\nord{(\mathrm{tr} \,a^2)^2} \Big|_{a\to \vphi+\bar\vphi}  &= 
4 :\!\left( \mathrm{tr}\,\vphi \bar\vphi\right)^2\!:+\,2:\!\mathrm{tr}\,\vphi^2\,\mathrm{tr}\,\bar\vphi^2\!:+\ldots
= V_{4}( \vphi,\bar\vphi)  +\ldots \\
(N^2+1) \nord{\mathrm{tr} \,a^2} \Big|_{a\to \vphi+\bar\vphi}  &= 2(N^2+1) :\mathrm{tr} \,\vphi\bar\vphi:
+\ldots=2 V_2(\vphi,\bar\vphi)+\ldots~, 
\end{aligned}
\label{S4V4V2}
\end{equation}
where the ellipses stand for the terms with different number of $\vphi$ and $\bar\vphi$ which 
give a vanishing contribution if inserted in the chiral/anti-chiral two-point function 
since they violate the U(1) charge conservation. Using these expressions and discarding these unbalanced terms, we therefore find
\begin{equation}
 -\big[S_4(a) - \langle S_4(a) \rangle_0\big]\Big|_{a\to \vphi+\bar\vphi}=
 -\frac{6\,\zeta(3)}{2}\,\Big[V_4(\vphi,\bar\vphi)
+2V_2(\vphi,\bar\vphi)\Big]
\end{equation}
in which it is easy to recognize precisely the same combination appearing in (\ref{Cijconf}).
This relation explains on general grounds the two-loop agreement between the matrix model and the field theory results for the two-point function of arbitrary operators ${O}_{(n)}^i$ and $\bar{{O}}_{(n)}^j$ in the superconformal case.

\subsection{Special correlators}
A very similar treatment can be made at two and three loops for the special correlators in which the
anti-chiral operator is chosen to be $\big(\mathrm{tr}(\bar\varphi)^2\big)^m$.
In this case, on the matrix model side the two-loop corrections arise simply 
from the insertion of  $-S_4(a)$.
Then, making the replacement $a\to \vphi +\bar \vphi$, we have
\begin{align}
-S_4(a)\Big|_{a\to \vphi+\bar\vphi} &= -\frac{\zeta(3)}{2}\,\Big[(2N-N_f) \mathrm{tr}\,a^4+
6\left( \mathrm{tr}\,a^2\right)^2\Big]\Big|_{a\to \vphi+\bar\vphi} \notag\\
&=-\frac{6\,\zeta(3)}{2}\,V_4(\vphi,\bar\vphi)+\ldots~,
\label{S4V4}
\end{align}
where the vertex $V_4$ is defined in (\ref{V4})\,\footnote{Note that
for the special correlators, the normal ordering in $V_4$ is irrelevant since the self-contractions
of $V_4$ give terms proportional to the tree-level correlators which vanish. The same applies to $V_6$ in (\ref{S6V6})}.
Here the ellipses stand again for terms with an unbalanced number of $\vphi$ and $\bar\vphi$ which
give a vanishing contribution, or for terms proportional to the structure constants arising from
commutators of $\vphi$ and $\bar\vphi$, which also give no contribution in the special correlators due to (\ref{zeroone}).  
This implies that the match with the localization formula is not limited to the operators we considered in Section~\ref{subsecn:2loop}, 
but can be extended to any special correlator involving an operator ${O}^i_{(2m)}$ with 
a traceless $R^i$ tensor and arbitrary $m$.

At three loops, the special correlators are obtained from the insertion of  $-S_6(a)$. Repeating the same
steps as above, we have
\begin{align}
-S_6(a)\Big|_{a\to \vphi+\bar\vphi} &= \frac{\zeta(5)}{3}\,\Big[(2N-N_f) \mathrm{tr}\,a^6+
30\,\mathrm{tr}\,a^4\,\mathrm{tr}\,a^2-20  \left( \mathrm{tr}\,a^3\right)^2\Big]\Big|_{a\to \vphi+\bar\vphi} \notag\\
&=\frac{20\,\zeta(5)}{3}\,V_6(\vphi,\bar\vphi)+\cdots~,
\label{S6V6}
\end{align}
where the vertex $V_6$ is defined in (\ref{vertexV6})
and the dots have the same meaning as before. 
This relation suggests that the field theory results should match those derived 
from the localization formula (\ref{loc03}) for any choice of operators $\widehat{{O}}^j_{(2m)}$, and not only for the 
ones with $m=3$ we have considered in Section~\ref{subsecn:3loop}. It would be nice to
explicitly verify this expectation.

\section{Concluding remarks}
\label{secn:concl}
The results presented in this paper provide some explicit evidence of a direct relation between matrix model correlators and field theory correlators. Of course many important issues remain to be addressed and clarified. From a technical point of view it would be interesting to extend our checks to three loops in the superconformal case and to four loops for the special correlators, and also to generalize the dictionary to three-point functions. From a more conceptual point of view it would be important to understand how to incorporate in this correspondence the effects of the renormalization of the gauge coupling constant and of the anomalous dimensions that arise in the non-superconformal theories when one considers generic operators, and also to include non-perturbative effects due to instantons. 
Moreover, it would be nice to thoroughly explore the connection between the original matrix model with the Hermitian matrix $a$  and the effective matrix model with the complex matrices 
$\vphi$ and $\bar\vphi$, which we have just highlighted in (\ref{S4V4V2}), (\ref{S4V4}) and (\ref{S6V6}). 

Once all these issues are clarified, the extreme simplicity of the matrix model approach and of localization could be fully exploited for field theory calculations. We leave the discussion of some of these important points to future work.

\acknowledgments
We would like to thank M. Baggio, Z. Komargodski, G. Korchemsky and especially M. Frau for many useful discussions.

\noindent
The work of M.B., A.L., F.F., J.F.M. and Ya.S.S. is partially supported by the MIUR PRIN Contract 2015MP2CX4 ``Non-perturbative Aspects Of Gauge Theories And Strings''.
The work of C.W. is partially supported by the U.S. Department of Energy, Office of Science, Office of High Energy Physics, under Award Number DE-SC0010255.

\appendix
\section{Notations and conventions}
\label{secn:conventions}
Our conventions are a ``Euclideanized'' version of those of Wess-Bagger \cite{Wess:1992cp}.
Chiral and anti-chiral spinor indices are denoted by $\alpha, \beta,\cdots$ and $\dot\alpha,\dot\beta,\cdots$, respectively. Spinors are contracted as follows
\begin{equation}
\psi \,\chi =\psi^\alpha\chi_\alpha~,\qquad
\bar\psi \,\bar\chi=\bar\psi_{\dot{\alpha}}\bar\chi^{\dot{\alpha}}~.
\end{equation}
We raise and lower the indices as follows
\begin{equation}
\psi^\alpha=\varepsilon^{\alpha\beta}\,\psi_\beta~,\qquad
\bar\psi_{\dot\alpha}= \varepsilon_{\dot\alpha\dot\beta}\,\bar\psi^{\dot\beta}~,
\end{equation}
where $\varepsilon^{12}=\varepsilon_{21}
=\varepsilon^{\dot 1\dot 2}=\varepsilon_{\dot 2\dot 1}=+1$.
{F}rom these rules it follows that
\begin{equation}
\psi^\alpha\,\psi^\beta\,=\,-\frac{1}{2}\,\varepsilon^{\alpha\beta}\,
\psi\,\psi~,\qquad
\bar\psi^{\dot\alpha}\,\bar\psi^{\dot\beta}\,
=\,\frac{1}{2}\,\varepsilon^{\dot\alpha\dot\beta}\,
\bar\psi\,\bar\psi~.
\label{psipsi}
\end{equation}
The matrices $(\sigma^\mu)_{\alpha\dot\beta}$ and
$(\bar\sigma^{\mu})^{\dot\alpha\beta}$ are defined by
\begin{equation}
\label{sigmas}
\sigma^\mu =
(\vec\tau,-\ii\mathbf{1})~,\qquad
\bar\sigma^\mu =
-\sigma_\mu^\dagger = (-\vec\tau,-\ii\mathbf{1})~,
\end{equation}
where $\vec\tau$ are the ordinary Pauli matrices. They are such that
\begin{equation}
(\bar\sigma^{\mu})^{\dot\alpha\alpha}=\epsilon^{\alpha\beta}\,\epsilon^{\dot{\alpha}\dot{\beta}}(\sigma^\mu)_{\beta\dot\beta}~,
\end{equation}
and satisfy the Clifford algebra
\begin{equation}
\label{cliff4}
\begin{aligned}
&\sigma_\mu\bar\sigma_\nu + \sigma_\nu\bar\sigma_\mu =
-2\delta_{\mu\nu}\,\mathbf{1}~.
\end{aligned}
\end{equation}
Using these matrices and the above rules, we can prove 
\begin{equation}
\psi\,\sigma^\mu\bar\psi\,\,\psi\,\sigma^\nu\bar\psi
\,=\,-\frac{1}{2}\,\psi\psi\,\bar\psi\bar\psi\,\delta^{\mu\nu}~.
\end{equation}
In Euclidean space, chiral and anti-chiral spinors satisfy the pseudoreality conditions
\begin{equation}
\psi_\alpha^\dagger = \psi^\alpha~,\qquad
\bar\psi_{\dot{\alpha}}^\dagger =\bar\psi^{\dot{\alpha}}~.
\label{hermitean}
\end{equation}
According to this, the chiral superspace coordinates $y^\mu=x^\mu+\ii\,\theta\sigma^\mu\bar\theta$
are invariant under this conjugation: $y^{\mu\,\dagger}=y^\mu$.
Similarly, the anti-chiral coordinates $\bar y^\mu=x^\mu-\ii\,\theta\sigma^\mu\bar\theta$
are such that $\bar y^{\mu\,\dagger}=\bar y^\mu$.
These coordinates satisfy 
\begin{equation}
\bar D_{\dot\alpha} y^\mu=0~,\qquad D_\alpha \bar y^\mu=0~,
\end{equation}
where $D_\alpha$ and $\bar D_{\dot\alpha}$ are defined in (\ref{DbarD}).

Finally, the integration over Grassmann variables is defined such that
\begin{equation}
\int d^2\theta\,\theta^2 =1~,\qquad \int d^2\bar\theta\,\bar\theta^2 =1~.
\label{inteGr}
\end{equation}

\section{The $\Phi^{(L)}$-functions}
\label{secn:appPhi}
In this appendix we review the derivation of some integrals required in the two 
and three loop computations.
We refer the reader to \cite{Usyukina:1993ch} for details.
 
The integrals we are interested in belong to an infinite sequence computing the 
$L$-loop contribution of ladder diagrams to the four-point function in $\phi^3$-theory. They are given by
\begin{equation}
D^{L}(x_i)=\int \frac{  d^4y_1\ldots   d^4y_L}{ y_{12}^2\, y_{23}^2\ldots y_{L-1,L}^2 \, (y_1-x_1)^2(y_L-x_4)^2 \, {\prod_{i=1}^L} (y_i-x_2)^2(y_i-x_3)^2 \phantom{\Big|}}
\end{equation}
with $y_{ij}=y_i-y_j$. These integrals have been computed in \cite{Usyukina:1993ch}
and the result can be written in the compact form as follows \footnote{This formula holds in Euclidean space. In Minkowski space-time the same formula holds after replacing $\pi^{2L}$ by $(\ii\pi^2)^L$.}
\begin{equation}
D^{L}(x_i) = \frac{\pi^{2L}}{x_{23}^{2L} \, x_{14}^2 } \,\Phi^{(L)}( u,v)~,
 \label{dlphi}
\end{equation}
where $u$ and $v$ are the cross ratios
\begin{equation}
 u=\frac{x_{12}^2\, x_{34}^2}{x_{23}^2 \, x_{14}^2 }~, \qquad
 v=\frac{x_{24}^2\, x_{13}^2}{x_{23}^2 \, x_{14}^2 }~,
\end{equation}
 and $\Phi^{(L)}(u,v)$ are the so-called Davydychev functions defined as
\begin{equation}
\begin{aligned}
\Phi^{(L)}(u,v) &= \sum_{j=L}^{2L} \frac{j! \ln^{2L-j}(v/u)}{\lambda\,L! (j-L)!(2L-j)! }
 \big[\mathrm{Li}_j(-\rho\, u)+ (-1)^j\mathrm{Li}_j(-\rho\, v)\big] \\
&~+ 
\mathop{\sum_{k,l=0}^L}_{k+l=\mathrm{even}}
\frac{ 2(k+l)! (1-2^{1-k-l} ) }{ \lambda\,  k! \,  l! \, (L-k)! \, (L-l)!  }\zeta(k+l) \,
\ln^{L-k}(\rho\, u) \ln^{L-l}(\rho\, v) ~.
\end{aligned} 
\label{PhiL}
\end{equation}
Here
\begin{equation}
\lambda=\sqrt{(1-u-v)^2-4uv } \quad  \mbox{and} \quad \rho = \frac{2}{1-u-v+\lambda}~,
\end{equation}
and $\mathrm{Li}_n$ is the index $n$ polylogarithm
\begin{equation}
\mathrm{Li}_n(z) = \sum_{k=1}^{\infty}  \frac{z^k}{k^n}~.
\label{Polylog}
\end{equation}
We are interested in the limit $x_4\to x_1$, which is equivalent to sending $v\to u$ and then $u\to \infty$.
Using (\ref{PhiL}), one finds
 \begin{equation}
 \mathop{\lim_{v\to u}}_{u\to \infty} \Phi^{(L)}  \left(u,v \right)  
 =\frac {(2L)! }{L!^2} \, \frac{\zeta (2L-1)}{u}+\cdots~,
 \end{equation}
where the dots stand for subleading terms. Then, inserting this in (\ref{dlphi}) one obtains
 \begin{equation}
 D^{L}(x_1,x_2,x_3,x_1) =\frac{ \pi^{2L}}{x_{23}^{2L-2}  \,  x_{12}^2 \, x_{13}^2 } \, 
 \frac{(2L)! }{ L!^2} \, \zeta (2L-1) ~.
 \end{equation}
In particular, for  $L=2,3$ we have \footnote{Here, for convenience we relabel $y_i \to x_{3+2i}$.}
\begin{align}
\int \frac{  d^4 x_5 \, d^4 x_7}{ x_{15}^2 \, x_{17}^2 \, x_{25}^2 \, x_{27}^2 \,   x_{35}^2  \, x_{37}^2 \,  \,  x_{57}^2} &= \frac{6\,\pi^4 \, \zeta(3)}{x_{12}^2\, x_{23}^2\, x_{13}^2 } ~,\label{intz5a} \\
 \int \frac{  d^4 x_5 \, d^4 x_7 \, d^4x_9}{  x_{15}^2 \, x_{19}^2  \,   x_{25}^2 \,    x_{27}^2 \, x_{29}^2 \,    x_{35}^2 \,  \, x_{37}^2  \, x_{39}^2 \, x_{57}^2\, x_{79}^2\,  } &= \frac{20\,\pi^6 \, \zeta(5)}{ 
x_{12}^2\, x_{23}^4\, x_{13}^2 } ~.\label{intz5} 
\end{align}
Since the integrals (\ref{intz5a}) and (\ref{intz5}) are conformal, we may send any 
of the external points to infinity, and obtain other integral identities
\begin{align}
\int \frac{  d^4 x_5 \, d^4 x_7}{ x_{15}^2 \, x_{17}^2 \, x_{25}^2 \, x_{27}^2   \,  x_{57}^2} &= 
\frac{6\,\pi^4 \, \zeta(3)}{x_{12}^2 }~, \label{intz6}\\
\int \frac{  d^4 x_5 \, d^4 x_7 \, d^4x_9}{  x_{15}^2 \, x_{19}^2  \,   x_{25}^2 \,    x_{27}^2 \, x_{29}^2 \,     x_{57}^2\, x_{79}^2\,  } &= \frac{20\,\pi^6 \, \zeta(5)}{x_{12}^2  } ~,
\label{intz6a} \\
\int \frac{  d^4 x_5 \, d^4 x_7 \, d^4x_9}{x_{25}^2 \,    x_{27}^2 \, x_{29}^2 \,    x_{35}^2 \,  \, x_{37}^2  \, x_{39}^2 \, x_{57}^2\, x_{79}^2\,  } &=\frac{20\,\pi^6 \, \zeta(5)}{x_{23}^4}
~.\label{intz6b}  
\end{align}

\section{Superdiagram calculations}

In this appendix we compute the two- and three-loop integrals contributing to the two-point functions considered in Section~\ref{secn:fieldtheory}.

We introduce the short-hand notation
\begin{equation}
\langle i  j \rangle =  \rme^{(\xi_{ii}+\xi_{jj}-2\xi_{ij}) \cdot \partial_i} \frac{1}{4\pi^2 x_{ij}^2}~,
\label{chiralP0}
\end{equation}
where $\xi_{ij}= \ii \,\theta_i\sigma\bar\theta_j$ denotes the chiral superfield propagator
in which the colour  and flavour indices are suppressed (see Fig.~1)
and 
\begin{equation}
\langle V(i) V(j) \rangle  =
- \frac{\theta_{ij}^2\, \bar \theta_{ij}^2}{8\pi^2 x_{ij}^2}
\label{vectorP0}
\end{equation}
denotes the vector superfield propagator without colour indices (see Fig.~1).

\label{secn:W}
\subsection{The integral $W_2$}
\label{secn:W2}
The first integral we consider is $W_2$ corresponding to the diagrams in Fig.~6. 
The two diagrams there differ only in the colour and flavour factors and have the same superspace structure. 
So for the computation of $W_2$ it is enough to consider just one of them, say the $H$-diagram, which we redraw in Fig.~9 without colour indices and where the labels indicate the points
where the external fields or the interaction vertices are inserted.

Using the notation (\ref{chiralP0}) and (\ref{vectorP0}), the integral 
$W_2$ corresponding to Fig.~9 is given by
\begin{equation}
\begin{aligned}
W_2(x_{12}) &=  \prod_{i=3}^{6} \int d^4 x_{i} d^2 \theta_i d^2 \bar \theta_i  
\,\bar\theta_5^2 \, \theta_6^2\\
&~~~\qquad\times \langle 16 \rangle \langle 36 \rangle \langle 46 \rangle \langle 53 \rangle
\langle 54 \rangle \langle 52 \rangle \langle V(3) V(4) \rangle ~.
\end{aligned}
\label{Prop2L1}
\end{equation}
\vspace{0.6cm}
\begin{fmffile}{w2}
 \begin{eqnarray}
 \hspace{-3.2cm}
 \parbox{20mm}{  \begin{fmfgraph*}(150,80)
 \fmfleft{w}\fmfright{e}
 \fmf{fermion}{w,vw}\fmf{fermion}{ve,e}
  \fmf{fermion,tension=.5}{ve,vn,vw}
  \fmf{fermion,tension=.5}{ve,vs,vw}
   \fmf{photon}{vn,vs}
 \fmffixed{(0,h)}{vn,vs}
\fmflabel{\small{$1$}}{w}
\fmflabel{\small{$2$}}{e}
\fmflabel{\small{$6$}}{vw}
\fmflabel{\small{$3$}}{vs}
\fmflabel{\small{$4$}}{vn}
\fmflabel{\small{$5$}}{ve}
 \end{fmfgraph*}
 } 
\nn
  \end{eqnarray}
   \end{fmffile}
\vskip 0.5cm
\centerline{\small{\bf{Figure~9}}: {\small{Irreducible two-loop box diagram contributing to $W_2$.}}}
\vskip .5cm
\noindent
Although $W_2$ is a two-loop contribution, it involves a four-fold coordinate space integral.
This is a common feature when computing Feynman diagrams in coordinate space
and one may ask why we prefer to use the coordinate space approach instead of the more common one in
momentum space where one naively would expect only a two-fold integral at two loops. 
There are two reasons for this. The first one is that this simple counting works only for correlators of fundamental fields. For composite operators, instead, at a fixed order in $g^2$, 
the number of momentum integrations grows linearly with the number of fields in the operator 
(see \cite{Penati:1999ba,Penati:2000zv}). The second reason is that even for finite 
coordinate space functions, the corresponding momentum space expressions may be ill defined (for example, 
think of the Fourier transform of $1/x^4$) and require regularization.  Moreover the coordinate space 
approach is easier to generalize to three and higher point functions.

Let us now evaluate the integrals in (\ref{Prop2L1}). Note that since the external fields at the 
points $x_1$ and $x_2$ are scalars, we can set $\theta_j = \bar \theta_j = 0$ for $j=1,2$.
Performing the $\theta$ and $\bar\theta$ integrations, and combining all terms we get
\begin{equation}
\begin{aligned}
W_2(x_{12}) &= - \frac{1}{2 (2 \pi)^{14}}
\int{\frac{ d^4 x_{3} d^4 x_{4} d^4 x_{5} d^4 x_{6} }{x_{16}^2 x_{25}^2 x_{34}^2}\,
\square_{6} \Big( \frac{1}{x_{36}^2 x_{46}^2}  \Big)\,
\square_{5} \Big( \frac{1}{x_{35}^2 x_{45}^2}  \Big) }\phantom{\bigg|}\\
&=  -\frac{1}{2 (2 \pi)^{14}}
\int{\frac{ d^4 x_{3} d^4 x_{4} d^4 x_{5} d^4 x_{6} }{ x_{34}^2 x_{35}^2 x_{45}^2 x_{36}^2 x_{46}^2}
\,\square_{6} \Big( \frac{1}{x_{16}^2}  \Big)\,
\square_{5} \Big( \frac{1}{ x_{25}^2}  \Big)} \phantom{\bigg|}\\
&=  -\frac{1}{2 (2 \pi)^{10}}
\int{\frac{ d^4 x_{3} d^4 x_{4}}{ x_{13}^2 x_{14}^2 x_{23}^2 x_{24}^2 x_{34}^2}} \, ,
\end{aligned}
\label{Prop2L2}
\end{equation}
where in the second line we integrated by parts, while in the third line we used
\begin{equation}
\square \Big( \frac{1}{x^2}\Big) = -4 \pi^2 \delta(x)~.
\label{box}
\end{equation}
The remaining last integral is given by (\ref{intz6}) in Appendix \ref{secn:appPhi}, so we finally obtain
\begin{equation}
W_2(x_{12}) =-\frac{3\,\zeta(3)}{\big(16\pi^2\big)^2}\,\frac{1}{\big(4\pi^2x_{12}^2\big)}~.
\label{Prop2L3}
\end{equation}
 
\subsection{The integral $W_4$}
\label{app:W4}
The second superspace integral we have to compute is $W_4$ corresponding to the diagrams in Fig.~7.
Again, the two diagrams there differ only in the colour and flavour factors and have the same superspace structure.  So for the computation of $W_4$ it is enough to consider just one of them, say the $H$-diagram. 
Furthermore, it is convenient to ``unfold'' the irreducible two-loop subdiagram and redraw it as shown in Fig.~10.
\vskip 1cm 
  \begin{fmffile}{w4}
\hspace{4.3cm}   \parbox{80mm}{
 \begin{fmfgraph*}(130,100)
\fmfleft{i1,i2} \fmfright{o1,o2}
\fmf{fermion}{i2,v1}
\fmf{fermion}{o1,v3}
 \fmf{fermion}{v2,o2}
\fmf{fermion}{v4,i1}
\fmf{fermion}{v4,v1}
\fmf{fermion}{v2,v1}
\fmf{fermion}{v2,v3}
\fmf{fermion}{v4,v3}
 \fmflabel{\small{$1$}}{i2}
 \fmfv{label=\small{$5$},label.angle=90}{v1}
\fmfv{label=\small{$6$},label.angle=90}{v2}
\fmfv{label=\small{$7$},label.angle=-90}{v3}
\fmfv{label=\small{$8$},label.angle=-90}{v4}
\fmflabel{\small{$3$}}{i1}
\fmflabel{\small{$4$}}{o1}
\fmflabel{\small{$2$}}{o2}
 \end{fmfgraph*}
 }
   \end{fmffile}
\vskip 1cm
\centerline{\small{\bf{Figure~10}}: {\small{Irreducible two-loop box diagram contributing to $W_4$.}}}
\vskip .5cm
\noindent
We write the associated superspace integral as
\begin{equation}
W_4(x_{12})  = \lim_{x_3 \rightarrow x_2} \lim_{x_4 \rightarrow x_1}
W_4(x_1,x_4; x_2,x_3) ~.
\label{ah4}
\end{equation}
This corresponds to a point-splitting regularization of the positions of the external legs that will be made to coincide only at the end of the computation. Indeed, although, the total superdiagram is finite, a regularization 
is necessary, since at intermediate steps one has to manipulate divergent or ambiguous (like $0/0$)
expressions.

Using the notation in (\ref{chiralP0}), the superspace integral $W_4$ can be written as
\begin{equation}
\begin{aligned}
W_4(x_1,x_4;x_2,x_3) &= \prod_{i=5}^{8} \int d^4 x_{i} d^2 \theta_i d^2 \bar \theta_i  ~
\theta_5^2  \bar \theta_{6}^2 \,  \theta_7^2  \bar \theta_{8}^2\,\\
&~~~\qquad\times
\langle 15 \rangle \langle 47 \rangle \langle 65 \rangle \langle 85 \rangle
\langle 67 \rangle \langle 87 \rangle \langle 62 \rangle \langle 83 \rangle ~.
\end{aligned}
\label{2Ldiagram1}
\end{equation}
Since the external fields in the positions $x_1,\dots, x_4$ are scalars, we have to set
$\theta_j = \bar \theta_j = 0$, $j=1,2,3,4$. This fact, together with the factors of $\theta^2$ and 
$\bar\theta^2$ explicitly present in (\ref{2Ldiagram1}), implies that
the superpropagators involving external fields reduce to the standard scalar propagators.
Hence in (\ref{2Ldiagram1}) we can put 
\begin{equation}
\langle 15 \rangle=\frac{1}{4\pi^2 x_{15}^2}~,
\quad
\langle 47 \rangle=\frac{1}{4\pi^2 x_{47}^2}~,
\quad
\langle 62 \rangle=\frac{1}{4\pi^2 x_{62}^2}~,
\quad
\langle 83 \rangle=\frac{1}{4\pi^2 x_{83}^2}
\end{equation}
and rewrite $W_4$ as
\begin{equation}
\begin{aligned}
W_4(x_1,x_4;x_2,x_3) &= \frac{1}{(2 \pi)^8}
\int  \frac{d^4 x_{5} d^4 x_{6} d^4 x_{7} d^4 x_{8} }{x_{15}^2 x_{47}^2 x_{26}^2 x_{38}^2} \\
&~~~\qquad\times
\int d^2 \theta_6\,d^2 \theta_8\,  d^2 \bar \theta_5 \,d^2 \bar \theta_7  \,
 \langle 65 \rangle^R  \langle 85 \rangle^R
\langle 67 \rangle^R \langle 87 \rangle^R~,
\label{2Ldiagram2}
\end{aligned}
\end{equation}
where the reduced propagator $\langle i j \rangle^R$ is defined as
\begin{equation}
\langle i j \rangle^R = \Big(
1 -2\ii\, \theta_i^\alpha (\sigma_\mu)_{\alpha \dot{\alpha}}  \bar \theta_j^{\dot{\alpha}}\, \partial_i^{\mu} 
+ \theta_i^2 { \bar \theta_j}^2 \,\square_{i}
\Big) \frac{1}{4\pi^2 x_{ij}^2} ~.
\label{reducedprop}
\end{equation}
After performing the $\theta$ and $\bar\theta$ integrations, the second line of (\ref{2Ldiagram2}) becomes
\begin{equation}
\begin{aligned}
\frac{1}{(2 \pi)^8} \bigg\{
- &\mathrm{ Tr}(\sigma_{\mu} \bar{\sigma}_{\nu} \sigma_{\rho} \bar{\sigma}_{\tau})\,
\partial^{\mu}_{5} \Big( \frac{1}{x_{56}^2} \Big)
\partial^{\nu}_{7} \Big( \frac{1}{x_{67}^2} \Big)
\partial^{\rho}_{7} \Big( \frac{1}{x_{78}^2} \Big)
\partial^{\tau}_{5} \Big( \frac{1}{x_{58}^2} \Big)\\
&+ \square_{5} \Big( \frac{1}{x_{56}^2} \Big) \,\frac{1}{x_{67}^2}\,
\square_{7} \Big( \frac{1}{x_{78}^2} \Big) \,\frac{1}{x_{58}^2}
+ \frac{1}{x_{56}^2} \,\square_{7} \Big( \frac{1}{x_{67}^2} \Big)\,
 \frac{1}{x_{78}^2}\, \square_{5} \Big( \frac{1}{x_{58}^2} \Big)\bigg\}~.
\end{aligned}
\label{2Ldiagrampart}
\end{equation}
Then one can perform the integrals in $x_6$ and $x_8$ by using (\ref{box}) and exploiting the identity
\begin{equation}
\sigma^\mu \bar\sigma^\nu \, \partial^\mu_{i} \,  \partial_{j}^\nu \int  \frac{ d^4 x_0 }{x^2_{i0} x^2_{j0} x^2_{k0}} =
 - 4 \pi^2 \sigma^\mu \bar\sigma^\nu \,   \frac{   x^\mu_{ik}  \,   x^\nu_{kj}}{x^2_{ik} x^2_{jk} x^2_{ij}} ~.\label{papa}
\end{equation}
This identity follows by differentiating the explicit expression for the integral in terms of the 
Davydychev $\Phi^{(1)}$ function introduced in Appendix~\ref{secn:appPhi}.
Using
\begin{equation}
\mathrm{Tr}(\sigma_{\mu} \bar{\sigma}_{\nu} \sigma_{\rho} \bar{\sigma}_{\tau})
= 2 \ (
\delta_{\mu \nu} \delta_{\rho \tau}
- \delta_{\mu \rho} \delta_{\nu \tau}
+\delta_{\mu \tau}  \delta_{\nu \rho} -\epsilon_{\mu\nu\rho\tau}) \, ,
\end{equation}
and simplifying the scalar products by means of
\begin{equation}
x_{ij} \cdot x_{k \ell} = \frac{1}{2} \left(
x_{i \ell}^2+x_{jk}^2-x_{ik}^2-x_{j \ell}^2 \right) ~, 
\label{xx}
\end{equation}
after some algebra we obtain
\begin{equation}
W_4(x_1,x_4;x_2,x_3) = \frac{x_{23}^2}{(2 \pi)^{12}} \int \frac{d^4x_5 d^4x_7}{x_{15}^2 x_{47}^2 x_{25}^2 x_{27}^2  x_{35}^2 x_{37}^2 x_{57}^2} ~.
\label{2LInt1}
\end{equation}
The integral in the right hand side has been computed in \cite{Usyukina:1993ch} (see also (\ref{intz5a})). 
It is finite in the limit $x_4 \to x_1$ but it is quadratically divergent in the limit $x_3 \to x_2$. However, taking into account also the factor of $x_{23}^2$ that is present in $W_4$ we find in the end
a finite result for the two point function. Indeed,
\begin{equation}
W_4(x_{12})= \lim_{x_3 \to x_2} \lim_{x_4 \to x_1}
W_4(x_1,x_4;x_2,x_3) =   \frac{ 6 \zeta(3)}{(16 \pi^2)^2}  \, \frac{1}{(4\pi^2 x_{12}^2)^2} \, .
\label{2LDiag2ptF}
\end{equation}

\subsection{The integral $W_6$}
\label{app:W6}
The last superspace integral we compute is $W_6$ corresponding to the diagrams in Fig.~8.
As in the previous cases, 
also here the only difference between the two diagrams is in the colour and flavour factors, while the
superspace structure is the same. Thus, in order to compute $W_6$ it is enough to consider just one of them, say the $H$-diagram, whose irreducible three-loop part is represented in its ``unfolded'' version in Fig.~11.

The relevant integral $W_6(x_{12})$ is six-fold and, as suggested by the hexagon diagram, it is regularized by point-splitting, namely
\begin{equation}
W_6(x_{12}) = \lim_{x_3\to x_1} \lim_{x_5\to x_1}\lim_{x_4\to x_2} \lim_{x_6\to x_2}
W_6(x_1,x_3,x_5;x_2,x_4,x_6)~,
\label{I6fromW}
\end{equation}
where
\begin{equation}
\begin{aligned}
W_6(x_1,x_3,x_5;&x_2,x_4,x_6)
=\int \prod_{i=7}^{9}  d^4 x_{i} d^4 x_{i'} d^2 \theta_i d^2 \bar \theta_i
 d^2 \theta_{i'} d^2 \bar \theta_{i'}\,\theta_i^2  \bar \theta_{i'}^2\\
 &\times
 \langle 17 \rangle \langle 38 \rangle \langle 59 \rangle
\langle 7' 2 \rangle \langle 8' 4 \rangle \langle 9' 6 \rangle
\langle 7' 7 \rangle \langle 7' 8 \rangle \langle 8' 7 \rangle
\langle 8' 9 \rangle \langle 9' 8 \rangle \langle 9' 9 \rangle ~.
\label{3Ldiagram1}
\end{aligned}
\end{equation}
\vskip 1cm
\begin{fmffile}{w6}
\hspace{4.4cm} \parbox{20mm}{
 \begin{fmfgraph*}(125,120)
\fmfleft{i1,i2,i3} \fmfright{o1,o2,o3}
\fmf{fermion}{i1,v1}
\fmf{fermion}{v2,v1}
 \fmf{fermion}{v2,o1}
\fmf{fermion}{v2,v3}
\fmf{fermion}{o2,v3}
\fmf{fermion}{v4,v3}
\fmf{fermion}{v4,o3}
\fmf{fermion}{v4,v5}
\fmf{fermion}{i3,v5}
\fmf{fermion}{v6,v5}
\fmf{fermion}{v6,i2}
\fmf{fermion}{v6,v1}
 \fmflabel{\small{$5$}}{i1}
\fmflabel{\small{$2$}}{o3}
\fmflabel{\small{$3$}}{o2}
\fmflabel{\small{$6$}}{o1}
\fmflabel{\small{$1$}}{i3}
\fmflabel{\small{$4$}}{i2}
\fmfv{label=\small{$9$},label.angle=180}{v1}
\fmfv{label=\small{$9^\prime$},label.angle=0}{v2}
\fmfv{label=\small{$8^\prime$},label.angle=120}{v6}
\fmfv{label=\small{$8$},label.angle=60}{v3}
\fmfv{label=\small{$7$},label.angle=180}{v5}
\fmfv{label=\small{$7^\prime$},label.angle=0}{v4}
 \end{fmfgraph*}
 }
   \end{fmffile}
\vskip 1.0cm
\centerline{\small{\bf{Figure~11}}: {\small{Irreducible three-loop hexagon diagram contributing to $W_6$.}}}
\vskip 0.5cm
\noindent
Since the external fields at the points $x_1,\ldots, x_6$ are all scalars, in
(\ref{3Ldiagram1}) we have to set  $\theta_j = \bar \theta_j = 0$ for $j=1 \dots 6$.
This fact and the $\theta$ and $\bar\theta$ factors that are explicitly present in the integrand,
allow us to rewrite $W_6$ as follows
\begin{equation}
\begin{aligned}
W_6(x_1,x_3,x_5;&x_2,x_4,x_6) =  \frac{1}{(2 \pi)^{12}}
\int \prod_{i=7}^{9} \frac{ d^4 x_{i} d^4 x_{i'}}{x_{17}^2 x_{38}^2 
x_{59}^2 x_{27'}^2 x_{48'}^2 x_{69'}^2}\\
&\times \int \prod_{j=7}^{9}d^2 \theta_{j'}  d^2 \bar \theta_j\,
\langle 7' 7 \rangle^R \langle 7' 8 \rangle^R \langle 8' 7 \rangle^R
\langle 8' 9 \rangle^R \langle 9' 8 \rangle^R \langle 9' 9 \rangle^R ~.
\end{aligned}
\label{3Ldiagram2}
\end{equation}
Performing the $\theta$ and $\bar\theta$ integrations, the second line of (\ref{3Ldiagram2})
becomes
\begin{equation}
\begin{aligned}
\frac{1}{(2 \pi)^{12}} \bigg\{
- &\mathrm{ Tr}(\sigma_{\mu} \bar{\sigma}_{\nu}
\sigma_{\rho} \bar{\sigma}_{\tau}
\sigma_{\kappa} \bar{\sigma}_{\lambda}
)  \\
&\times \partial^{\mu}_{8} \Big( \frac{1}{x_{7'8}^2} \Big)
\partial^{\nu}_{7} \Big( \frac{1}{x_{7'7}^2} \Big)
\partial^{\rho}_{7} \Big( \frac{1}{x_{8'7}^2} \Big)
\partial^{\tau}_{9} \Big( \frac{1}{x_{8'9}^2} \Big)
\partial^{\kappa}_{9} \Big( \frac{1}{x_{9'9}^2} \Big)
\partial^{\lambda}_{8} \Big( \frac{1}{x_{9'8}^2} \Big) \\
&+
 \frac{1}{x_{7'7}^2 x_{8'9}^2 x_{9'8}^2}\,
\square_{8} \Big( \frac{1}{x_{7'8}^2} \Big)\,
\square_{7} \Big( \frac{1}{x_{8'7}^2} \Big)\,
\square_{9} \Big( \frac{1}{x_{9'9}^2} \Big)
\\
&+
 \frac{1}{x_{9'9}^2 x_{7'8}^2 x_{8'7}^2}\,
\square_{7} \Big( \frac{1}{x_{7'7}^2} \Big)\,
\square_{9} \Big( \frac{1}{x_{8'9}^2} \Big)\,
\square_{8} \Big( \frac{1}{x_{9'8}^2} \Big) \bigg\} ~.
\end{aligned}
\label{res6}
\end{equation}
Inserting this expression in (\ref{3Ldiagram2}),we can perform three of the integrations with the help of (\ref{papa}). We choose them to be those over $x_{7'}, x_{8'}, x_{9'}$, but equivalently we could have
integrated in $x_{7}, x_{8}, x_9$. Evaluating the trace of the sigma matrices and using (\ref{xx}) 
to express everything in terms of squares, we find
\begin{equation}
W_6(x_1,x_1,x_1;x_2,x_4,x_6) = -\frac{1}{2(4 \pi^2)^9}
\left(A + x_{24}^2 B_{24} + x_{46}^2 B_{46} + x_{26}^2 B_{26} \right) ~. 
\label{3LGen1}
\end{equation}
Here we have already taken the limits ${x_3\to x_1}$ and $x_5\to x_1$ which are safe in this expression
and simplify the result. The integrals in (\ref{3LGen1}) are \footnote{To be very precise, $A$ contains also
terms proportional to the $\epsilon$-symbol that arise from the trace of the six sigma matrices in (\ref{res6}). However, due to the antisymmetry of the $\epsilon$-symbols, these terms are identically zero in the coincidence limit (\ref{I6fromW}) and hence we do
not write them explicitly.}
\begin{align}
A &=
\int \frac{ d^4x_7 d^4x_8 d^4x_9} {x_{78}^2 x_{79}^2 x_{89}^2 x_{17}^2 x_{18}^2 x_{19}^2}
\bigg(
 \frac{x_{67}^2 x_{48}^2 x_{29}^2}{x_{27}^2 x_{28}^2 x_{47}^2 x_{49}^2 x_{68}^2 x_{69}^2 }
-\frac{ x_{48}^2}{ x_{28}^2 x_{47}^2 x_{49}^2 x_{68}^2}\phantom{\Bigg|}\notag\\
& ~~~\quad -\frac{x_{67}^2}{x_{27}^2  x_{47}^2 x_{68}^2 x_{69}^2 }
+ \frac{1}{x_{28}^2 x_{47}^2 x_{69}^2 }
- \frac{ x_{29}^2}{x_{27}^2 x_{28}^2  x_{49}^2 x_{69}^2 }
+ \frac{1}{x_{27}^2 x_{49}^2 x_{68}^2 }
\bigg)~,\phantom{\Bigg|}\\
B_{24} &=\int \frac{ d^4x_7 d^4x_8 d^4x_9} { x_{17}^2 x_{18}^2 x_{19}^2
x_{78}^2 x_{79}^2} \bigg(\frac{1}{x_{27}^2 x_{28}^2 x_{48}^2 x_{49}^2 x_{69}^2 }\phantom{\Bigg|}
\notag\\
& ~~~\qquad\quad +  \frac{1}{x_{28}^2 x_{29}^2 x_{47}^2 x_{48}^2 x_{69}^2 }
- \frac{x_{67}^2}{x_{27}^2 x_{28}^2  x_{47}^2 x_{49}^2 x_{68}^2 x_{69}^2 }\bigg)
~,\phantom{\Bigg|}
\label{3LAB}
\end{align}
while $B_{26}$ ($B_{46}$) is obtained from $B_{24}$ by exchanging $x_4$ with $x_6$ ($x_2$ with $x_6$).

Given these expressions, it is difficult to directly perform {\it simultaneously} the further coincidence limits  $x_6 \to x_2$ and $x_4 \to x_2$ as required by (\ref{I6fromW}), but under the reasonable assumption that the result is single valued, we can perform these limits in a consecutive. 
We choose first to send $x_6 \to x_2$, then send $x_4 \rightarrow x_2$; however, 
due to the manifest permutation symmetry of (\ref{3LGen1}), it is immediate to realize
that any order will give the same result.

Let us first consider the integral $A$. In it we can take the limit $x_6 \to x_2$ in the integrand, obtaining a vanishing result (indeed the terms in the integrand with opposite signs cancel pairwise in this limit). Hence also the integral $A$ vanishes in this limit.

Let us now consider the integral $B_{24}$. We notice first that in the limit  $x_6 \to x_2$, the last two terms in (\ref{3LAB}) cancel against each other after renaming the dummy variables 
$x_8 \leftrightarrow x_9$ in the second term. In the remaining first term, the limit $x_6 \to x_2$ can be performed directly in the integrand and the result is given by (\ref{intz5}). More explicitly, we have
\begin{equation}
\lim_{x_6 \to x_2} B_{24}  =    \int \frac{   d^4x_7 d^4x_8 d^4 x_9  } {
 x_{17}^2 x_{18}^2  x_{19}^2   x_{27}^2    x_{28}^2   x_{29}^2  x_{48}^2  x_{49}^2  x_{78}^2  x_{79}^2   }  =\frac{20 \pi^6 \zeta(5)}{x_{12}^4 x_{14}^2 x_{24}^2} ~.
\label{B241n2}
\end{equation}
Using this, we then find
\begin{equation}
\lim_{x_4\to x_2} \Big(
\lim_{x_6 \to x_2} x_{24}^2 B_{24}\Big)   = \frac{20 \pi^6 \zeta(5)}{x_{12}^6} ~.
\label{B241n4}
\end{equation}
Following exactly these same steps (and recalling that $B_{46}$ is simply obtained from $B_{24}$  by exchanging $x_2$ with $x_6$) we obtain
\begin{equation}
\lim_{x_4\to x_2} \Big(
\lim_{x_6 \to x_2} x_{46}^2 B_{46}\Big)  = \frac{20 \pi^6 \zeta(5)}{x_{12}^6}~. 
\label{B461n4}
\end{equation}
The term proportional to $B_{26}$, instead, gives no contribution in the limit $x_6\to x_2$.
Indeed, all three integrals in this case (the first two are essentially equivalent) diverge when  $x_6 \to 
x_2$, but the divergence is only logarithmic, namely in this limit $B_{26}$ behaves as 
${\log}^3(x_{26}^2)$ + less singular terms \footnote{The first two integrals have been computed in \cite{Rossi:2008ay} for the special configuration where all points are aligned on a line, which is enough to capture all the singularities of the integral. The relevant equation there is Eq.~(35). The singularities of the last
integral have been studied in \cite{Eden:2011we,Eden:2012rr}. The relevant equation is Eq.~(47) of
\cite{Eden:2012rr}.}. Therefore, it is suppressed by the overall $x_{26}^2$ factor which multiplies
$B_{26}$ in (\ref{3LGen1}), and in the end we have
\begin{equation}
\lim_{x_6 \to x_2} x_{26}^2 B_{26}  = 0 ~.
\label{B261n1}
\end{equation}
Putting everything together in (\ref{3LGen1}) and (\ref{I6fromW}), we therefore find 
\begin{equation}
W_6(x_{12}) =-\frac{20\,\zeta(5)}{\big(16\pi^2\big)^3}\,\frac{1}{\big(4\pi^2x_{12}^2\big)^3}~.
\label{3LDiag2ptF}
\end{equation}

\providecommand{\href}[2]{#2}\begingroup\raggedright\endgroup

\end{document}

%% file: fig7.pdf_tex
%% Creator: Inkscape inkscape 0.91, www.inkscape.org
%% Accompanies image file 'fig7.pdf' (pdf, eps, ps)
\begingroup%
  \makeatletter%
  \providecommand\color[2][]{%
    \errmessage{(Inkscape) Color is used for the text in Inkscape, but the package 'color.sty' is not loaded}%
    \renewcommand\color[2][]{}%
  }%
  \providecommand\transparent[1]{%
    \errmessage{(Inkscape) Transparency is used (non-zero) for the text in Inkscape, but the package 'transparent.sty' is not loaded}%
    \renewcommand\transparent[1]{}%
  }%
  \providecommand\rotatebox[2]{#2}%
  \ifx\svgwidth\undefined%
    \setlength{\unitlength}{480bp}%
    \ifx\svgscale\undefined%
      \relax%
    \else%
      \setlength{\unitlength}{\unitlength * \real{\svgscale}}%
    \fi%
  \else%
    \setlength{\unitlength}{\svgwidth}%
  \fi%
  \global\let\svgwidth\undefined%
  \global\let\svgscale\undefined%
  \makeatother%
  \begin{picture}(1,0.370292)%
    %\put(0,0){\includegraphics[width=\unitlength,page=1]{fig7.pdf}}%
    \put(0.00125234,0.20148048){\makebox(0,0)[lb]{\smash{{\small $a_{n-1}$}}}}%
    \put(0,0){\includegraphics[width=\unitlength,page=2]{fig7.pdf}}%
    \put(0.00125234,0.15556127){ \makebox(0,0)[lb]{\smash{{\small $a_{n}$}}}}%
    \put(0.34485932,0.20148048){ \makebox(0,0)[lb]{\smash{{\small $b_{n-1}$}}}}%
    \put(0.34485932,0.15556127){ \makebox(0,0)[lb]{\smash{{\small $b_{n}$}}}}%
    \put(0.18356044,0.25617359){ \makebox(0,0)[lb]{\smash{{\small $\vdots $}}}}%
    \put(0,0){\includegraphics[width=\unitlength,page=3]{fig7.pdf}}%
    \put(0.00187342,0.35154608){ \makebox(0,0)[lb]{\smash{{\small $a_1$}}}}%
    \put(-0.00091996,0.30566474){ \makebox(0,0)[lb]{\smash{{\small $a_2$}}}}%
    \put(0.34532585,0.35228379){ \makebox(0,0)[lb]{\smash{{\small $b_1$}}}}%
    \put(0.34629685,0.30527703){ \makebox(0,0)[lb]{\smash{{\small $b_2$}}}}%
    \put(0.47392774,0.20148048){ \makebox(0,0)[lb]{\smash{{\small $a_{n-1}$}}}}%
    \put(0,0){\includegraphics[width=\unitlength,page=4]{fig7.pdf}}%
    \put(0.47392774,0.15556127){ \makebox(0,0)[lb]{\smash{{\small $a_{n}$}}}}%
    \put(0.81753471,0.20148048){ \makebox(0,0)[lb]{\smash{{\small $b_{n-1}$}}}}%
    \put(0.81753471,0.15556127){ \makebox(0,0)[lb]{\smash{{\small $b_{n}$}}}}%
    \put(0.65623582,0.25617359){ \makebox(0,0)[lb]{\smash{{\small $\vdots $}}}}%
    \put(0,0){\includegraphics[width=\unitlength,page=5]{fig7.pdf}}%
    \put(0.47454882,0.35154608){ \makebox(0,0)[lb]{\smash{{\small $a_1$}}}}%
    \put(0.47175547,0.30566474){ \makebox(0,0)[lb]{\smash{{\small $a_2$}}}}%
    \put(0.81800126,0.35228379){ \makebox(0,0)[lb]{\smash{{\small $b_1$}}}}%
    \put(0.81897226,0.30527703){ \makebox(0,0)[lb]{\smash{{\small $b_2$}}}}%
    \put(0.4110821,0.25663576){ \makebox(0,0)[lb]{\smash{{\small , }}}}%
  \end{picture}%
\endgroup%

%% file: fig8.pdf_tex
%% Creator: Inkscape inkscape 0.91, www.inkscape.org
%% Accompanies image file 'fig8.pdf' (pdf, eps, ps)
\begingroup%
  \makeatletter%
  \providecommand\color[2][]{%
    \errmessage{(Inkscape) Color is used for the text in Inkscape, but the package 'color.sty' is not loaded}%
    \renewcommand\color[2][]{}%
  }%
  \providecommand\transparent[1]{%
    \errmessage{(Inkscape) Transparency is used (non-zero) for the text in Inkscape, but the package 'transparent.sty' is not loaded}%
    \renewcommand\transparent[1]{}%
  }%
  \providecommand\rotatebox[2]{#2}%
  \ifx\svgwidth\undefined%
    \setlength{\unitlength}{480bp}%
    \ifx\svgscale\undefined%
      \relax%
    \else%
      \setlength{\unitlength}{\unitlength * \real{\svgscale}}%
    \fi%
  \else%
    \setlength{\unitlength}{\svgwidth}%
  \fi%
  \global\let\svgwidth\undefined%
  \global\let\svgscale\undefined%
  \makeatother%
  \begin{picture}(1,0.41682836)%
    %\put(0,0){\includegraphics[width=\unitlength,page=1]{fig8.pdf}}%
    \put(-0.00092227,0.20198583){\makebox(0,0)[lb]{\smash{{\small $a_{n-1}$}}}}%
    \put(0,0){\includegraphics[width=\unitlength,page=2]{fig8.pdf}}%
    \put(-0.00092227,0.15595144){ \makebox(0,0)[lb]{\smash{{\small $a_{n}$}}}}%
    \put(0.34354655,0.20198583){ \makebox(0,0)[lb]{\smash{{\small $b_{n-1}$}}}}%
    \put(0.34354655,0.15595144){ \makebox(0,0)[lb]{\smash{{\small $b_{n}$}}}}%
    \put(0.1818431,0.25681611){ \makebox(0,0)[lb]{\smash{{\small $\vdots $}}}}%
    \put(0,0){\includegraphics[width=\unitlength,page=3]{fig8.pdf}}%
    \put(-0.00092227,0.35242782){ \makebox(0,0)[lb]{\smash{{\small $a_2$}}}}%
    \put(-0.00092227,0.3064314){ \makebox(0,0)[lb]{\smash{{\small $a_3$}}}}%
    \put(0.34401424,0.35316738){ \makebox(0,0)[lb]{\smash{{\small $b_2$}}}}%
    \put(0.34498767,0.30604272){ \makebox(0,0)[lb]{\smash{{\small $b_3$}}}}%
    \put(0.40993542,0.25727944){ \makebox(0,0)[lb]{\smash{{\small , }}}}%
    \put(0,0){\includegraphics[width=\unitlength,page=4]{fig8.pdf}}%
    \put(-0.00092227,0.39865816){ \makebox(0,0)[lb]{\smash{{\small $a_1$}}}}%
    \put(0.34296484,0.39877499){ \makebox(0,0)[lb]{\smash{{\small $b_1$}}}}%
    \put(0.47260826,0.20198583){ \makebox(0,0)[lb]{\smash{{\small $a_{n-1}$}}}}%
    \put(0,0){\includegraphics[width=\unitlength,page=5]{fig8.pdf}}%
    \put(0.47260826,0.15595144){ \makebox(0,0)[lb]{\smash{{\small $a_{n}$}}}}%
    \put(0.81707705,0.20198583){ \makebox(0,0)[lb]{\smash{{\small $b_{n-1}$}}}}%
    \put(0.81707705,0.15595144){ \makebox(0,0)[lb]{\smash{{\small $b_{n}$}}}}%
    \put(0.65537361,0.25681611){ \makebox(0,0)[lb]{\smash{{\small $\vdots $}}}}%
    \put(0,0){\includegraphics[width=\unitlength,page=6]{fig8.pdf}}%
    \put(0.47260826,0.35242782){ \makebox(0,0)[lb]{\smash{{\small $a_2$}}}}%
    \put(0.47260826,0.3064314){ \makebox(0,0)[lb]{\smash{{\small $a_3$}}}}%
    \put(0.81754477,0.35316738){ \makebox(0,0)[lb]{\smash{{\small $b_2$}}}}%
    \put(0.81851821,0.30604272){ \makebox(0,0)[lb]{\smash{{\small $b_3$}}}}%
    %\put(0.88346596,0.25727944){ \makebox(0,0)[lb]{\smash{{\small , }}}}%
    \put(0,0){\includegraphics[width=\unitlength,page=7]{fig8.pdf}}%
    \put(0.47260826,0.39865816){ \makebox(0,0)[lb]{\smash{{\small $a_1$}}}}%
    \put(0.81649535,0.39877499){ \makebox(0,0)[lb]{\smash{{\small $b_1$}}}}%
  \end{picture}%
\endgroup%

%% file: 2-point_correlators_arXiV.bbl
\begin{thebibliography}{10}

\bibitem{Teschner:2014oja}
J.~Teschner, \href{http://dx.doi.org/10.1007/978-3-319-18769-3_1}{\emph{{Exact
  Results on $\mathcal{N} =$ 2 Supersymmetric Gauge Theories}},} in {\em New
  Dualities of Supersymmetric Gauge Theories}, J.~Teschner editor, Springer (2016)
\href{http://arxiv.org/abs/1412.7145}{{\tt arXiv:1412.7145 [hep-th]}}.
\newblock
%%CITATION = ARXIV:1412.7145;%%.

\bibitem{Pestun:2016zxk}
V.~Pestun {\em et al.}, \emph{{Localization techniques in quantum field
  theories}},
\href{http://arxiv.org/abs/1608.02952}{{\tt arXiv:1608.02952 [hep-th]}}.
%%CITATION = ARXIV:1608.02952;%%.

\bibitem{berline}
B.~Berline, E.~Getzler, and M.~Vergne, {\em {Heat kernels and Dirac
  operators}}, Springer (1992).

\bibitem{Nekrasov:2002qd}
N.~Nekrasov, \emph{{Seiberg-Witten prepotential from instanton counting}}, Adv.
  Theor. Math. Phys. {\bf 7} (2004)  831--864,
\href{http://arxiv.org/abs/hep-th/0206161}{{\tt arXiv:hep-th/0206161}}.
%%CITATION = HEP-TH/0206161;%%.

\bibitem{Nekrasov:2003rj}
N.~Nekrasov and A.~Okounkov, \emph{{Seiberg-Witten theory and random
  partitions}}, \href{http://dx.doi.org/10.1007/0-8176-4467-9_15}{Prog. Math.
  {\bf 244} (2006)  525--596},
\href{http://arxiv.org/abs/hep-th/0306238}{{\tt arXiv:hep-th/0306238
  [hep-th]}}.
%%CITATION = HEP-TH/0306238;%%.

\bibitem{Pestun:2007rz}
V.~Pestun, \emph{{Localization of gauge theory on a four-sphere and
  supersymmetric Wilson loops}},
  \href{http://dx.doi.org/10.1007/s00220-012-1485-0}{Commun.Math.Phys. {\bf
  313} (2012)  71--129},
\href{http://arxiv.org/abs/0712.2824}{{\tt arXiv:0712.2824 [hep-th]}}.
%%CITATION = ARXIV:0712.2824;%%.

\bibitem{Erickson:2000af}
J.~K. Erickson, G.~W. Semenoff, and K.~Zarembo, \emph{{Wilson loops in N=4
  supersymmetric Yang-Mills theory}},
  \href{http://dx.doi.org/10.1016/S0550-3213(00)00300-X}{Nucl. Phys. {\bf B582}
  (2000)  155--175},
\href{http://arxiv.org/abs/hep-th/0003055}{{\tt arXiv:hep-th/0003055
  [hep-th]}}.
%%CITATION = HEP-TH/0003055;%%.

\bibitem{Berenstein:1998ij}
D.~E. Berenstein, R.~Corrado, W.~Fischler, and J.~M. Maldacena, \emph{{The
  Operator product expansion for Wilson loops and surfaces in the large N
  limit}}, \href{http://dx.doi.org/10.1103/PhysRevD.59.105023}{Phys. Rev. {\bf
  D59} (1999)  105023},
\href{http://arxiv.org/abs/hep-th/9809188}{{\tt arXiv:hep-th/9809188
  [hep-th]}}.
%%CITATION = HEP-TH/9809188;%%.

\bibitem{Semenoff:2001xp}
G.~W. Semenoff and K.~Zarembo, \emph{{More exact predictions of SUSYM for
  string theory}}, \href{http://dx.doi.org/10.1016/S0550-3213(01)00455-2}{Nucl.
  Phys. {\bf B616} (2001)  34--46},
\href{http://arxiv.org/abs/hep-th/0106015}{{\tt arXiv:hep-th/0106015
  [hep-th]}}.
%%CITATION = HEP-TH/0106015;%%.

\bibitem{Fucito:2015ofa}
F.~Fucito, J.~F. Morales, and R.~Poghossian, \emph{{Wilson loops and chiral
  correlators on squashed spheres}},
  \href{http://dx.doi.org/10.1007/JHEP11(2015)064}{JHEP {\bf 11} (2015)  064},
\href{http://arxiv.org/abs/1507.05426}{{\tt arXiv:1507.05426 [hep-th]}}.
%%CITATION = ARXIV:1507.05426;%%.

\bibitem{Andree:2010na}
R.~Andree and D.~Young, \emph{{Wilson Loops in N=2 Superconformal Yang-Mills
  Theory}}, \href{http://dx.doi.org/10.1007/JHEP09(2010)095}{JHEP {\bf 09}
  (2010)  095},
\href{http://arxiv.org/abs/1007.4923}{{\tt arXiv:1007.4923 [hep-th]}}.
%%CITATION = ARXIV:1007.4923;%%.

\bibitem{Baggio:2014sna}
M.~Baggio, V.~Niarchos, and K.~Papadodimas, \emph{{Exact correlation functions
  in $SU(2) \mathcal N=2$ superconformal QCD}},
  \href{http://dx.doi.org/10.1103/PhysRevLett.113.251601}{Phys. Rev. Lett. {\bf
  113} (2014) no.~25, 251601},
\href{http://arxiv.org/abs/1409.4217}{{\tt arXiv:1409.4217 [hep-th]}}.
%%CITATION = ARXIV:1409.4217;%%.

\bibitem{Baggio:2014ioa}
M.~Baggio, V.~Niarchos, and K.~Papadodimas, \emph{{tt$^{*}$ equations,
  localization and exact chiral rings in 4d $ \mathcal{N} $ =2 SCFTs}},
  \href{http://dx.doi.org/10.1007/JHEP02(2015)122}{JHEP {\bf 02} (2015)  122},
\href{http://arxiv.org/abs/1409.4212}{{\tt arXiv:1409.4212 [hep-th]}}.
%%CITATION = ARXIV:1409.4212;%%.

\bibitem{Baggio:2015vxa}
M.~Baggio, V.~Niarchos, and K.~Papadodimas, \emph{{On exact correlation
  functions in SU(N) $ \mathcal{N}=2 $ superconformal QCD}},
  \href{http://dx.doi.org/10.1007/JHEP11(2015)198}{JHEP {\bf 11} (2015)  198},
\href{http://arxiv.org/abs/1508.03077}{{\tt arXiv:1508.03077 [hep-th]}}.
%%CITATION = ARXIV:1508.03077;%%.

\bibitem{Gerchkovitz:2016gxx}
E.~Gerchkovitz, J.~Gomis, N.~Ishtiaque, A.~Karasik, Z.~Komargodski, and S.~S.
  Pufu, \emph{{Correlation Functions of Coulomb Branch Operators}},
  \href{http://dx.doi.org/10.1007/JHEP01(2017)103}{JHEP {\bf 01} (2017)  103},
\href{http://arxiv.org/abs/1602.05971}{{\tt arXiv:1602.05971 [hep-th]}}.
%%CITATION = ARXIV:1602.05971;%%.

\bibitem{Rodriguez-Gomez:2016ijh}
D.~Rodriguez-Gomez and J.~G. Russo, \emph{{Large N Correlation Functions in
  Superconformal Field Theories}},
  \href{http://dx.doi.org/10.1007/JHEP06(2016)109}{JHEP {\bf 06} (2016)  109},
\href{http://arxiv.org/abs/1604.07416}{{\tt arXiv:1604.07416 [hep-th]}}.
%%CITATION = ARXIV:1604.07416;%%.

\bibitem{Rodriguez-Gomez:2016cem}
D.~Rodriguez-Gomez and J.~G. Russo, \emph{{Operator mixing in large $N$
  superconformal field theories on S$^{4}$ and correlators with Wilson loops}},
  \href{http://dx.doi.org/10.1007/JHEP12(2016)120}{JHEP {\bf 12} (2016)  120},
\href{http://arxiv.org/abs/1607.07878}{{\tt arXiv:1607.07878 [hep-th]}}.
%%CITATION = ARXIV:1607.07878;%%.

\bibitem{Baggio:2016skg}
M.~Baggio, V.~Niarchos, K.~Papadodimas, and G.~Vos, \emph{{Large-N correlation
  functions in $ \mathcal{N} $ = 2 superconformal QCD}},
  \href{http://dx.doi.org/10.1007/JHEP01(2017)101}{JHEP {\bf 01} (2017)  101},
\href{http://arxiv.org/abs/1610.07612}{{\tt arXiv:1610.07612 [hep-th]}}.
%%CITATION = ARXIV:1610.07612;%%.

\bibitem{Papadodimas:2009eu}
K.~Papadodimas,
\emph{Topological Anti-Topological Fusion in Four-Dimensional Superconformal Field Theories},
 \href{http://dx.doi.org/10.1007/JHEP08(2010)118}{JHEP {\bf 1008} (2010) 118},
\href{http://arxiv.org/abs/0910.4963}{\tt arXiv:0910.4963 [hep-th]}.
%%CITATION = ARXIV:0910.4963;%%.

\bibitem{Gerchkovitz:2014gta}
  E.~Gerchkovitz, J.~Gomis and Z.~Komargodski,
  \emph{Sphere Partition Functions and the Zamolodchikov Metric},
  \href{http://dx.doi.org/10.1007/JHEP11(2014)001}{JHEP {\bf 1411} (2014) 001},
  \href{http://arxiv.org/abs/1405.7271 }{\tt arXiv:1405.7271 [hep-th]}.
%%CITATION = doi:10.1007/JHEP11(2014)001;%%

\bibitem{Wess:1992cp}
J.~Wess and J.~Bagger, {\em {Supersymmetry and supergravity}},
Princeton Series in Physics (1992).
%%CITATION = INSPIRE-350988;%%.

\bibitem{Usyukina:1993ch}
N.~I. Usyukina and A.~I. Davydychev, \emph{{Exact results for three and four
  point ladder diagrams with an arbitrary number of rungs}},
\href{http://dx.doi.org/10.1016/0370-2693(93)91118-7}{Phys. Lett. {\bf B305}
  (1993)  136--143}.
%%CITATION = PHLTA,B305,136;%%.

\bibitem{Penati:1999ba}
S.~Penati, A.~Santambrogio, and D.~Zanon, \emph{{Two-point functions of chiral
  operators in N=4 SYM at order $g^4$}},
  \href{http://dx.doi.org/10.1088/1126-6708/1999/12/006}{JHEP {\bf 12} (1999)
  006},
\href{http://arxiv.org/abs/hep-th/9910197}{{\tt arXiv:hep-th/9910197
  [hep-th]}}.
%%CITATION = HEP-TH/9910197;%%.

\bibitem{Penati:2000zv}
S.~Penati, A.~Santambrogio, and D.~Zanon, \emph{{More on correlators and
  contact terms in N=4 SYM at order $g^4$}},
  \href{http://dx.doi.org/10.1016/S0550-3213(00)00633-7}{Nucl. Phys. {\bf B593}
  (2001)  651--670},
\href{http://arxiv.org/abs/hep-th/0005223}{{\tt arXiv:hep-th/0005223
  [hep-th]}}.
%%CITATION = HEP-TH/0005223;%%.

\bibitem{Rossi:2008ay}
G.~C. Rossi and {\relax Ya}.~S. Stanev, \emph{{The massless supersymmetric
  ladder with L rungs}},
  \href{http://dx.doi.org/10.1016/j.nuclphysb.2008.07.030}{Nucl. Phys. {\bf
  B807} (2009)  534--546},
\href{http://arxiv.org/abs/0807.1964}{{\tt arXiv:0807.1964 [hep-th]}}.
%%CITATION = ARXIV:0807.1964;%%.

\bibitem{Eden:2011we}
B.~Eden, P.~Heslop, G.~P. Korchemsky, and E.~Sokatchev, \emph{{Hidden symmetry
  of four-point correlation functions and amplitudes in N=4 SYM}},
  \href{http://dx.doi.org/10.1016/j.nuclphysb.2012.04.007}{Nucl. Phys. {\bf
  B862} (2012)  193--231},
\href{http://arxiv.org/abs/1108.3557}{{\tt arXiv:1108.3557 [hep-th]}}.
%%CITATION = ARXIV:1108.3557;%%.

\bibitem{Eden:2012rr}
B.~Eden, \emph{{Three-loop universal structure constants in N=4 susy Yang-Mills
  theory}},
\href{http://arxiv.org/abs/1207.3112}{{\tt arXiv:1207.3112 [hep-th]}}.
%%CITATION = ARXIV:1207.3112;%%.

\end{thebibliography}
